\documentclass[rmp,aps,nofootinbib,twocolumn,endfloat]{revtex4-1}

\usepackage[final]{graphicx}
\usepackage{amsmath}
\usepackage{amsfonts}
\usepackage{amssymb}
\usepackage{color}

\newcommand{\tpsi}{\tilde{\psi}}
\newcommand{\td}{\tilde{d}}
\newcommand{\tPsi}{\tilde{\Psi}}

\newcommand{\trho}{\tilde{\rho}}

\newcommand{\expct}[1]{\left\langle #1 \right\rangle}
\newcommand{\expcts}[1]{\langle #1 \rangle}

\newcommand{\bra}[1]{\langle #1 |}
\newcommand{\ket}[1]{| #1 \rangle}
\newcommand{\ie}{{\it i.e.}}
\newcommand{\eg}{{\it e.g.}}
\newcommand{\ttheta}{\tilde{\theta}}
\newcommand{\tphi}{\tilde{\phi}}
\renewcommand{\th}{{\rm th}}
\newcommand{\VJ}{\vec{J}}

\renewcommand{\Re}{\operatorname{Re}} 
\renewcommand{\Im}{\operatorname{Im}}

\begin{document}

\title{One-Dimensional Quantum Liquids: Beyond the Luttinger Liquid Paradigm}

\author{Adilet~Imambekov}
 \affiliation{Department of Physics and
Astronomy, Rice University, Houston, TX, 77005, USA}

\author{Thomas~L.~Schmidt}
\email{thomas@thoschmidt.de}
 \affiliation{Department of Physics, Yale
University, New Haven, Connecticut 06520, USA}

\author{Leonid~I.~Glazman}
\email{leonid.glazman@yale.edu}
 \affiliation{Department of Physics, Yale
University, New Haven, Connecticut 06520, USA}
\date{\today}

\begin{abstract}

For many years, the Luttinger liquid theory has served as a useful paradigm for the
description of one-dimensional (1D) quantum fluids in the limit of low energies. This
theory is based on a linearization of the dispersion relation of the particles constituting
the fluid. We review the recent progress in understanding 1D quantum fluids beyond
the low-energy limit, where the nonlinearity of the dispersion relation becomes essential.
The novel methods which have been developed to tackle such systems combine
phenomenology built on the ideas of the Fermi edge singularity and the Fermi liquid theory, perturbation theory in the interaction strength, and new ways of treating finite-size properties of integrable models. These methods can be applied to a wide variety of 1D fluids, from 1D spin liquids to electrons
in quantum wires to cold atoms confined by 1D traps. We review existing results for
various dynamic correlation functions, in particular the dynamic structure factor and the
spectral function. Moreover, we show how a dispersion nonlinearity leads to finite particle
lifetimes, and its impact on the transport properties of 1D systems at finite
temperatures is discussed. The conventional Luttinger liquid theory is a special limit of the new
theory, and we explain the relation between the two.

\end{abstract}

\date{\today}

\maketitle
\tableofcontents

\section{INTRODUCTION}
\label{sec:intro}

The conventional description of the low-energy properties of quantum
condensed matter uses the notion of quasiparticles: elementary
excitations behaving as free quantum particles with some energy
spectrum which depends on the microscopic interactions. The low-energy properties of
interacting electrons in normal metals, for example, are
well represented by the theory of a Fermi liquid~\cite{nozieres97}. Its
elementary excitations are similar to free fermions. One may view the
quasiparticle states as those evolving from free fermions when
adiabatically turning on the interactions.

Quasiparticle states are labeled by their momenta, but their
dispersion relation (\ie, the quasiparticle energy $\xi$ measured from the Fermi level as a function of the
momentum $\bf k$) differs from the one for free fermions. These states
combine the best of the two worlds -- on the one hand, their overlap with
free fermions is significant. On the other hand, their lifetimes $\tau$
become infinitely long as their energy $\xi$
vanishes: in the absence of static disorder, $\tau\propto\xi^{-2}$. An
electron easily tunnels into a metal, ``dressing up'' in the tunneling
process to become a quasiparticle.  Neglecting the small relaxation
rate $1/\tau$, an electron entering a Fermi liquid with momentum $\bf
k$ creates a single quasiparticle with a well-defined energy $\xi({\bf k})$.
In an inverse process, an electron may tunnel out, leaving behind
a hole with well-defined energy.
In either case, the
tunneling probability per unit time of an electron with
given momentum ${\bf k}$ and energy $\varepsilon$ or, more precisely,
its spectral function $A({\bf k},\varepsilon)$ is close to
a delta-function, $A({\bf k},\varepsilon)\propto\delta
(\varepsilon-\xi({\bf k}))$. Residual interactions between the
quasiparticles can be readily accounted for within conventional
perturbation theory. A perturbative evaluation of the
quasiparticle's self-energy leads to a finite relaxation rate
$1/\tau$ and to a slight broadening of the spectral function,
transforming it into a Lorentzian with a width
$\delta\varepsilon\sim 1/\tau$.

If the momentum is not conserved in the tunneling event (this happens, for
example, if the electron tunnels through a point contact), then the
electron extracted from the Fermi liquid leaves behind a
superposition of holes, each of them having the same energy
$\varepsilon$. The area of the constant-energy surface in momentum
space defined by the equation $\xi({\bf k})=\varepsilon$ determines the
density of states $\nu(\varepsilon)\propto\int d{\bf k} A({\bf k},
\varepsilon)$ available for tunneling~\cite{nozieres97} at energy
$\varepsilon$. Similar to the case of free fermions, the tunneling density of
states at the Fermi level $\nu(0)$ in the Fermi liquid is
finite.

One can find a more subtle example of a quasiparticle description in
the Bogoliubov treatment of the excitations of a Bose gas with
weak inter-particle repulsion~\cite{pitaevskii03}. Bogoliubov quasiparticles are the
bosonic excitations above the Bose condensate; their spectrum differs
qualitatively from the spectrum of ``bare'' bosons: due to the
interaction between the ``bare'' particles, the Bogoliubov
quasiparticles are characterized by a sound-like spectrum, $\xi ({\bf
  k}) \propto k$, at low energies. Still, a particle entering the
system of interacting bosons easily ``dresses up'' to become a
quasiparticle. The spectral function $A({\bf k},\varepsilon)$ of a
boson with momentum ${\bf k}$ is close to a delta-function centered at
energy $\varepsilon=\xi({\bf k})$. At small energies, the lifetime of a
Bogoliubov quasiparticle~\cite{beliaev58} with energy $\xi$ diverges
as $\tau\propto 1/\xi^5$, making the quasiparticle states
well-defined. The rate $1/\tau $ determines the broadening of the
spectral function.

In the above examples, the affinity between the free particle and an
elementary excitation of the many-body system is exemplified by the
narrow energy width of the spectral function $A({\bf k},\varepsilon)$:
the spectral weight is concentrated around the quasiparticle energy
$\varepsilon=\xi({\bf k})$ within a region
$\delta\varepsilon\ll\xi({\bf k})$. While working well in higher
dimensions, this picture fails in the case of a one-dimensional (1D)
gas of quantum particles. This is clearly visible, \eg, for 1D
fermions in the presence of even a weak interaction between them. The
correction to the fermionic spectral function to the second (lowest nonvanishing) order in the interaction potential decays as
$1/[\varepsilon-\xi (k)]$~\cite{dzyaloshinskii74}, transferring the spectral weight away from the quasiparticle mass shell
$\varepsilon=\xi(k)$.  Along with the slow decay of the spectral
function, the second-order corrections to the tunneling density of
states $\nu (\varepsilon)$ at $\varepsilon=0$ and to the momentum
distribution function $n(k)\propto\int^0_{-\infty}d\varepsilon
A(k,\varepsilon)$ at the Fermi points $k=\pm k_F$, are singular.

The ``magic bullet'' effective in resolving many of the difficulties
of the 1D quantum many-body problem was suggested in a seminal paper
by Tomonaga~\cite{tomonaga50}. It was noticed there that replacing the generic dispersion relation $\xi(k)$ of 1D fermions
with a linear one, $\xi(k)=\pm v_F(k\mp k_F)$, immensely simplifies
finding the many-body dynamics of the system (here the upper/lower
signs correspond to the right-/left-moving particles, and $k_F$ is the Fermi momentum). For free fermions with
a linear spectrum, the energy $E$ of a right-moving excitation
consisting of an arbitrary number of particle-hole pairs with a given
total momentum $q$ depends only on that total momentum, $E=v_Fq$. This
degeneracy allowed Tomonaga to encode the excitations of the Fermi gas
(at fixed numbers of left- and right-movers) into the excitation
spectrum of free bosons. These 1D bosons are nothing but quantized
waves of density of a 1D Fermi gas, and their description is identical
to that of acoustic phonons~\cite{ziman60}. The beauty of the encoding is
that it puts the free-fermion Hamiltonian and
density-density interactions on an equal footing. Indeed, the full
Hamiltonian of interacting 1D fermions now becomes a bilinear form in
bosonic creation-annihilation operators. Its diagonalization is
standard and not different from the corresponding procedure for
phonons~\cite{ziman60}. This way, the Hamiltonian for interacting 1D
fermions with linear spectrum is diagonalized and cast in terms of
free bosons with linear dispersion relation, $\omega(q)=v|q|$ (we use units $\hbar=1$ throughout the text);
the velocity $v$ differs from $v_F$ because of the interactions.

The bosonic representation introduced by Tomonaga makes the calculation
of the density correlation function of 1D fermions straightforward. Indeed,
the density fluctuation operator $\rho(x,t)$ is linear in boson
creation/annihilation operators. According to the diagonalized form
of the Hamiltonian, those represent excitations which propagate freely
with a constant velocity $v$. This is reflected in the dynamic
structure factor (DSF), which is defined as the probability per unit time to excite a density fluctuation by an external source coupled to $\rho$. The DSF
takes the form $S(q,\omega)\propto |q|\delta(\omega-v|q|)$.

The evaluation of the propagator of a fermion, its spectral function
$A(k,\varepsilon)$, the tunneling density of states $\nu(\varepsilon)$,
or the distribution function $n(k)$ is somewhat more complicated.
\textcite{luttinger63} attempted to evaluate the distribution function
using the fermionic representation; some aspects of that calculation
were clarified later by~\textcite{mattis65}. The spin-$1/2$ fermion
propagator $G(x,t)$ in  space and time domain was evaluated for the
Tomonaga-Luttinger model by~\textcite{dzyaloshinskii74}. Their
diagrammatic technique heavily relied on the linear dispersion
relation of fermions. \textcite{luther74} evaluated $G(x,t)$ with the
help of the bosonic representation of the Tomonaga-Luttinger model~\cite{luther74,mattis74},
and also showed that the tunneling density of states
$\nu(\varepsilon)$ displays a power-law behavior at energies
$\varepsilon$ much smaller that the Fermi energy. Similar
techniques have been used to calculate the long-distance behavior of correlation
functions for 1D bosonic systems~\cite{efetov75}.

At low energies, the excitations of
noninteracting fermions are particles and holes with momenta $k$ in
the vicinity of $\pm k_F$. The energy of, say, a right-mover is $\xi(k)=v_F(k-k_F)+(k-k_F)^2/(2m^*)$. The second term here may be
considered as the lowest-order expansion of a general nonlinearity of
the dispersion relation (in case of particle-hole symmetry, such
an expansion would start from the cubic $(k-k_F)^3$ term).  For
noninteracting fermions with Galilean-invariant spectrum, $m^*$
is equal to the bare mass. The quadratic term here scales as
$\xi^2/(m^*v_F^2)$ at small $\xi$: by a power-counting argument, this
term is irrelevant. Indeed, it has been shown by~\textcite{haldane81a}
that the spectrum nonlinearity does not significantly affect the
long-range behavior of the fermion propagator $G(x,0)$ at fixed time
($t=0$). That gives an incentive to dispense with the nonlinearity
of the dispersion relation. After that, a wide variety of 1D systems
with gapless excitation spectra can be mapped, at low energies, on the
Tomonaga-Luttinger model.
Thus, the Tomonaga-Luttinger model provided the foundation for the concept of
the Luttinger liquid (LL), a phenomenological description of the low-energy
excitations of interacting quantum particles confined to
1D~\cite{haldane81a,haldane81b}.
In addition to liquids of fermions or
bosons, these include also the low-energy excitations of
half-integer spin chains, i.e., spin liquids~\cite{haldane80}.

After a real system is replaced by the corresponding LL,
the ``good'' low-energy excitations appear to be waves of {\it
density} of the corresponding liquid which have a linear spectrum
$\omega(q)=v|q|$. These excitations propagate along the $x$-axis with fixed velocities $\pm
v$ and without any dispersion: a perturbation
created at some point $x$ propagates without changing its shape to
the points $x\pm vt$. The presence of the formally  irrelevant term
$(k-k_F)^2/(2m^*)$ in the spectrum results in a dispersion of the propagating
perturbation. In analogy with single-particle quantum mechanics, the width of the perturbation
grows with time $\propto\sqrt{t/m^*}$.
The irrelevance of the quadratic term in $\xi(k)$ means
that the growth is slow compared to the rate of linear displacement of
the perturbation: $\sqrt{t/m^*}/(vt)\to 0$ at $t\to\infty$. The
infinitely ``sharp'' DSF of the Tomonaga-Luttinger model,
$S(q,\omega)\propto |q|\delta (\omega-v|q|)$, reflects the
propagation of the density perturbation with fixed velocities $\pm
v$. The $\propto\sqrt{t/m^*}$ dispersion of the perturbation
corresponds to some kind of broadening of the DSF: at given
$|q|\approx\omega/v$ the characteristic width of the DSF is
$\delta\omega\sim\omega^2/(m^*v^2)$.

It may look like the designation of density waves in 1D as the proper
excitations parallels the Fermi-liquid idea in higher dimensions.
Indeed, dispensing with irrelevant perturbations in these two systems,
 one finds a delta-function structure of $S(q,\omega)$ for 1D bosons and
similarly of $A({\bf k},
\varepsilon)$ for Fermi quasiparticles in higher
dimensions. From a dimensional analysis, one expects
the irrelevant terms to broaden these delta-functions by
$\propto\omega^2$ and $\varepsilon^2$, respectively, at given $q$ and ${\bf k}$.
However, the similarity stops there. In a Fermi liquid, irrelevant
interactions lead to a
self-energy with small imaginary part in the quasiparticle Green function, which transforms $A({\bf
  k},\varepsilon)$ to a Lorentzian. The self-energy can then be
evaluated within perturbation theory in the irrelevant
terms~\cite{abrikosov63}. In contrast, a na\"ive attempt to use perturbation theory
to evaluate the self-energy of bosons and thus ``broaden up'' $S(q,\omega)$ in the LL
is doomed~\cite{andreev80,samokhin98,punk06,pirooznia07}. The symmetry (in fact, Lorentz-invariance)
introduced by the linearization results in a degeneracy of the
excitation spectrum. The terms describing the curvature of
the dispersion relation break that symmetry. The very same
simplification which allowed~\textcite{tomonaga50}  to find the exact solution of the
problem makes the perturbation theory in the curvature strongly
degenerate. The nominally irrelevant terms remove the degeneracy,
lead to the emergence of new qualitative behavior of the
dynamic correlation functions, and to relaxation processes which do not exist in the linear
LL.

The perturbation theory in curvature is plagued by on-shell divergencies, and is in need of a proper resummation procedure, yet to be developed~\cite{samokhin98,aristov07}. A possible way of building a perturbation theory in boson representation may involve studying the
responses in an off-shell domain and comparing the results to the known limits, such as the free fermions~\cite{pereira06,pereira07,teber06,teber07}.

A progress, however, was achieved along a different route, via linking
the properties of a nonlinear LL to the well-known Fermi edge
singularity effect~\cite{mahan81,nozieres69,ohtaka90}. That connection was quite easy to
notice for weakly-interacting fermions with a generic (nonlinear)
spectrum~\cite{pustilnik06a}. Further development of that
relation put the problem of edge singularities in
a nonlinear LL into the class of so-called ``quantum impurity'' models~\cite{khodas07a,khodas07b,balents00,lamacraft08,lamacraft09,pereira08,pereira09,pereira10,cheianov08,imambekov08,affleck09,imambekov09a,imambekov09b,schmidt10a,schmidt10b,zvonarev09a,zvonarev09b,tsukamoto98,sorella96,castella93,castroneto94,castroneto96}.

The use of the field-theoretical approach based on quantum impurity models has led eventually to a
new phenomenological theory of nonlinear LLs. Remarkably, the threshold power-law singularities in the dynamic responses [$S(q,\omega)$ and $A(k,\varepsilon)$ are the examples] occur not only
in the vicinity of special points ($k=\pm k_F$ for the spectral
function), but at arbitrary momenta. The nonlinear LL phenomenology
relates the exponents characterizing the singularities to the properties of the threshold
spectra, thus establishing a relation between two sets of
independently measurable quantities.

The phenomenology also provides effective tools for the evaluation of
the exponents characterizing the dynamic responses: one may find first the
energy spectra
and then use the phenomenological relations to find the
exponents. A class of systems for which such a program is
especially attractive are the models exactly solvable by the Bethe ansatz.
The thermodynamic Bethe ansatz is well suited for the evaluation of the spectra but not for the dynamic responses. However, the corresponding exponents
can be found exactly by combining the thermodynamic Bethe ansatz with a field-theoretical description~\cite{imambekov08,pereira08,pereira09, cheianov08,essler10}.
Exactly solvable models provide stringent nontrivial tests of the field-theoretical approaches.

The linear LL theory does not discriminate between integrable and generic 1D systems: in either case, the original system is replaced by free particles devoid of any relaxation mechanisms.
Methods emerging within the nonlinear theory pave a way of studying the kinetics of a 1D quantum liquid and see the differences between generic and integrable systems.

In summary, understanding 1D quantum liquids outside the sector of low-energy excitations requires breaking the spell of linearization. The emerging theory, which accounts for the nonlinear energy spectrum of particles forming the liquid, answers that challenge. We review a number of methods of the nonlinear LL
theory and expose relations between them. These methods have already led to a progress in understanding the dynamic responses and relaxation of 1D quantum liquids. The approaches we describe are controllable, yet versatile enough for application to a broad class of systems, from electrons in quantum wires \cite{deshpande10}, to spin liquids and cold atoms confined to 1D.

This review is organized as follows. In Sec.~\ref{sec:singular}, we introduce and develop in detail the general field-theoretical approach to describe the singularities of
dynamic response functions in the momentum-energy plane. That approach is based on the phenomenology of effective mobile impurities moving in LLs. In Sec.~\ref{sec:exact}
we combine this field-theoretical approach with the analysis of exactly solvable models, which allows one to obtain a plethora of new results for the latter,
and provides stringent non-perturbative checks of the phenomenological approaches. In Sec.~\ref{sec:kinetics}, we illustrate the importance of the physics beyond the linear LL
theory for the kinetics and the transport in 1D quantum liquids.

\section{SINGULARITIES OF THE DYNAMIC RESPONSE FUNCTIONS}
\label{sec:singular}

An adequate description of a quantum many-body system not only requires
an understanding of the ground state, but also a characterization of
its excitations. One of the most natural ways to probe the excitation spectrum
is to measure the dynamic responses of the system to external
fields, such as electromagnetic radiation of a given momentum and energy.
Within Fermi's golden rule, the scattering rate of such external fields is
related to various dynamic response functions, such as the DSF or the spectral function (see precise definitions below).
This motivates the interest in studying qualitative features of the dynamic response functions of low-dimensional systems.
In particular, we will mostly be interested in their behavior near
the spectrum of elementary excitations. Many experimental techniques can be applied to
probe the dynamic responses of 1D systems, such as neutron  scattering~\cite{lake05,nagler91,lake09,thielemann09,tennant09,masuda06,zheludev08,ruegg05,zheludev02,stone03}, angle resolved photoemission spectroscopy (ARPES)~\cite{claessen02,sing03,hionkis05,kim06,wang06,wang09,kondo10,blumenstein11}, and various forms of
Bragg spectroscopy~\cite{stamper-kurn99,clement09,fabbri09,ernst10} and photoemission spectroscopy~\cite{stewart08,gaebler10,dao07}.
Additional interest in the response functions of 1D systems is driven by a rapid progress in their numerical evaluation~\cite{white08,feiguin09,barthel09,kokalj09,kohno10a} based on time-dependent extensions of Density Matrix Renormalization Group (DMRG) techniques~\cite{white92,dechiara08,schollwoeck05,vidal03,vidal04}.

A nonlinear dispersion relation and interactions between particles forming a 1D quantum
liquid modify in a nontrivial way all dynamical responses of the
liquid, resolved in energies and momenta. The DSF
provides one of the examples. It is defined as
\begin{equation}
S(q,\omega)
=\int_{-\infty}^\infty dt \int_{-\infty}^\infty dx e^{i(\omega
  t-qx)}\langle\rho(x,t)\rho(0,0)\rangle.
\label{eq:dsf}
\end{equation}
Here, $\rho (x,t)$ is the density operator and
the averaging $\langle\dots\rangle$ is performed over the Gibbs ensemble
or the ground state in the case of finite or zero temperature,
respectively. In the Tomonaga-Luttinger model, at small wavevectors
$q$ the DSF takes the form $S_{\rm LL}\propto |q|\delta(\omega-v|q|)$
at any temperature. The dispersion results in a ``broadening'' of the
delta-function.
Accounting for a finite width of $S(q,\omega)$ even for small $q$ is important~\cite{pustilnik03}
for understanding Coulomb drag experiments~\cite{debray01,debray02,yamamoto02,yamamoto06,laroche10}.
The broadening occurs even at zero temperature ($T=0$), and we will
concentrate on that case.

To illustrate the origin of the structure arising in
$S(q,\omega)$ due to the dispersion, let us consider first the simplest
case of free spinless fermions with a quadratic dispersion relation,
\begin{equation}
\xi(k)=\frac{k^2-k_F^2}{2m}\,.
\label{eq:freefermions}
\end{equation}
At zero temperature, the structure factor can be thought of as an
absorption coefficient, \ie, the dissipative part of the
linear susceptibility with respect to a perturbation $\delta{\cal
  H}=U(x,t)\rho(x,t)$ by a potential $U(x,t)$ varying in space
and time with the wave vector $q$ and frequency $\omega$,
respectively. In the case of free fermions, dissipation is caused by
creation of particle-hole pairs by the perturbing potential, see Fig.~\ref{fig:DSF}.
At $q < 2k_F$, a simple evaluation of Eq.~(\ref{eq:dsf}) yields
\begin{equation}
S_0(q,\omega)=
(m/q)\theta (q^2/(2m)-|\omega-v_Fq|)
\label{eq:dsf-freefermions}
\end{equation}
with the Fermi velocity $v_F=k_F/m$. The two thresholds for absorption
correspond to two special configurations in the momentum space of the
particle-hole pairs, see Fig.~\ref{fig:DSF}. Specifically, the lower boundary,
$\omega_-(q)=v_Fq-q^2/(2m),$ corresponds to a particle just above the
Fermi level, and a ``deep'' hole with momentum $k_F - q$, moving with velocity $v_h=v_F-q/m$,
smaller than $v_F$. Equation (\ref{eq:dsf-freefermions}) and Fig.~\ref{fig:DSF} allow us to make three
interesting observations:

\begin{figure}
\includegraphics[width=8cm]{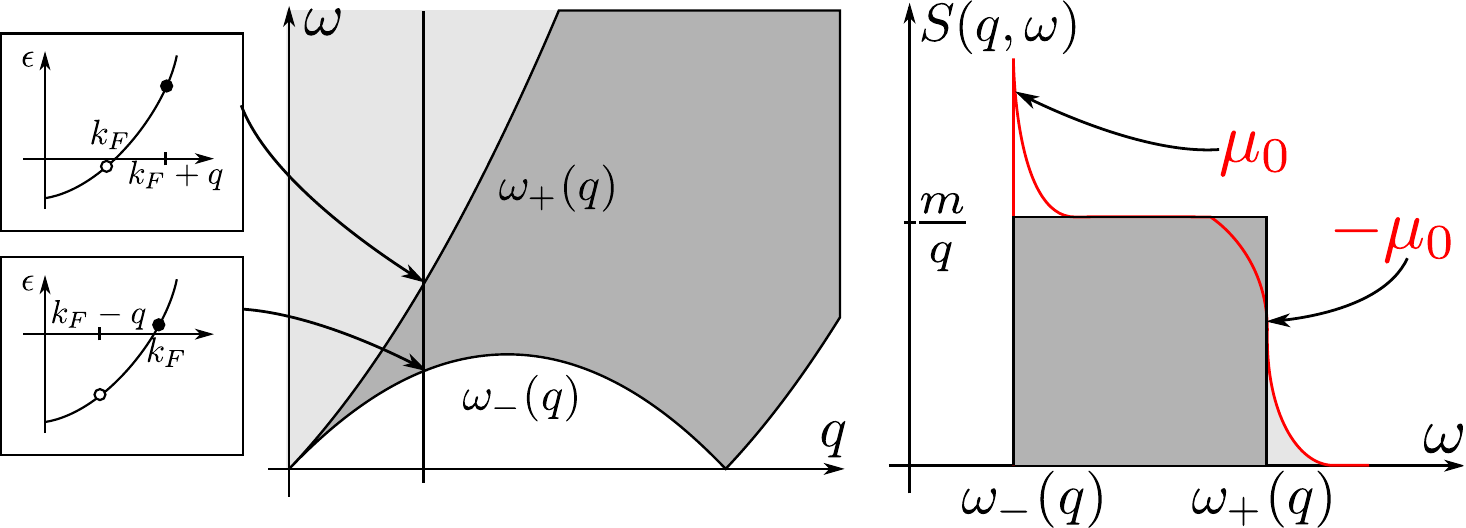}
\caption{\label{fig:DSF} (Color online) Density structure factor $S(q,\omega)$ for noninteracting fermions [see Eq.~(\ref{eq:dsf-freefermions})] and weakly interacting fermions [see Eqs.~(\ref{eq:perturb_S1}), (\ref{eq:perturb_S2}) and (\ref{eq:S_perturb_offshell})]. In the noninteracting case, $S_0(q,\omega)$ is constant in the dark shaded region and vanishes otherwise. \emph{Left: } Particle-hole configurations responsible for the upper and lower thresholds. \emph{ Right: } $S(q,\omega)$ for fixed $0 < q < 2 k_F$. In the noninteracting case, $S(q,\omega)$ has a rectangular shape. Interactions turn the steps into power-law singularities, and $S(q,\omega)$ becomes nonzero in the light shaded region. }
\end{figure}

{\it First}, at fixed $q \ll  2k_F$, the width of the structure factor in the
frequency domain $\delta\omega\sim\omega^2/(mv_F^2)$ scales as
$\propto\omega^2$. This is consistent with the power-counting argument
for the irrelevant curvature term in the spectrum $\xi (k)$. The limit
of linear spectrum (which is also a trivial limit of an LL
at zero inter-particle interaction) corresponds to taking $m\to\infty$
at fixed value of $v_F$. In this limit, $\delta\omega\to 0$.

{\it Second}, the dependence of $S_0(q,\omega)$ on its arguments is not
analytic; besides, if one allows an arbitrary sign of the mass $m$, it
becomes immediately clear that $\delta\omega \propto 1/|m|$ at a fixed
value of the Fermi velocity $v_F$. Each of these two facts kills the
hope to develop a simple perturbation theory in the irrelevant
perturbation, \ie, the curvature of the particles dispersion
relation. Indeed, it is clear from the above discussion that the
perturbing part of the Hamiltonian associated with the curvature is
proportional to $1/m$. To obtain a broadened structure factor, one
needs to evaluate the self-energy of the density propagator. The form
of Eq.~(\ref{eq:dsf-freefermions}) tells us that the imaginary part of
self-energy is $\propto 1/|m|$ and may result only from a summation of
some infinite series in $1/m$.

{\it Third}, the specific structure of the particle-hole pair corresponding
to the edge $\omega_-(q)$ gives us a hint at how weak interactions may
modify Eq.~(\ref{eq:dsf-freefermions}). Indeed, suppose fermions weakly
repel each other. Then, the created particle would be attracted to the
hole it left upon excitation. This interaction would lead to the Mahan
(excitonic) singularity in the absorption coefficient~\cite{mahan81}. That is, the
step-like threshold at $\omega=\omega_-(q)$ would transform into a
divergent power-law function with an exponent dependent on the
inter-particle repulsion strength.

The above simple picture establishes the relation, which is central for this section, of the nonlinear
LL problem to the well-studied problem of Fermi-edge singularities~\cite{nozieres69}. The latter is reviewed in great detail elsewhere~\cite{mahan81,gogolin98,ohtaka90}. We will see that the power-law
asymptote of the DSF at energies close to the threshold
is a robust feature valid at arbitrary interaction
strength and arbitrary $q$. We will also develop ways to evaluate the
corresponding exponent, which does depend on these parameters.

A nonlinear dispersion relation of interacting quantum particles
confined to 1D affects also their spectral
function $A(k,\varepsilon)$. The latter is defined as
\begin{equation}
A(k,\varepsilon)=-\frac{1}{\pi}{\rm Im}G(k,\varepsilon){\rm sign}\varepsilon
\label{eq:sf}
\end{equation}
with the Green's function~\cite{abrikosov63}
\begin{equation}
G(k,\varepsilon)= -i\int_{-\infty}^\infty dt \int_{-\infty}^\infty dx e^{i(\varepsilon t-kx)}\langle T[\Psi
(x,t)\Psi^{\dagger} (0,0)]\rangle.
\label{eq:gf}
\end{equation}
Here $\Psi(x,t)$ and $\Psi^{\dagger}(x,t)$ are the particle (fermion or boson) annihilation and creation operators, respectively, $T$ denotes the time ordering, and
the energy $\varepsilon$ is measured from the chemical potential.

The spectral function may be thought of as a tunneling density of
states: the probability for a particle (hole) with given momentum $k$ and
energy $\varepsilon > 0$ ($\varepsilon < 0$) to tunnel into a system is proportional to
$A(k,\varepsilon)$. It can be measured using ARPES~\cite{kim06,wang06,wang09,kondo10} for solid state systems or photoemission spectroscopy~\cite{stewart08,gaebler10}
for low-dimensional ultracold atomic systems~\cite{gorlitz01,paredes04,kinoshita04,kinoshita06}. It also determines electronic transport in systems with momentum and energy conserving
tunneling~\cite{auslaender02,auslaender05,jompol09,barak10}. In the absence of interactions, a particle with
a given momentum may tunnel only if its energy fits the dispersion
relation of the particles constituting the system,
$A_0(k,\varepsilon)=\delta[\varepsilon-\xi(k)]$. The right-hand side
here is the density of single-particle eigenstates with given energy
and momentum.

\begin{figure}[t]
\includegraphics[width=8cm]{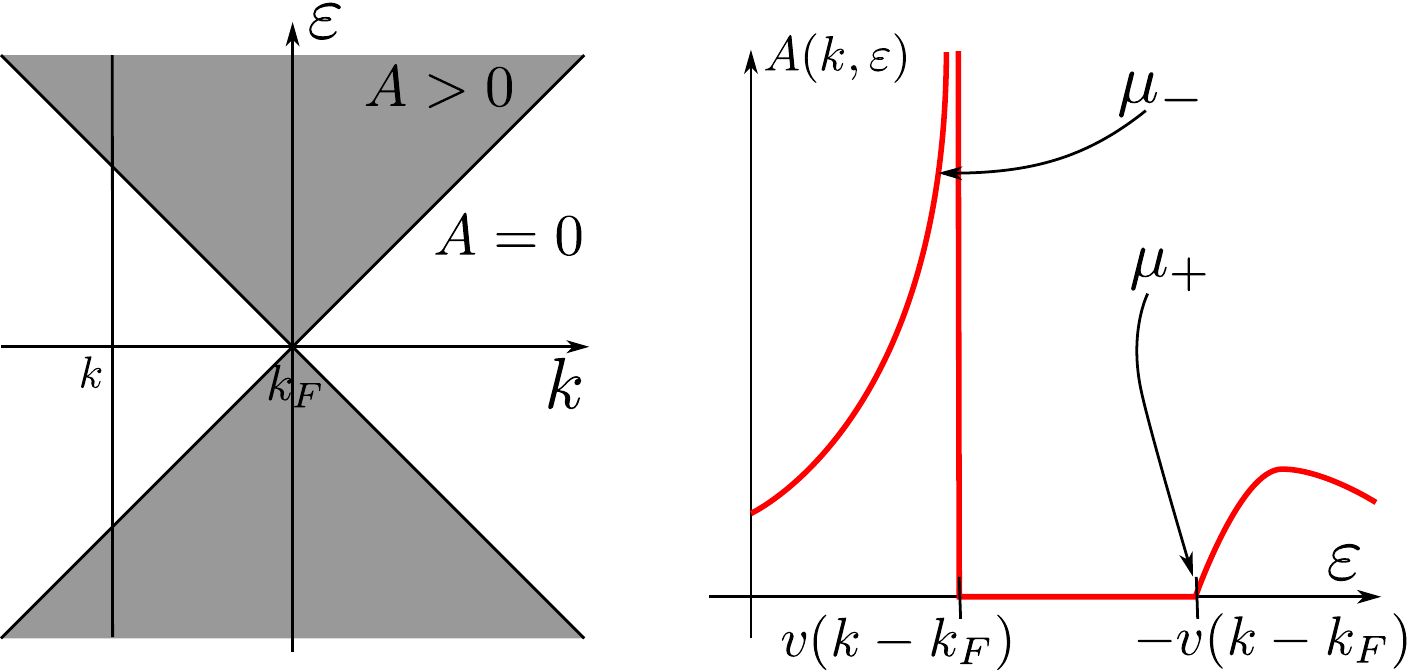}
\caption{\label{fig:spectral} (Color online) Spectral function $A(k \approx k_F,\varepsilon)$ for a Luttinger liquid (LL) with linearized spectrum. \emph{Left: } $A(k,\varepsilon)$ vanishes in the white regions. \emph{Right: } $A(k,\varepsilon)$ for fixed $k \lessapprox k_F$. Interactions cause a power-law divergence at the mass shell $\varepsilon \approx v(k - k_F)$ with exponent $\mu_- = 1 - \left( K + K^{-1} - 2 \right)/4$. A convergent power-law cusp with exponent $\mu_+ = - \left( K + K^{-1} - 2 \right)/4$ emerges at the inverted mass shell $\varepsilon \approx - v(k - k_F)$.
}
\end{figure}

Before considering the effects of a nonlinear dispersion, let us recall
the behavior of $A(k,\varepsilon)$ in a fermionic LL.
In the absence of interactions, the tunneling density of states
for, say, right-movers is
$A(k,\varepsilon)=\delta[\varepsilon-v(k-k_F)]$. Interactions between
the particles forming the LL broaden the spectrum of
energies at which tunneling is possible. In the vicinity of the Fermi
point $+k_F$ one has~\cite{luther74,schonhammer92,voit93a,voit93b,voit95}
\begin{eqnarray}
A(k,\varepsilon)&\propto&{\rm sign}(\varepsilon)\frac{\theta[\varepsilon^2-v^2(k-k_F)^2]}
{\varepsilon-v(k-k_F)}
\nonumber\\
&\times&\left[\varepsilon^2-v^2(k-k_F)^2\right]^{\frac{1}{4}\left(K+\frac{1}{K}\right)-\frac{1}{2}}\,.
\label{eq:sf-luttinger}
\end{eqnarray}
The shape of $A(k,\varepsilon)$ for a linear LL is shown in Fig.~\ref{fig:spectral}.
Here, the LL parameter $K$ depends on the interaction
strength ($K<1$ for repulsion); free fermions correspond to the limit
$K\to 1$. The delta-function in the tunneling density of
states of free particles got transformed into a power-law, divergent
at the line $\varepsilon=v(k-k_F)$ if the interaction is not too
strong, see Eq.~(\ref{eq:sf-luttinger}). Note that an important feature of the linear LL
result is the particle-hole symmetry under the transformation
\begin{align}
\varepsilon &\to -\varepsilon, \notag  \\
(k-k_F) &\to -(k-k_F),\label{eq:phsymmetry}
\end{align}
which is a necessary consequence of the spectrum linearization.

\begin{figure}[t]
\includegraphics[width=0.99\columnwidth]{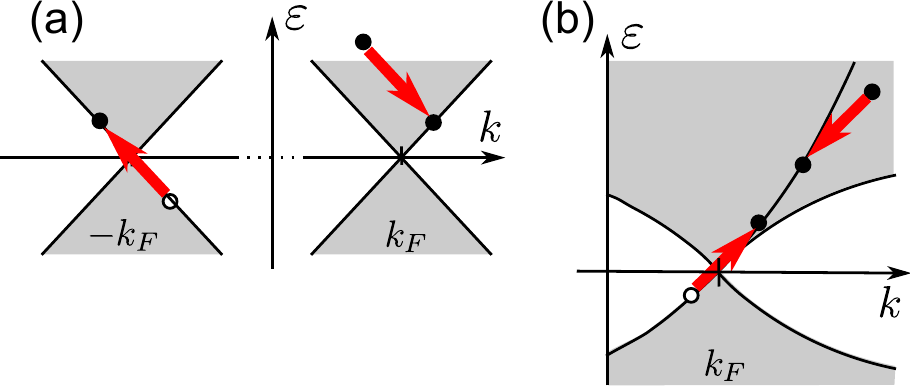}
\caption{\label{fig:shadow} (Color online) (a) An incoming particle with momentum $k \approx k_F$ and energy above the mass shell can tunnel into the system by creating a particle with momentum near $k_F$ on the mass shell and a low-energy particle-hole pair near the \emph{opposite} Fermi point. (b) If the spectrum is curved, an incoming particle with momentum $k \gtrapprox k_F$ and energy \emph{below} the mass shell can tunnel into the system by creating a particle on mass shell and a low-energy particle-hole pair near the \emph{same} Fermi point.
}
\end{figure}

The nonzero values of $A(k,\varepsilon)$ outside the line
$\varepsilon=v(k-k_F)$ may be understood within the perturbation
theory if one invokes the notion of tunneling probability. In the
presence of interactions between particles, the tunneling free particle
with energy $\varepsilon>v|k-k_F|$ may spend the extra energy on
the creation of a particle-hole pair on the branch of left-movers
and land on the mass shell $\xi(k)=v(k-k_F)$ for the
right-moving particles. This process is depicted in
Fig.~\ref{fig:shadow}a. Similarly, the tunneling of a particle at energy $-\varepsilon>v|k-k_F|$ {\it out}
of the system creates a right-moving hole
and left-moving particle-hole pair. To summarize, in this
perturbative picture tunneling of a particle into (out of) the
system creates three excitations: a right-moving particle (hole) and a
particle-hole pair on the {\it opposite} branch.

The above perturbative consideration is easy to carry over to
fermions with the nonlinear dispersion relation~(\ref{eq:freefermions}).
We see immediately the difference between the particle-like ($k>k_F$)
and hole-like parts of the spectrum. The hole-like part of the
spectrum becomes the energy threshold $\xi(k)<0$ for
tunneling at $|k|<k_F$. Indeed, because the hole's velocity $|k/m|<v_F$,
it is not capable of emitting ``Cherenkov radiation'' of low-energy particle-hole pairs in
either of the two allowed directions. At energies
$\varepsilon<\xi(k)<0,$ the tunneling of a hole is
accompanied by the creation of particle-hole pairs at both Fermi points.

However, as shown in Fig.~\ref{fig:shadow}, the particle-like part of the spectrum~(\ref{eq:freefermions})
falls into a continuum of energies available for tunneling at given
$|k|>k_F$. To see this, we may again concentrate on the tunneling of a
right-moving free particle with momentum $k>k_F$. Unlike in the case of
a linear spectrum, now a particle with energy $0<\varepsilon<\xi(k)$
may tunnel by creating a comoving particle and a particle-hole
pair. The total of three excitations should have the momentum $k$, but the
sum of their energies is {\it less} than $\xi(k)$ if all three momenta
are within the region of width $k-k_F$ around $k_F$. Within
perturbation theory, we find the threshold in this region at  $-\xi(k-2k_F)$
for $k_F<k<3k_F$, see Fig.~\ref{fig:shadow}b. Consequently, states at the
free-particle mass shell $\xi(k)$ at $k>k_F$ are not protected by
kinematics: particles move fast enough ($|k/m|>v_F$) to allow
Cherenkov radiation of particle-hole pairs.

A comparison of the perturbative pictures for the linear and nonlinear
dispersion relations reveals some substantial ramifications introduced
by the nonlinearity. The nonlinearity destroys the particle-hole symmetry, Eq.~(\ref{eq:phsymmetry}),
which existed in the LL. The hole-like part of the
threshold morphs from a straight line
into a nonlinear function; the nature of excitations created by a
tunneling hole is not changed by the introduction of the
curvature. However, the particle-like part of the threshold changes
drastically; the excitations defining it have no counterpart in the
linear LL. In Sec.~\ref{sec:universal} we will review a universal nonlinear
LL theory valid in the vicinities of Fermi points~\cite{imambekov09a}.

We should emphasize that at $k\to\pm k_F$ the range of energies in
which $A(k,\varepsilon)$ is substantially modified compared to linear LL
theory becomes narrow, since it scales as $|\varepsilon-\xi(k)|\propto (k-k_F)^2/m$. At energies $|\varepsilon-\xi(k)|\gg
(k-k_F)^2/m$ the linear LL theory does indeed describe the structure of spectral function. This is consistent with the curvature being an irrelevant perturbation~\cite{haldane81a}.  However, the true threshold behavior of the spectral function is controlled by the nonlinear spectrum at {\it any} wavevector $k,$ close to or far away from $\pm k_F$.

The above discussion based on the perturbation theory tells us that the presence of thresholds in the dynamic response functions is protected by kinematics: a slow-moving excitation (a hole in the above consideration) can not decay in 1D. An inspection of Figs.~\ref{fig:DSF} and \ref{fig:shadow}b may raise a question, whether the response functions of a nonlinear LL
have singularities within the spectral continuum. The answer is: ``it depends''. In the case of integrable models, these singularities do survive, while in a generic liquid they are broadened and get progressively washed out if one moves away from the Fermi points. We discuss the singularities in the continuum in Sec.~\ref{sec:exact} devoted to the integrable models, and the broadening of singularities by relaxation processes in Sec.~\ref{sec:kinetics} devoted to the kinetics of nonlinear LLs.

The remainder of this section is organized as follows. In Sec.~\ref{sec:perturbative} we make the above perturbative considerations quantitative and derive the effective mobile-impurity Hamiltonian for weakly interacting spinless fermions. The perturbation
theory gives a clear hint on how to proceed with the
calculation of the threshold singularities at arbitrary interaction
strength. In Sec.~\ref{sec:universal} we explain the theory of a
nonlinear LL at arbitrary interactions in the vicinity of the points $k=\pm k_F$,
considering in detail the crossover between the generic threshold
behavior and the linear LL asymptote. The adequate apparatus based on a mobile quantum impurity
moving in a linear LL is extended further, and
we develop the phenomenology of the threshold behavior of the dynamic
responses for spinless fermions (Sec.~\ref{sec:phenomenology}), spin liquids (Sec.~\ref{sec:spinchains}), bosonic systems (Sec.~\ref{sec:bosonic}), as well as for spinful fermionic systems (Sec.~\ref{sec:spinful}). Next, Sec.~\ref{sec:finite} is devoted to effects which arise due to a finite system size and finite temperatures. Finally, in Sec.~\ref{sec:realspace} we discuss the implications of the threshold singularities for correlation functions in the space-time domain and for the breakdown of conformal invariance.

\subsection{Perturbative treatment of interactions}

\label{sec:perturbative}

\subsubsection{Dynamic structure factor (DSF)}

\label{sec:perturbative_DSF}

The DSF $S(q,\omega)$ characterizes the linear response of the density to an external field which couples to the density. The absorption rate for quanta of momentum $q$ and energy $\omega$ of such a field (\eg, photons) is proportional to $S(q,\omega)$. The rate is given by Fermi's golden rule,
\begin{align}\label{eq:S_FGR}
 S(q,\omega) &= \frac{2\pi}{L} \sum_{\ket{f}} | \bra{f} \rho^\dag_q \ket{0} |^2 \delta(\omega - \epsilon_f).
\end{align}
The equivalence of Eqs.~(\ref{eq:dsf}) and (\ref{eq:S_FGR}) can be demonstrated using the Lehmann spectral representation. Here, the system length is denoted by $L$, and the Fourier components of the
density are defined by $\rho^\dag_q = \sum_k \Psi^\dag_{k+q} \Psi_k$.

In noninteracting systems, the ground state $\ket{0}$ at zero temperature consists of a Fermi sea filled up to the Fermi momentum $k_F$. A nonzero matrix element $\bra{f} \rho^\dag_q \ket{0}$ emerges only for final states $\ket{f}$ which contain exactly one particle-hole pair with momentum $q$. All possible final states can thus be parameterized by the particle momentum $k_p$ and the hole momentum $k_h$,
\begin{align}
 \ket{f} = \Psi^\dag_{k_p} \Psi_{k_h} \ket{0},
\end{align}
where $|k_p| > k_F$, $|k_h| < k_F,$ and $k_p - k_h = q$. For the quadratic spectrum (\ref{eq:freefermions}), the energy of such a state is $\epsilon_f = (k_p^2 - k_h^2)/(2m)$. The evaluation of the DSF according to Eq.~(\ref{eq:S_FGR}) is then straightforward and yields
for $q < 2 k_F$,
\begin{equation}
S_0(q,\omega) =
\frac{m}{|q|}\theta [\omega - \omega_-(q)] \theta[\omega_+(q) - \omega].
\label{eq:dsf-freefermions2}
\end{equation}
It turns out that $S_0(q,\omega)$ is nonzero only within the interval $\omega_-(q) < \omega < \omega_+(q)$. The upper and lower edges of support physically correspond to final states $\ket{f}$ which have the highest and lowest possible energies, respectively, for a given momentum $q$. As shown in Fig.~\ref{fig:DSF}, for $q > 0$ the final state with the maximum energy contains a hole with momentum $k_h = k_F$ and a particle with momentum $k_p = k_F + q$. The energy of this state is $\omega_+(q) = v_F q + q^2/(2m)$. Similarly, the density excitation with the minimum energy for $q > 0$ contains a hole with momentum $k_h = k_F - q$ and a particle near the Fermi point, $k_p = k_F$. This state has the lower threshold energy $\omega_-(q) = v_F q - q^2/(2m)$. The width of support of $S_0(q,\omega)$ for fixed $q$ is therefore
\begin{align}\label{eq:perturb_deltaomega}
 \delta\omega(q) = \omega_+(q) - \omega_-(q) = \frac{q^2}{m}.
\end{align}
In view of a perturbative analysis, it is convenient to express the noninteracting DSF in terms of fermionic Green's functions. Using the fluctuation-dissipation theorem, the DSF at zero temperature can be related to the susceptibility~\cite{doniach98}, $S(q,\omega > 0) = -2 \Im \chi(q,\omega)$, where $\chi(q,\omega)$ is the Fourier transform of
\begin{align}
 \chi(x,t) = -i \theta(t) \expct{ [ \rho(x,t), \rho(0,0)]}{}.
\end{align}
For $\omega > 0$, the imaginary part of the retarded density-density correlation function $\chi(q,\omega)$ coincides with the imaginary part of the polarization diagram,
\begin{align}
    \mathcal{P}_0(q,\omega) = \raisebox{-4mm}{\includegraphics{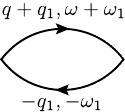}},
\end{align}
where solid lines denote time-ordered fermion Green's functions, and the internal momentum $q_1$ and the internal energy $\omega_1$ are integrated over. Therefore, for the noninteracting system $S_0(q,\omega > 0) = -2 \Im \mathcal{P}_0(q,\omega)$, yielding again Eq.~(\ref{eq:dsf-freefermions2}).

Next, let us calculate the correction to the DSF for a weak density-density interaction,
\begin{align}
 H_{int} = \frac{1}{2} \int dx dy \rho(x) V(x-y) \rho(y).
\end{align}
The first-order term $\delta S^{(1)}(q,\omega)$ can be cast into an RPA-like diagram and a vertex correction,
\begin{align}
 \delta S^{(1)}(q,\omega) &\propto \Im \bigg[\  \raisebox{-2mm}{\includegraphics{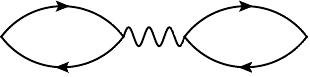}} \ \bigg] \notag \\
 &+
\Im \bigg[\  \raisebox{-2mm}{\includegraphics{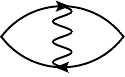}} \ \bigg].
\end{align}
The wiggly lines denote the interaction. Let us start with the first diagram and estimate its contribution towards the lower edge of support, $\omega \approx \omega_-(q)$. Its imaginary part is proportional to $V_q \Im \mathcal{P}_0(q,\omega) \Re \mathcal{P}_0(q,\omega)$. However, we showed that $\Im \mathcal{P}_0(q,\omega)$ is proportional to the noninteracting DSF, so $\Im \mathcal{P}_0(q,\omega) \propto \theta[ \omega - \omega_-(q)]$ contains a threshold. The Kramers-Kronig relation~\cite{landau80} therefore predicts a logarithm in the real part, $\Re \mathcal{P}_0(q,\omega) \propto \ln\{ [\omega - \omega_-(q)]/\delta \omega(q)\}$. So, the first diagram diverges logarithmically towards the edge. The second diagram can also be calculated and leads to an identical asymptote but with a prefactor $-V_0$. As a result, the total first order correction near the threshold reads
\begin{align}\label{eq:deltaS1_perturb}
 \delta S^{(1)}(q,\omega) = \frac{m^2(V_q - V_0)}{\pi |q|^2} \theta[\omega - \omega_-(q)] \ln\left[ \frac{\omega - \omega_-(q)}{\delta \omega(q)} \right].
\end{align}
Hence, a straightforward calculation of the first-order correction $\delta S^{(1)}(q,\omega)$ leads to a logarithmic threshold divergence. The underlying physical mechanism is reminiscent of the Fermi edge singularity problem \cite{mahan67,anderson67,nozieres69,schotte69,mahan81,ohtaka90,gogolin98}: the final state $\ket{f}$ for $\omega = \omega_-(q)$ and $0 < q < 2k_F$ contains a hole at momentum $k_F-q$ which generates a scattering potential. The infrared divergence is produced by scattering of particles near the Fermi points with small momentum exchange. In order to obtain a viable result, a partial resummation of the perturbation series is needed.

The analogy between the physics of the Fermi edge singularity and the threshold behavior of the DSF can be exploited \cite{pustilnik06a} by using a method which is familiar from the solution of the Fermi edge singularity problem by \textcite{schotte69}. In the calculation of $S(q,\omega)$ near the threshold $\omega \approx \omega_-(q)$, the hole with momentum $k_F - q$ assumes the role of a ``deep'' hole. The energy left for additional excitations $|\omega - \omega_-(q)|$ is small, so all particle-hole pairs created due to the interaction are restricted to a small window of width $k_0 \ll q$ around the Fermi points. The Hamiltonian can then be projected onto three subbands of width $k_0$, one centered around $k_F - q$ containing the deep hole and two containing the Fermi points $\pm k_F$. We note that this projection into subbands is in fact very similar to a conventional procedure employed in the Fermi edge singularity problem. In the latter, a full fermionic operator is split into contributions from non-overlapping conduction and core-hole bands. In our case, all fermions are in the same band, but since the momenta of the important states do not overlap due to kinematic constraints, splitting the fermionic operator into subbands is a legitimate procedure. In contrast to the original Fermi edge problem, the deep hole is mobile but this does not destroy the edge singularity \cite{ogawa92,balents00}.

Let us illustrate the calculation for $0 < q <2 k_F$. In this case, the fermion annihilation operator $\Psi(x)$ is projected onto the three subbands using
\begin{align}\label{eq:Psi_projection}
 \Psi(x) \to e^{i k_F x} \psi_R(x) + e^{-i k_F x} \psi_L(x) + e^{i k x} d(x),
\end{align}
where $k=k_F - q$ lies within the Fermi sea and the projected operators $d(x)$ and $\psi_{R,L}(x)$ have nonzero Fourier components only within the narrow bandwidth $k_0$. Using this projection in the definition of the DSF and retaining only Fourier components close to $q$ leads to
\begin{align}\label{eq:Sqomega_projected}
& S(q,\omega)
= \int dx dt e^{i \omega t - i q x} \expcts{ \Psi^\dag(x,t) \Psi(x,t) \Psi^\dag(0,0) \Psi(0,0)}{} \notag \\
&= \int dx dt e^{i \omega t} \expcts{ d^\dag(x,t) \psi_R(x,t) \psi^\dag_R(0,0) d(0,0) }{}.
\end{align}

The next step is to project the interacting system Hamiltonian onto the narrow subbands. Because the reduced bandwidth $k_0$ is small, the spectrum within each of the subbands can be linearized. This makes it convenient to bosonize $\psi_{R,L}$ using
\begin{align}\label{eq:perturb_bosonize}
 \psi_{R,L}(x) \propto e^{-i [\pm \phi(x) - \theta(x)]},
\end{align}
and to write the corresponding terms in the Hamiltonian in the bosonic basis. In Eq.~(\ref{eq:perturb_bosonize}), the conventional bosonic fields $\theta$ and $\phi$ satisfy a canonical commutation relation (we use the notations of \textcite{giamarchi04} thoughout the text),
\begin{align}\label{eq:canonical_commutation}
  [\phi(x), \nabla \theta(x')] = i \pi \delta(x-x').
\end{align}
After a projection of the microscopic interactions onto subbands and bosonization, the Hamiltonian becomes $H = H_0 + H_d + H_{int}$, where
\begin{align}\label{eq:A_MI}
H_0 &= \frac{v_F}{2\pi} \int dx [ (\nabla \theta)^2 + (\nabla \phi)^2 ], \\
 H_d &= \int dx d^\dag(x) [\xi(k) - i v_d \nabla ] d(x), \notag \\
 H_{int} &= \int dx \left[ (V_{k-k_F} - V_0) \rho_R + (V_{k + k_F} - V_0) \rho_L \right] d d^\dag.\notag
\end{align}
The term $H_0$ describes the kinetic energy of the particles near the Fermi points. The energy of the impurity $d$ is close to $\xi(k),$  its motion is described by $H_d$, and its velocity reads
\begin{align}
 v_d = \frac{\partial \xi(k)}{\partial k} =\frac{k}{m}= v_F - \frac{q}{m}.
\end{align}
Last but not least, the term $H_{int}$ contains the density-density interactions between the impurity and the particles near the Fermi points. The densities of right- and left-movers are  given by
\begin{align}\label{eq:LL_rho_boson}
 \rho_{\alpha}(x) = \frac{1}{2\pi} \nabla (- \phi + \alpha \theta),
\end{align}
where $\alpha=R,L=+,-.$
Interactions lead to the formation of low-energy particle-hole pairs and are thus crucial for the shape of the
DSF. \textcite{schotte69} showed that $H_{int}$ can be removed using a unitary transformation. Indeed, introducing the unitary operator
\begin{align}\label{eq:U_perturb}
 U = \exp\left\{ i \int dx \left(\frac{\delta_+}{2\pi}  [\theta - \phi] -\frac{\delta_-}{2\pi}  [\theta + \phi]\right) d d^\dag \right\}
\end{align}
one finds $U^\dag (H_0 + H_d + H_{int}) U = H_0 + H_d$ by using the phase shifts
\begin{align}
 \delta_+ = \frac{V_0-V_{k-k_F} }{v_d - v_F}, \; \delta_- = \frac{V_0-V_{k+k_F} }{v_d + v_F}. \label{eq:deltas_pert}
\end{align}
Physically, the values of $\delta_\pm(k)$ correspond to the scattering phase shifts between the deep hole and the low-energy particles near the right and left Fermi points in the Born approximation. In order to calculate $S(q,\omega)$ using Eq.~(\ref{eq:Sqomega_projected}), the same unitary transformation has to be applied to the operators $\psi_R(x)$ and $d(x)$. The impurity operator acquires a phase shift in the rotation,
\begin{align}\label{eq:Udag_d_U}
 U^\dag d(x) U = e^{i \left (\frac{\delta_+}{2\pi} [ \phi(x) - \theta(x)]+\frac{\delta_-}{2\pi} [ \phi(x) + \theta(x)] \right)} d(x).
\end{align}
Note the similarity to the bosonization formula (\ref{eq:perturb_bosonize}). Indeed, the ``shake-up'' of the particles at the right Fermi point caused by the interaction with the deep hole manifests itself as an additional phase in the bosonic representation of the operator $\psi_R(x),$  and similarly for the left Fermi point.

After bosonization and rotation, the expectation value in Eq.~(\ref{eq:Sqomega_projected}) thus factorizes into a term containing the bosonic fields $\phi$ and $\theta$, and a term containing the impurity operator $d$. The dynamics of the fields $\phi(x)$ and $\theta(x)$ is governed by $H_0$ and is therefore linear. Hence, the expectation values of exponentials of bosonic operators can be calculated straightforwardly \cite{giamarchi04}. The impurity dynamics is governed by $H_d$ and leads to $\expcts{d^\dag(x,t) d(0,0)}{} = e^{-i \omega_-(q) t} \delta(x - v_d t)$. Upon Fourier transformation of the time-dependent correlation function, the result for the DSF at its lower edge of support reads
\begin{align}\label{eq:perturb_S1}
 \frac{S(q,\omega)}{m/q} =  \left[ \frac{\delta \omega(q)}{\omega - \omega_-(q)} \right]^{\mu_0(q)} \; \text{for }\delta\omega(q) \gg\omega-\omega_-(q)>0.
\end{align}
The threshold exponent of $S(q,\omega)$ depends on momentum and interaction strength and reads
\begin{align}
 \mu_0(q) &= 1-\left(1+\frac{\delta_+}{2\pi}\right)^2-\left(\frac{\delta_-}{2\pi}\right)^2 \notag \\
&\approx -\frac{\delta_+}{\pi}=  \frac{m}{\pi |q|} ( V_0 - V_q ).
\label{eq:mu_DSF_perturb}
\end{align}
For generic repulsive interaction potentials $\mu_0(q) >0 $, so the DSF has a power-law divergence at the lower threshold.
The expansion of Eq.~(\ref{eq:perturb_S1}) for $\mu_0(q) \ln\left\{ [\omega - \omega_-(q)]/\delta \omega(q) \right\} \ll 1$  coincides with the leading order logarithmic result  (\ref{eq:deltaS1_perturb}). The present calculation corresponds to a resummation of the leading logarithmic divergences to each order in the perturbation series and thus yields the exponent $\mu_0(q)$ to first order in the interaction potential $V_q$.

For momenta $|q| \approx 2 k_F$, the exponent (\ref{eq:mu_DSF_perturb}) evaluated in the leading order of perturbation theory in the interaction potential coincides with the corresponding limit of the linear LL exponents. Indeed, at $q \to 2k_F$, the LL theory predicts a power-law divergence, $S(q,\omega) \propto \theta[\omega - |v(q - 2k_F)|] [\omega - |v(q - 2k_F)|]^{K-1}$, with an exponent $K - 1$. For weak interaction, $K$ can be calculated perturbatively, and one finds~\cite{giamarchi04}
\begin{align}\label{eq:K_perturb}
 K \approx 1-(V_0 - V_{2k_F})/(2 \pi v_F),
\end{align}
thus reproducing the prediction (\ref{eq:mu_DSF_perturb}). At $q \to 0$  and $V(x)$ decaying faster than $\propto 1/x^2, (V_q-V_0)/q \to 0.$ Then  the exponent (\ref{eq:mu_DSF_perturb}) vanishes, and for a fixed $q$ a rectangular shape of $S(q,\omega)$ is recovered. The latter has a width $\delta \omega(q) \propto q^2,$  height $m/q,$ and is located at the mass shell $\omega = v_F q$. In the limit $q \to 0$, this peak indeed acquires a delta-shape as predicted by the LL theory. Later we will see that the relation $\mu_0(0)=0$ for a short-range potential holds beyond the perturbation theory; the rectangular shape of $S(q,\omega)$ at small fixed $q$ is quite generic.

The procedure that led to the DSF near its lower threshold exemplifies a rather versatile framework for the perturbative calculation of various dynamic response functions near singular thresholds. Generally, singularities appear whenever conservation laws allow that the entire energy of an incoming density- or single-particle excitation is transferred to a \emph{single} particle or hole in the system. As illustrated above, the general procedure consists of the following three steps: (i) Identification of the ``deep hole'' configuration responsible for the singular behavior at the threshold of interest. This configuration always follows from momentum and energy conservation. (ii) Projection of the Hamiltonian onto a reduced band structure containing narrow bands around the deep hole and the Fermi points. (iii) Determination of the phase shifts due to the interactions between the deep hole and particles at the Fermi points by applying a unitary transformation.

The shape of the threshold singularity of $S(q,\omega)$ for $\omega \approx \omega_+(q)$ can be obtained similarly. For $q>0$, the configuration giving rise to this singularity contains a hole near the right Fermi point as well as a particle near $k_F + q$ (see Fig.~\ref{fig:DSF}). Projecting the Hamiltonian onto narrow bands around the Fermi points and a narrow band around $k_F + q$, and following essentially the same procedure as before, it was found that for $\delta\omega(q) \gg|\omega-\omega_+(q)|$ \cite{pustilnik06a}
\begin{align}\label{eq:perturb_S2}
 \frac{S(q,\omega)}{m/q} = \begin{cases}
                                \frac{\nu(q)}{\mu_0(q)} + \left[ \frac{ \delta \omega(q) }{\omega_+(q) - \omega} \right]^{-\mu_0(q)} & \omega < \omega_+(q), \\
                               \frac{\nu(q)}{\mu_0(q)}\left(1- \left[ \frac{ \delta \omega(q) }{\omega - \omega_+(q)} \right]^{-\mu_0(q)} \right) &  \omega > \omega_+(q).
                              \end{cases}
\end{align}
where
\begin{align}\label{eq:perturb_nu}
 \nu(q) = \left( \frac{q}{4 m v_F} \right)^2 \left( \frac{V_0-V_{2k_F}}{2 \pi v_F} \right)^2.
\end{align}
Hence, power-law singularities with identical exponents $-\mu_0(q)$ appear on both sides of the threshold $\omega_+(q)$. In stark contrast to the noninteracting limit, $S(q,\omega)$ no longer vanishes above the upper threshold. Also note that the prefactors are different on both flanks.

The DSF $S(q,\omega)$ remains nonzero even above the upper threshold because any excess energy $\omega - \omega_+(q)$ can be used for the creation of an additional particle-hole pair on the left branch. As the excess energy increases, the momenta of the two particles and two holes may be increasingly far away from the Fermi points. Hence, ordinary second-order perturbation theory works well for $\omega - \omega_+(q) \gg \delta \omega.$ In this range, one finds
\begin{align}\label{eq:S_perturb_offshell}
 S(q,\omega) = 2 \nu(q) \frac{v_F q^2}{\omega^2 - v_F^2 q^2}.
\end{align}
The DSF for a weakly interacting system is depicted in Fig.~\ref{fig:DSF}.

In conclusion, interactions lead to notable changes in $S(q,\omega)$. Instead of the rectangular shape of $S(q,\omega)$ for any given $q$ in the noninteracting case, interactions lead to the appearance of power-law singularities at the thresholds $\omega_\pm(q)$. Moreover, the function no longer vanishes above the upper edge $\omega_+(q)$.

\subsubsection{Spectral function}

\label{sec:perturbative_spectral}

The procedure employed for the perturbative calculation of the DSF can also be used for the calculation of the edge singularities of the spectral function \cite{khodas07a}. In contrast to $S(q,\omega)$, the spectral function $A(k,\varepsilon)$ characterizes the response of the system to the addition of a \emph{single} particle or hole. It determines the probability for a particle with energy $\varepsilon$ and momentum $k$ to enter (or, at $\varepsilon<0$, emerge from) the system in a momentum-conserving tunneling event; particle and hole sectors correspond to $\varepsilon>0$ and $\varepsilon<0$, respectively. In the absence of interactions, $A(k,\varepsilon) = \delta[\varepsilon - \xi(k)]$, because a particle or hole with momentum $k$ can only be absorbed if its energy is on the mass shell $\xi(k)$. In an interacting system, on the other hand, the spectral function will generally become
nonzero even away from the mass shell $\xi(k)$, because incoming particles or holes may give up part of their energy and momentum to excite additional particle-hole pairs.

\begin{figure}
\includegraphics[width=0.99\columnwidth]{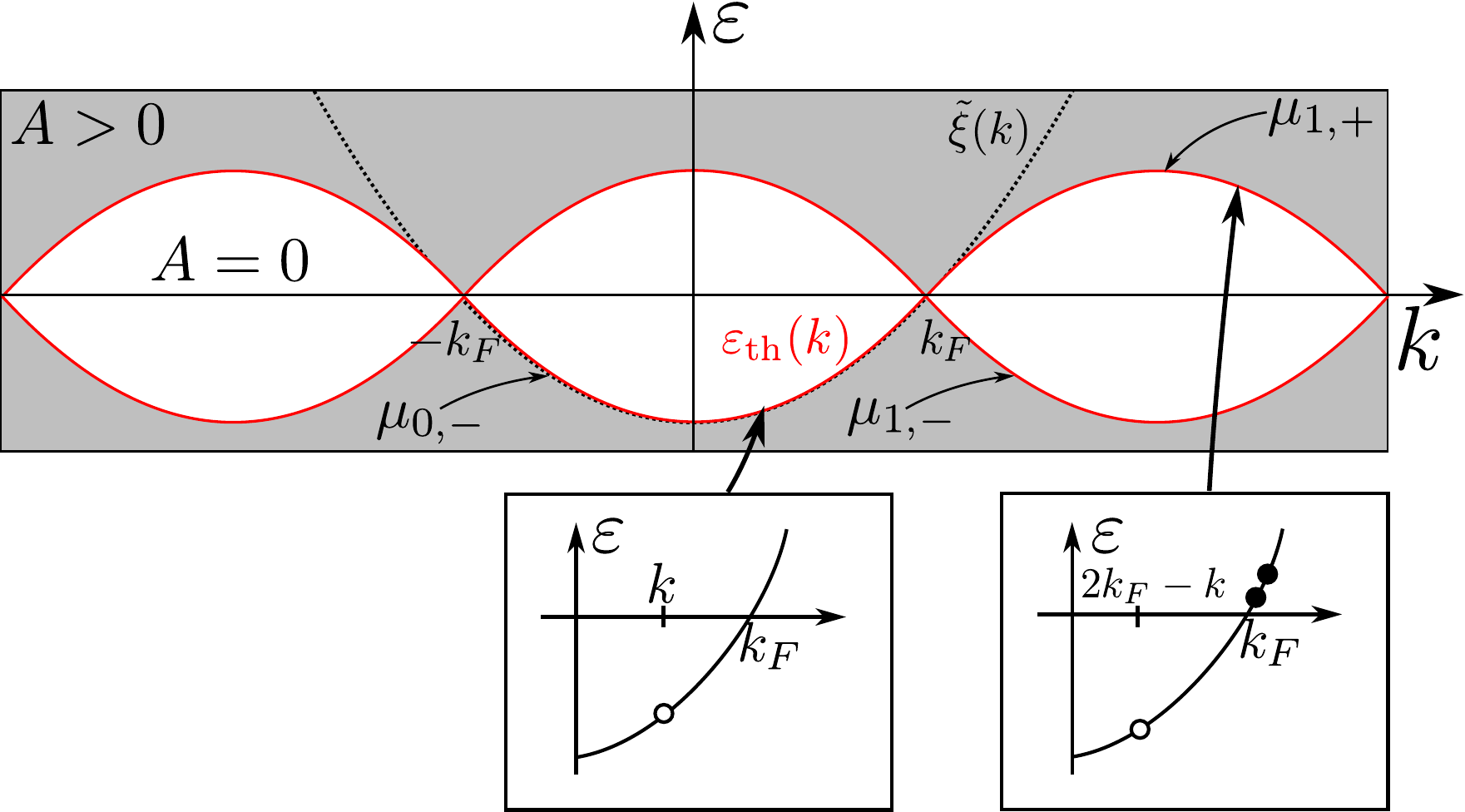}
\caption{\label{fig:AWeaklyInteracting} (Color online) Spectral function $A(k,\varepsilon)$ for interacting fermions, with notations for the edge exponents. The configurations determining the edge exponents $\mu_{0,-}$ and $\mu_{1,+}$ are indicated. In the weakly interacting case, the edge of support $\varepsilon_{\rm th}(k)$ at $|k|<k_F$ coincides with the mass shell $\xi(k)$, see Eq.~(\ref{eq:freefermions}). In the noninteracting case, the spectral function $A(k,\varepsilon)$ is a delta-function at the mass shell, $A_0(k,\varepsilon)=\delta[\varepsilon-\xi(k)].$ }
\end{figure}

For $|k| < k_F$ and $\varepsilon < 0$, the edge of support of $A(k,\varepsilon)$ coincides with the fermion mass shell. In this case, the calculation of the edge singularity is very analogous to the previous section. If a particle with momentum $k$ and energy $\varepsilon \approx \xi(k)$ is extracted from the system, it leaves behind a hole with momentum near $k$ on mass shell, as well as low-energy particle-hole pairs near either of the Fermi points. Therefore, it is again sufficient to retain narrow bands of widths $k_0 \ll |k|$ around $k$ and $\pm k_F$, and project the fermion operator as   in Eq.~(\ref{eq:Psi_projection}). The remaining calculations closely follow Sec.~\ref{sec:perturbative_DSF}.

The spectral function in the vicinity of $\xi(k)<0$ can be calculated as
\begin{align}\label{eq:perturb_A1}
 A(k,\varepsilon) &\propto \int dt dx e^{-i \varepsilon t} \expct{d^\dag(x,t) d(0,0)}_{H_0 + H_d + H_{int}} \notag \\
 &\propto \theta[\xi(k) - \varepsilon] [\xi(k) - \varepsilon]^{-\mu_{0,-}}.
\end{align}
The exponent to lowest order in the interaction strength reads
\begin{align}\label{eq:mu_k}
   \mu_{0,-} = 1-\left(\frac{\delta_-}{2\pi}\right)^2-\left(\frac{\delta_+}{2\pi}\right)^2.
\end{align}
The phase shifts $\delta_{\pm}$ are defined in Eq.~(\ref{eq:deltas_pert}). For $k \to k_F$, the phase shift $\delta_+$ vanishes linearly for interaction potentials decaying faster than $1/x^2$ in real space. On the other hand,  $\delta_-$ at $k\to k_F$ remains finite, $\delta_- =-(V_{2 k_F}-V_0)/(2v_F).$ Therefore, in the limit $k \to k_F$, the exponent $\mu_{0,-}$ coincides with the Luttinger model prediction for weak interactions \cite{luther74}.

Note that to lowest order the exponent is quadratic in the interaction potential $V_k$, which is in contrast to the result  (\ref{eq:mu_DSF_perturb}) for the DSF. Moreover, the interactions of the impurity with left-movers and right-movers are equally important.

Outside the regime $\varepsilon < 0$ and $|k| < k_F$, the edge of support of the spectral function no longer coincides with the mass shell, but it is still determined by kinematic considerations: the injection of a particle or hole and the ensuing creation of particle-hole pairs due to the interactions must respect momentum and energy conservation. For weak interactions, the edge of support can be determined quantitatively by using the Lehmann spectral representation~\cite{abrikosov63}. Let us focus on the particle sector ($\varepsilon > 0$) and the momentum range $k_F< k <3 k_F$, where
\begin{align}
 A(k,\varepsilon) &= \sum_{\ket{f}} | \bra{f} \Psi^\dag_{k} \ket{0} |^2 \delta_{k - P_f, 0} \delta(\varepsilon - E_f).
\end{align}
The initial state $\ket{0}$ corresponds to the ground state of the system and the sum runs over a complete basis $\{ \ket{f} \}$ of the Fock space. The energies and momenta of the final states are denoted by $E_f$ and $P_f$, respectively. In the absence of interactions, the ground state $\ket{0}$ is the filled Fermi sea and the only final state with nonzero overlap is $\ket{f} = \Psi^\dag_{k} \ket{0}$. This immediately leads to $A(k,\varepsilon) = \delta[\varepsilon - \xi(k)]$.

In an interacting system, on the other hand, the ground state $\ket{0}$, written in the basis generated by the operators $\Psi_{k}$ and $\Psi^\dag_k$, may contain particle-hole pairs. Therefore, a nonzero overlap can also be achieved for final states which contain additional excitations. The simplest set of such final states is
\begin{align}\label{eq:PT_ket_f}
 \ket{f} = \Psi^\dag_{k_1} \Psi^\dag_{k_2} \Psi_{k_3} \ket{0},
\end{align}
and it can be parametrized by the momenta $k_1, k_2, k_3$. The edge of support of $A(k,\varepsilon)$ can be determined by enforcing momentum and energy conservation and finding the configuration with lowest excitation energy $E_f $. Within perturbation theory, the energy $E_f$ is determined using the noninteracting Hamiltonian. The result for the particle sector ($\varepsilon>0$) at $k>k_F$ is $k_1 = k_2 = k_F$ and $k_3 = 2k_F - k$, \ie, at the edge of support the entire energy is carried by a single hole with momentum $2 k_F - k$. The energy of this configuration and thus the edge of support of $A(k,\varepsilon)$ is given by
\begin{align}
\label{eq:PT_threshold}
- \xi(2k_F - k) = v_F (k - k_F) - \frac{(k - k_F)^2}{2m}.
\end{align}
The configuration yielding the threshold (\ref{eq:PT_threshold}) remains the lowest energy state with total momentum $k_F<k <3k_F$ even if compared to states with a higher number of particle-hole pairs than Eq.~(\ref{eq:PT_ket_f}).
Hence, the edge of support in this region coincides with the shifted and inverted mass shell $-\xi(2k_F - k)$.

The threshold configuration consists of a hole at momentum $2k_F - k$ and two particles with momenta infinitesimally close to $k_F$. Hence, the Hamiltonian must be projected onto narrow bands (of widths $k_0 \ll k - k_F$) around these momenta. However, the initial particle with momentum $k$ is outside this band structure, so the operator $\Psi^\dag_k$ would vanish in a na\"ive projection of the form (\ref{eq:Psi_projection}). The solution is to use a Schrieffer-Wolff transformation \cite{schrieffer66} to derive a projection which is of first order in the interactions. Then, it can be shown explicitly that the particle at momentum $k$ creates the threshold configuration \cite{khodas07a},
\begin{align}\label{eq:PT_Psik_project}
 \Psi^\dag_k \propto \sum_{k_1,k_2} \psi^\dag_{R,k_1} \psi^\dag_{R,k_2} d_{k_1+k_2},
\end{align}
where $k_1, k_2 \ll k_0$. The impurity operator $d$ creates a hole near momentum $2k_F - k$, whereas the operators $\psi^\dag_{R,k_{1,2}}$ create particles close to the right Fermi point. The fact that a single incoming particle now has to form three excitations in order to be able to tunnel into the system opens up the phase space available for the process, due to the free variables $k_1$ and $k_2$ in the projection (\ref{eq:PT_Psik_project}). The spectral function in the vicinity of the edge  is proportional to the Fourier transform of $\expcts{ \Psi_k(t) \Psi^\dag_k(0) }{}$. Neglecting the interactions between the subbands, a direct calculation of this correlator using Eq.~(\ref{eq:PT_Psik_project}) leads to $A(k,\varepsilon) \propto [\varepsilon + \xi(2k_F - k) ]^3$, \ie, a convergent threshold behavior. An increased number of excitations in the projection of the physical fermion operators, as required at  momenta $|k| > k_F$, generally leads to more convergent power-laws. It is a recurring feature that all thresholds are characterized by configurations in which the entire energy is carried by a single particle or hole, while additional particle-hole pairs reside close to the Fermi points.

Interactions between the impurity $d$ and the particles near the Fermi points again lead to a correction to this exponent. Using a mobile impurity Hamiltonian, the ensuing calculation is analogous  to the previous discussion. The operator $\Psi^\dag_k$ can be bosonized as $\Psi^\dag_k \propto e^{2i (\phi - \theta)} d$, and the spectral function near the edge behaves as
\begin{align}
 A(k,\varepsilon)  \propto \theta[\varepsilon + \xi(2k_F - k)] [\varepsilon + \xi(2k_F - k)]^{-\mu_{1,+}},
\end{align}
where
\begin{align}
\mu_{1,+} = 1- \left(2+\frac{\delta_+}{2\pi}\right)^2-\left(\frac{\delta_-}{2\pi}\right)^2\approx -3-\frac{2\delta_+}{\pi}.
\end{align}

The thresholds in the other sectors of the $(k,\varepsilon)$-plane can be derived similarly. The support of the spectral function in the weakly interacting limit and the lowest-energy configurations at the respective edges are displayed in Fig.~\ref{fig:AWeaklyInteracting}.

The configurations which give rise to singularities at the edges of support are always stable: they represent the excitations of lowest energy for a given momentum and are thus protected from decay by conservation laws. On the other hand and in striking contrast to the noninteracting case, the mass shell $\xi(k)$ no longer forms the edge of support for momenta $k > k_F$. Instead, it now lies within a continuum of excitations. Therefore, particles on the mass shell are generally subject to decay. In the spectral function, this will give rise to a broadening of the singularity at $\varepsilon = \xi(k)$. This will be discussed more in detail in Sec.~\ref{sec:QP_Relax}.

\subsection{The universal limit of nonlinear Luttinger liquids}
\label{sec:universal}

The analysis presented in the previous section predicts the dynamic response functions for weakly interacting Fermi systems at arbitrary momenta. In many realistic systems, however, the interaction energy can be of the same order as the kinetic energy, thus making perturbation theory inapplicable. To treat such systems, having a theory which accounts for the interactions exactly would be desirable. If the fermionic spectrum is strictly linear, all dynamic response functions at low energies can be calculated exactly~\cite{luther74,dzyaloshinskii74}. Close to Fermi points, it is tempting to consider the band curvature as a small perturbation to the linear spectrum. The resulting corrections to single-variable correlation functions (for example in the fermion distribution function $n_k = \expcts{\Psi^\dag_k \Psi_k}$) are indeed uniformly small. This is not the case, however, for the dynamic response functions. We will see here that the true values of the threshold exponents are different from the predictions of the linear LL theory even in the limit $|k|\to k_F$~\cite{imambekov09a}. The frequency domain near the threshold where these strong deviations take place,
narrows down as $(|k|-k_F)^2$.

Phenomenological bosonization is the obvious approach to tackle a strongly interacting 1D system in the low-energy regime~\cite{efetov75, haldane81a, haldane81b}. This approach is based on rephrasing the fermionic problem in a bosonic language using the bosonization identities (\ref{eq:perturb_bosonize}) and (\ref{eq:LL_rho_boson}).
Using this basis offers the advantage that a density-density interaction between the physical fermions produces a quadratic term in the bosonic variables. At low energies,  where only degrees of freedom close to the Fermi points are involved, bosonization allows an exact treatment of the interaction.

However, for the quadratic spectrum (\ref{eq:freefermions}) the kinetic energy becomes more complicated when expressed using the bosonic fields $\theta(x)$ and $\phi(x)$. The kinetic energy density of an ideal gas in the ground state can be calculated by integrating the spectrum $\xi(k)$ over $k \in [-k_F, k_F]$. Local fluctuations of the left and right particle densities shift the Fermi points, $k_{F}^{R,L}(x) = \pm [k_F + \pi \rho_{R,L}(x)]$, and therefore change this energy density. By expressing the density fluctuations using Eq.~(\ref{eq:LL_rho_boson}), the kinetic Hamiltonian can be derived. Near the two Fermi points, the spectrum  for noninteracting fermions can be expanded as $\xi(k) \approx v_F(\pm k - k_F) + (k \mp k_F)^2/(2m)$. Its linear component together with the interaction term produces the conventional LL Hamiltonian,
\begin{align}\label{eq:HLL}
 H_{\rm LL} &= \frac{v}{2\pi} \int dx \left[ K (\nabla \theta)^2 + \frac{1}{K} (\nabla \phi)^2\right],
\end{align}
where $v$ denotes the renormalized Fermi velocity, and $K$ is the Luttinger parameter, which is in the interval $0 < K < 1$ for repulsive interactions. For the noninteracting system, $K=1$ and $v = v_F$. The quadratic component of the spectrum, on the other hand, leads to cubic terms in the bosonic fields~\cite{haldane81a},
\begin{align}\label{eq:Hnl}
 H_{nl} &= - \frac{1}{6 \pi m} \int dx \left[ (\nabla \phi)^3 + 3 (\nabla \phi) (\nabla \theta)^2 \right].
\end{align}
As soon as the cubic band curvature terms are taken into account, an exact diagonalization of the Hamiltonian $H_{\rm LL} + H_{nl}$ in terms of the fields $\phi$ and $\theta$ is no longer possible. The most obvious route is to treat $H_{nl}$, which is proportional to $1/m$, as a perturbation. Diagrammatically, the terms in Eq.~(\ref{eq:Hnl}) correspond to three-boson interaction vertices. It turns out, however, that such an endeavor is far from trivial because the bosonic self-energy diverges at the mass shell $\omega = v k$ \cite{samokhin98}. The physical reason is the linear spectrum of the bosonic modes: all bosonic excitations propagate with the same velocity $v$ independent of their momentum and therefore - semiclassically speaking - have an infinite time to interact.

This difficulty precludes a straightforward calculation of the DSF near the mass shell. Far away from the mass shell ($\omega \gg v q$), on the other hand, the perturbation theory in the band curvature is convergent and a high-energy tail in $S(q,\omega)$ emerges in the order $(1/m)^2$ \cite{pereira07}. For small nonzero interactions, this agrees with the perturbative result~(\ref{eq:S_perturb_offshell}).

An expansion of the free-fermion result (\ref{eq:dsf-freefermions2}) in orders of $1/m$ reveals that each individual term of the series diverges at the mass shell. Therefore, in order to access $S(k,\omega)$ close to the mass shell, an efficient resummation scheme is required. Standard procedures like the Born approximation fail because they still produce a divergence at $\omega = v q$. Various approximate schemes have been developed \cite{aristov07,schoenhammer07,teber07,pirooznia07,pirooznia08}, but even the free-fermion result has been reproduced in the bosonic basis only up to the order $(1/m)^4$ \cite{pereira07}.

Many of the complications which plague the bosonic perturbation theory can be avoided by using a basis of fermionic quasiparticles~\cite{mattis65,rozhkov05}. Similar to the bosonic fields, the quasiparticles remain free in the case of a strictly linear spectrum whereas a nonzero band curvature of the underlying particles, together with the inter-particle interactions, leads to interactions between the fermionic quasiparticles. In contrast to the bosonic theory, however, the spectrum nonlinearity of the physical fermions also entails a band curvature of the quasiparticles. Hence, quasiparticles with different momenta propagate at different velocities. We will see that the scattering between them at momenta close to the Fermi points can be treated, for instance, within the Born approximation. Moreover, the use of a fermionic basis allows again the introduction of a mobile impurity Hamiltonian and thus connects to the method employed in the perturbative calculation.

To illustrate the origin of the fermionic  quasiparticle representation of the Luttinger model, let us diagonalize the
Hamiltonian (\ref{eq:HLL})  by introducing the rescaled fields
\begin{align}  \label{eq:rescaling}
\ttheta(x) = \sqrt{K} \theta(x), \quad
\tphi(x) = \phi(x)/\sqrt{K}.
\end{align}
Since this is a canonical Bogoliubov transformation, the fields $\ttheta(x)$ and $\tphi(x)$ are still canonically conjugate. In the new variables $\ttheta$ and $\tphi$, the Hamiltonian is indistinguishable from the bosonized version of the Hamiltonian of free fermions with linear spectrum. These free left- and right-moving fermionic quasiparticles can be defined by using the bosonization identity \cite{mattis74,luther74,haldane81a} on the rescaled fields,
\begin{align}\label{eq:LL_tPsi}
 \tPsi_{\alpha}(x) \propto \exp\{ -i [ \alpha \tphi(x) - \ttheta(x)]\},
\end{align}
for $\alpha = R,L = +,-$. Here $\tPsi^{\dagger}_{R(L)}, \tPsi_{R(L)}$ are creation and annihilation operators for quasiparticles on the right (left) branch, satisfying usual fermionic commutation relations [as usually, we didn't write out the Klein factors explicitly~\cite{giamarchi04}]. In terms of quasiparticles, $H_{\rm LL}$ becomes
\begin{align}\label{eq:LL_HLL}
 H_{\rm LL} = -i v \int dx \left[ : \tPsi_R^\dag(x) \nabla \tPsi_R(x) : - : \tPsi_L^\dag(x) \nabla \tPsi_L(x) : \right].
\end{align}
Colons indicate the normal ordering with respect to filled Fermi seas: for the right (left) branch all states with negative (positive) momenta are occupied. The relations between $\trho_{R(L)}$ and $\rho_{R(L)}$ are linear, and follow from Eqs. (\ref{eq:LL_rho_boson}) and (\ref{eq:rescaling}).

To proceed further, we need relations between $\tPsi_{R(L)}$ and $\Psi_{R(L)}.$ Since both $\tPsi_{R}$ and $\Psi_{R} $ carry momentum $+k_F$ and change the total number of particles by one, $\Psi_{R}$ should contain $\tPsi_{R}$ and low-energy particle-hole pairs. Using the bosonization formula for expressing $\Psi_{R,L}$ in terms of  $\tphi$ and $\ttheta$, and using Eq.~(\ref{eq:LL_tPsi}) to ``pull out'' an operator $\tPsi_{R}$, one finds for right-movers
\begin{align}\label{eq:LL_PsiR}
    \Psi_{R}(x) = F_R(x) \tPsi_R(x).
\end{align}
An analogous expression holds for the left-movers. The string operator $F_R(x)$ is an exponential of the left-moving and right-moving quasiparticle densities (\ref{eq:LL_rho_boson}),
\begin{align}\label{eq:LL_string}
    F_R(x) = \exp\left\{ -i \int_{-\infty}^x dy [ \delta_+ \trho_R(y) + \delta_- \trho_L(y) ] \right\}.
\end{align}
The described refermionization procedure defines uniquely the two parameters
\begin{equation}\label{eq:LL_phases}
    \frac{\delta_+}{2\pi} = 1 - \frac{1}{2\sqrt{K}} - \frac{\sqrt{K}}{2}<0,\;
    \frac{\delta_-}{2\pi} = \frac{1}{2\sqrt{K}} - \frac{\sqrt{K}}{2}.
\end{equation}
According to Eq.~(\ref{eq:LL_PsiR}), the annihilation of a physical right-moving fermion in the liquid causes the annihilation of a right-moving quasiparticle. In addition it leads to a shake-up of the Fermi seas of the left-moving and right-moving quasiparticles which is described by $F_R$. The parameters $\delta_\pm$ can be interpreted as the phase shifts for the scattering of quasiparticles at $\pm k_F$ off the right-moving hole in the quasiparticle distribution.

The mapping between the interacting physical fermions and the noninteracting fermionic quasiparticles can also be performed directly via a unitary transformation~\cite{mattis65,rozhkov05}, without invoking bosonization as an intermediate step, and produces identical results. The Hamiltonian (\ref{eq:LL_HLL}) together with Eqs.~(\ref{eq:LL_string}) and (\ref{eq:LL_phases}) reproduces the usual results for the fermionic Green's function~\cite{rozhkov05}.

The quadratic spectrum (\ref{eq:freefermions}) of the physical fermions leads to additional terms in the quasiparticle Hamiltonian. Most importantly, a quadratic term emerges in the quasiparticle spectrum \cite{rozhkov06,rozhkov08,rozhkov09}. It is reflected in the Hamiltonian
\begin{align}\label{eq:LL_Hnl1}
    H'_{nl} = \frac{1}{2 \tilde{m}} \int dx \left[  :(\nabla \tPsi_R^\dag \nabla \tPsi_R) : + : (\nabla \tPsi_L^\dag \nabla \tPsi_L) : \right].
\end{align}
The effective mass $\tilde{m}$ depends on interactions, and using the methods described in Sec.~\ref{sec:phenomenology}, it is possible to express it via low energy properties as~\cite{pereira06}
\begin{align}\label{eq:LL_meff}
   \frac{1}{\tilde{m}} = \frac{v}{K} \frac{\partial}{\partial\mu} (v \sqrt{K}),
 \end{align}
where $\mu$ is the chemical potential. From  Eqs.~(\ref{eq:LL_HLL}) and (\ref{eq:LL_Hnl1}), the quasiparticle mass shell for $k \approx k_F$ is given by
\begin{align}\label{eq:LL_xik}
 \tilde \xi(k) = v(k - k_F) + \frac{(k - k_F)^2}{2\tilde{m}}.
\end{align}

In addition, a spectrum curvature of and interactions between the physical fermions generally lead to interactions between the fermionic quasiparticles. One such term describes an interaction between quasiparticles on opposite branches \cite{rozhkov06},
\begin{align}\label{eq:LL_Hint1}
    H'_{int} = i \tilde{g} \sum_{\alpha=R,L} \int dx \trho_{-\alpha} \left[ :  \tPsi^\dag_\alpha (\nabla \tPsi_\alpha): -   : (\nabla \tPsi^\dag_\alpha) \tPsi_\alpha: \right].
\end{align}
In order to understand the effect of this term, we consider the scattering phase shift between two quasiparticles with momenta $p+ k_F$ ($|p|\ll k_F$) and $-k_F$ as in Sec.\ref{sec:perturbative}. Fourier transforming Eq.~(\ref{eq:LL_Hint1}) reveals that for small $p$, the interaction potential is proportional to  $p$. This has to be compared with the difference in velocities, which according to Eq.~(\ref{eq:LL_xik}) is finite and close to $2v_F$. Therefore, the term (\ref{eq:LL_Hint1}) produces only small additional phase shifts of order $p/k_F \ll 1.$

Another interaction term can appear as a consequence of a momentum dependence of the physical interaction potential. It corresponds to a density-density interaction between quasiparticles on the same branch \cite{imambekov09a},
\begin{align}\label{eq:LL_Hint2}
    H''_{int} = \sum_{\alpha=R,L} \int_{|p|\ll k_F} dp \tilde{W}(p) \trho_{\alpha}(p)  \trho_{\alpha}(-p).
\end{align}
The Luttinger parameter $K$ already accounts for interaction processes with momentum exchange close to zero, so the additional quasiparticle interaction potential must fulfill $\tilde{W}(0) = 0$. Moreover, it should be symmetric $\tilde{W}(p) = \tilde{W}(-p)$. For small $p$, the scattering phase shift between two particles with momenta $p + k_F$ and $k_F$ is now of order $\tilde{m} \tilde{W}(p)/p$. This phase shift therefore becomes small for generic interaction potentials which fulfill $\tilde{W}(p) \propto p^2$ for $p \to 0$. This requirement is only violated  for real-space potentials decaying as $\propto 1/x^2$ or slower. In particular, this excludes Haldane-Shastry~\cite{haldane88, shastry88} or Calogero-Sutherland~\cite{calogero69,calogero71,sutherland71,sutherland04} type models.

The above arguments establish that  Eqs.~(\ref{eq:LL_HLL}) and (\ref{eq:LL_Hnl1}),  written in terms of fermionic quasiparticles, serve as the universal low-energy Hamiltonian  which captures the leading role of the spectrum nonlinearity. The crucial advantage of the quasiparticle representation is that unlike in bosonic extensions of the LL theory, the
universal Hamiltonian does not contain interactions. The nonlinearity of the spectrum lifts the ``accidental'' degeneracies existing in the Luttinger model, and allows to treat small
interaction terms (such as given in Eqs.~(\ref{eq:LL_Hint1}) and (\ref{eq:LL_Hint2}))  in  addition to the universal Hamiltonian within convergent perturbation theory.  Surprisingly,  the only additional phenomenological parameter  is the effective mass $\tilde{m}$, which sets the energy scale for nonlinear effects. Below we will work out universal predictions for the low-energy  DSF and the spectral function beyond the linear spectrum approximation. We will show that the true values of the threshold exponents differ from predictions of the linear LL theory, but nevertheless can be expressed as functions of the
LL parameter $K$ only.

The relation between the total physical particle density and the total quasiparticle density is linear, $\rho_L + \rho_R = \sqrt{K} [\trho_L + \trho_R]$. Therefore, the shape of the DSF $S(q,\omega)$ for $q \ll k_F$ remains identical to the noninteracting result (\ref{eq:dsf-freefermions2}), albeit with a renormalized mass $\tilde{m}$,
\begin{equation}\label{eq:LL_S}
S(q,\omega) =
\frac{K \tilde{m}}{|q|} \theta\left(\frac{q^2}{2 \tilde{m}} - |\omega - v q|\right)
\end{equation}
This agrees with the perturbative results (\ref{eq:perturb_S1}) and (\ref{eq:perturb_S2}) in the limit $q \to 0$. Power-law singularities at the upper and lower thresholds $\omega_\pm(q) = v q \pm q^2/(2 \tilde{m})$, respectively, as well as a high-energy tail would be produced by the quasiparticle interaction term (\ref{eq:LL_Hint1}) which becomes relevant only beyond the limit $q/k_F \ll 1$.

Let us first investigate the spectral function $A(k,\varepsilon)$ in the presence of band curvature in the hole sector ($0<k_F-k \ll k_F$ and $\varepsilon < 0$). In this region, the kinematic edge of support of $A(k,\varepsilon)$ coincides with the mass shell (\ref{eq:LL_xik}), so we assume $\varepsilon \approx\tilde  \xi(k)$. According to Eq.~(\ref{eq:LL_PsiR}), the extraction of a physical particle with momentum $k$ from the system leads to the formation of a quasiparticle hole with momentum near $k$ on mass shell, and the excess energy is used for the formation of excitations near the Fermi points. In analogy to the previous section, the Hamiltonian (\ref{eq:LL_HLL})-(\ref{eq:LL_Hnl1}) as well as the single-particle operator (\ref{eq:LL_PsiR}) should be projected onto small bands around the momenta $\pm k_F$ and $k$. The right-moving quasiparticle operator is projected as
\begin{align}\label{eq:LL_proj_psiR}
 e^{i k_F x} \tPsi_R(x) \to e^{i k_F x} \tpsi_R(x) + e^{i k x} \td(x)
\end{align}
and $\tPsi_L(x) \to \tpsi_L(x)$ is used for left-movers; here, $\tpsi_{R,L}(x)$ denote quasiparticles within a narrow momentum range around the Fermi points~\footnote{Hereinafter, we reserve the upper-case symbols $\tPsi_{L,R}(x)$ and the lower-case symbols $\tpsi_{L,R}(x)$ for operators before and after the projection, respectively}, while $\td(x)$ denotes an impurity with momentum close to $k$. Projecting the string operators $F_R(x)$ requires a projection of the quasiparticle densities $\trho_\alpha(x)= \tPsi^\dag_\alpha(x) \tPsi_\alpha(x)$ in Eq.~(\ref{eq:LL_string}). This produces terms containing the quasiparticle densities in the narrow bands around the Fermi points, $\tpsi^\dag_\alpha(x) \tpsi_\alpha(x)$. In addition, the projection leads to impurity terms of the form $\td^\dag(x) \td(x)$ and mixed terms $\tpsi_{\alpha}^\dag(x) \td(x)$. However, the former is a constant, and the latter can be neglected close to the edges because they require a higher energy. Therefore, projecting the string operators corresponds to replacing $\trho_\alpha(x)$ in Eq.~(\ref{eq:LL_string}) with $\tpsi^\dag_\alpha(x) \tpsi_\alpha(x)$.

\begin{figure}
\includegraphics[width=8cm]{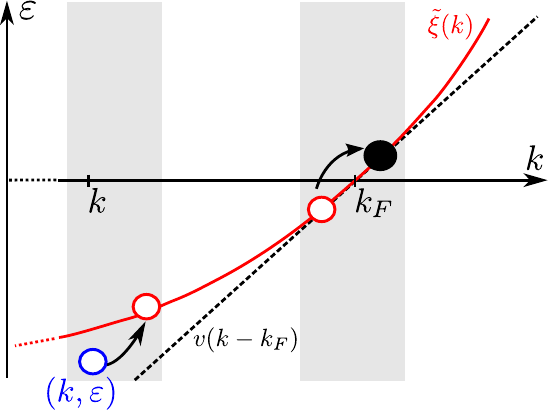}
\caption{\label{fig:ImpurityBand} (Color online) Band structure for right-movers used for the calculation of $A(k,\varepsilon)$ for $0 < k_F - k \ll k_F$ and $\varepsilon < 0$, see Eq.~(\ref{eq:LL_proj_psiR}). It contains the impurity band near momentum $k$ and a low-energy band near the right Fermi point $+k_F$. The injection of a hole with momentum $k$ and energy $\varepsilon$ into the system leads to the formation of a hole (empty circle) with momentum near $k$ on the mass shell, as well as a particle-hole pair near the Fermi point. For $|\varepsilon - \tilde{\xi}(k)| \ll (k - k_F)^2/(2 \tilde{m})$, the subbands are well separated and the mobile-impurity Hamiltonian can be applied.}
\end{figure}

A projection of the kinetic Hamiltonian $H_{\rm LL} + H'_{nl}$ onto the three subbands leads to the mobile impurity Hamiltonian $H_0 + H_d$, where
\begin{align}\label{eq:LL_imp}
 H_0 &= -i v \int dx \left[ : \tpsi_R^\dag(x) \nabla \tpsi_R(x) : - : \tpsi_L^\dag(x) \nabla \tpsi_L(x) : \right], \notag \\
 H_d &= \int dx \td^\dag(x) \left[\tilde  \xi(k) - i v_d \nabla \right] \td(x),
\end{align}
where the impurity velocity is $v_d = \partial \tilde \xi(k)/\partial k$. Note that the requirement that the impurity band be separated from the low-energy bands means that this mobile impurity Hamiltonian can be used to calculate the spectral function at energies $\varepsilon$ satisfying $|\varepsilon - \tilde \xi(k)| \ll (k - k_F)^2/(2 \tilde{m})$. This is illustrated in Fig.~\ref{fig:ImpurityBand}.

Then, the spectral function in the region $0<k_F-k \ll k_F$ and $\varepsilon < 0$ becomes
\begin{align}
 A(k,\varepsilon) &= \int dt dx e^{-i \varepsilon t} \expcts{\td^\dag(x,t) \td(0,0)} \notag \\
&\times \expcts{F_R^\dag(x,t) F_R(0,0)}
\end{align}
The impurity correlation function can easily be calculated using the noninteracting Hamiltonian $H_d$. The free string correlation function can most conveniently be derived by bosonizing $H_0$ and $F_R$. One finds that $A(k,\varepsilon) \propto \theta(\tilde \xi(k) - \varepsilon) [\varepsilon -\tilde  \xi(k)]^{-\mu_{0,-}}$, where
\begin{align}\label{eq:LL_mu0}
 \mu_{0,-} = 1 - \left(\frac{\delta_-}{2\pi}\right)^2 - \left(\frac{\delta_+}{2\pi}\right)^2.
\end{align}

As in the previous section, one of the consequences of a nonlinear spectrum with $\tilde{m} > 0$ is that the edge of support of the spectral function in the particle sector no longer coincides with the quasiparticle mass shell $\tilde \xi(k)$ but rather with the shifted and inverted quasiparticle spectrum. For instance, for $0<k-k_F\ll k_F$, the threshold is
given by $- \tilde  \xi(2k_F - k).$ The threshold configuration generated at this edge contains a quasihole with momentum $2 k_F - k < k_F$, as well as two quasiparticles with total momentum $2k_F$. The power-law at this threshold can be derived using the same mobile-impurity Hamiltonian (\ref{eq:LL_imp}) but the projection of the quasiparticle operator has to be done analogous to Eq.~(\ref{eq:PT_Psik_project}). Due to the creation of additional particles, the spectral function at this edge is convergent. One finds $A(k,\varepsilon) \propto \theta(\varepsilon +\tilde  \xi(2k_F - k)) [ \varepsilon+\tilde  \xi(2k_F - k)]^{-\mu_{1,+}}$, where
\begin{align}\label{eq:LL_mu1}
 \mu_{1,+} = 1 - \left(\frac{\delta_-}{2\pi}\right)^2 - \left(2 + \frac{\delta_+}{2\pi}\right)^2.
\end{align}
Similar to the weakly interacting case, the singularity at the mass shell for $k > k_F$ and $\varepsilon \approx \tilde \xi(k)$ now lies within a continuum of excitations. Since interactions neglected in the universal Hamiltonian generally lead to a nonzero decay rate for quasiparticles on mass shell, one may expect that the singularity may be smeared. Within perturbation theory, the decay rate close to the Fermi points scales as $\Gamma \propto (k - k_F)^8$ \cite{khodas07a}, and will be discussed in detail in Sec.~\ref{sec:QP_Relax}. However, the power-law behavior is expected to be observable within an energy window of width $(k - k_F)^2/(2 \tilde{m})$ around the mass shell \cite{imambekov09a}. Therefore, the singularity is indeed resolved for $k \to k_F$. Using the mobile-impurity Hamiltonian, one finds power laws with identical exponents on both sides of the mass shell, $A(k,\varepsilon) \propto |\varepsilon -\tilde \xi(k)|^{-\mu_{0,-}}$,  where the exponent is given by Eq.~(\ref{eq:LL_mu0}).
Despite the fact that the exponents on both sides of the singularity are identical, the prefactors are not. It can be shown that \cite{imambekov09a}
\begin{align}\label{eq:LL_shoulder}
 \lim_{\delta \varepsilon \to 0} \frac{A[k,\tilde{\xi}(k) + \delta \varepsilon]}{A[k,\tilde{\xi}(k) - \delta \varepsilon]}
= \frac{\sin\Big[ \pi \left(\frac{\delta_-}{2\pi}\right)^2 \Big]}{\sin\Big[ \pi \left(\frac{\delta_+}{2\pi}\right)^2 \Big]}.
\end{align}
The exponents at the mass shell are identical in the particle and hole sectors, and they differ from the predictions of the linear LL theory. As has been established in Sec. \ref{sec:perturbative_spectral}, the expansion of the exponent (\ref{eq:LL_mu0}) to leading order in $K-1$ coincides with the Luttinger model prediction (\ref{eq:LL_muLL}). However, a difference to the LL exponent emerges in the order $(K-1)^4,$ and is of order one  for strong interactions ($|K-1|\sim 1$).

The result of the linear LL theory is recovered for energies further above the mass shell, $\varepsilon - \tilde \xi(k) \gg (k - k_F)^2/(2 \tilde{m})$. As illustrated in Fig.~\ref{fig:ImpurityBand}, in this regime the momentum bands encountered in the projection start to overlap, and the mobile-impurity Hamiltonian ceases to be applicable. From the point of view of an incoming particle, the band curvature becomes irrelevant at these energy scales. Therefore, the spectral function can be calculated using
\begin{align}\label{eq:LL_ALL}
 A(k,\varepsilon) &\propto \int dt dx e^{i \varepsilon t} e^{-i k x} \notag \\
&\times \expcts{\tpsi_R^\dag(x,t) F^\dag_R(x,t) F_R(0,0) \tpsi_R(0,0)}{},
\end{align}
and by assuming that the quasiparticle spectrum is linear. In this case, the time evolution of the quasiparticle densities becomes simple, $\trho_\alpha(x,t) = \trho_\alpha(x - \alpha v t)$ and allows for a calculation of the dynamics of the string operator $F_R(x,t)$. This makes it possible to calculate Eq.~(\ref{eq:LL_ALL}) in the basis of fermionic quasiparticles \cite{rozhkov05}. Alternatively, Eq.~(\ref{eq:LL_ALL}) with a linear spectrum may be calculated by bosonizing it. As a result, for $\varepsilon - \tilde \xi(k) \gg (k - k_F)^2/(2 \tilde{m})$ one finds $A(k,\varepsilon) \propto [\varepsilon - \tilde \xi(k)]^{-\mu_{\rm LL}}$ with the LL exponent
\begin{align}\label{eq:LL_muLL}
 \mu_{\rm LL} = 1 - \left(\frac{\delta_-}{2\pi}\right)^2 = 1 - \frac{1}{4} \left( K + \frac{1}{K} - 2 \right).
\end{align}
Compared to the exponent $\mu_{0,-}$ near the mass shell, see Eq.~(\ref{eq:LL_mu0}), the $\delta_+$ term is missing. The phase $\delta_+$ is the scattering phase shift for interactions between the impurity at momentum $k \approx k_F$ and particles near the right Fermi point. Because the impurity band around $k$ and the band around $+k_F$ start to merge for energies further away from the mass shell, this phase is absent in Eq.~(\ref{eq:LL_muLL}). The phase $\delta_-$, in contrast, is brought about by interactions with particles near the left Fermi point, and continues to be present.

\begin{figure}
\includegraphics[width=8cm]{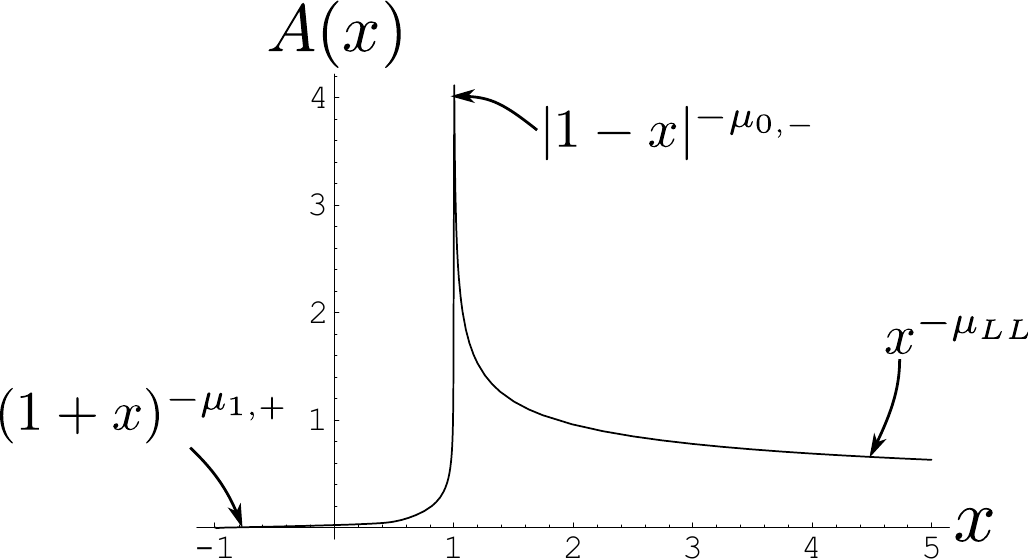}
\caption{\label{fig:aefig} The numerically evaluated function $A(x)$, see Eq.~(\ref{eq:LL_A}), for a fixed value of the Luttinger parameter $K= 4.54$ (corresponding to $\delta_+/(2\pi)=-0.3$). This function determines the behavior of $A(k,\varepsilon)$ in the crossover region between the threshold (left end) and the linear Luttinger liquid (LL) asymptotes.}
\end{figure}

Understanding the crossover between regions with exponents (\ref{eq:LL_mu1}), (\ref{eq:LL_mu0}) and (\ref{eq:LL_muLL}) requires a calculation of $A(k,\varepsilon)$ at intermediate energies. Due to kinematic constraints, only the nonlinearity of the spectrum of right-movers is important for $k > k_F$ and $\varepsilon > \tilde \xi(k)$. Therefore, an analysis of the exact dynamics of the string operators $F_R(x)$ is needed. Similar correlators have attracted attention recently in connection with the nonlinear quantum shock wave dynamics of a free Fermi gas \cite{bettelheim06,bettelmann06b,bettelheim07,bettelheim08}. Even though the quasiparticles are noninteracting, this is a highly nontrivial problem due to the nonlinear spectrum $\tilde \xi(k)$. However,  since it is essentially a single-particle problem, it can be mapped onto a calculation of certain infinite-size determinants~\cite{abanin04,abanin05} and then tackled numerically~ \cite{imambekov09a}. The spectral function for momenta $k \approx k_F$ at arbitrary energies $\varepsilon$ can be written as a function of a single variable,
\begin{align} \label{eq:LL_A}
 A(k,\varepsilon) \propto A(x), \quad \text{where } x = \frac{ \varepsilon - v(k - k_F) }{(k - k_F)^2/(2 \tilde{m})},
\end{align}
where the function $A(x)$  depends only on the Luttinger parameter $K$. In this sense, for momenta close to the Fermi points, the universality present in the linear LL theory persists even in the presence of band curvature. The results of a numerical evaluation of $A(x)$ for a particular value of $K$ are shown in Fig.~\ref{fig:aefig}.

Note that the width of the energy window where the exponent differs from the predictions of the linear LL theory is proportional to $(k - k_F)^2$. The fact that this width vanishes quadratically for $k \to k_F$ is consistent with the irrelevance of the band curvature in the universal limit $k \to k_F$.

\subsection{Phenomenology beyond the
low-energy limit: the mobile quantum impurity model}

\label{sec:phenomenology}

In Sec.~\ref{sec:perturbative}, the dynamic response functions were calculated perturbatively for weak interactions and at arbitrary momenta. A complementary result, valid for arbitrary interactions but only close to the Fermi points, was derived in Sec.~\ref{sec:universal}. In this section, the basic ideas of the two approaches, refermionization and the use of a quantum impurity model, will be combined in order to obtain phenomenological relations determining the exponents of the threshold singularities at arbitrary interaction strengths and momenta.

To be specific, let us discuss the edge of support of the spectral function in the momentum region $|k| < k_F$ in the hole sector, \ie, for energies $\varepsilon < 0$. For a given $k$, the creation of a hole will generally require a nonzero minimum energy. At zero temperature, this entails the existence of a sharp edge of support which we will denote by $\varepsilon_\th(k),$ see Fig.~\ref{fig:AWeaklyInteracting} for an illustration. The extraction of a physical particle is impossible for $|\varepsilon| < |\varepsilon_\th(k)|$. For a generic system with arbitrary interaction potential, the exact shape of $\varepsilon_\th(k)$ is not known, because it is related to the exact eigenenergies which crucially depend on the microscopic nature of the interactions.

For noninteracting systems at zero temperature, the Fermi momentum $k_F$ is defined as the momentum of the highest occupied single-particle state in the Fermi sea. The concept of a Fermi momentum can be extended to gapless interacting systems by defining $k_F$ as the smallest positive momentum $k$ for which $A(k,\varepsilon = 0) \neq 0$, \ie, at momentum $\pm k_F$, the system is capable of absorbing even infinitesimal quanta of energy. Applied to this definition, the Luttinger theorem \cite{luttinger60,yamanaka97,blagoev97} ensures that the value of $k_F$ is independent of the interaction strength. Therefore, the edge of support still satisfies $\varepsilon_\th(\pm k_F) = 0$ even in an interacting system.

In Sec.~\ref{sec:perturbative} and  Sec.~\ref{sec:universal}, it was established for weak interactions and for momenta close to Fermi points for arbitrary interactions,
that the behavior of the spectral function near the edges of support can be understood by introducing effective models of impurities moving in LLs. It is important to emphasize that, similar to Fermi liquid quasiparticles, such impurities have a {\it finite} overlap with the original fermionic operators,
and possess the same quantum numbers. Extracting a physical fermion with momentum $k$ and energy close to $\varepsilon_\th(k)$ creates an impurity with momentum near $k$ which carries almost the entire energy, as well as additional low-energy excitations near Fermi points. It is not the overlap between the impurity and the fermion, but the ``shake-up'' of low-energy modes near Fermi points which causes the power-law divergences in the spectral function as a function of the excess energy $\varepsilon_\th(k)-\varepsilon>0.$

Combining these two limits, we may expect that even away from the low-energy limit for arbitrary interactions, the threshold properties of response functions near the edges of support can be described in terms of mobile-impurity Hamiltonians~\cite{pereira08}.
For this purpose, the interacting Hamiltonian is projected onto a band structure which contains two subbands around the Fermi points $\pm k_F$ and one mobile-impurity subband around the momentum $k$. By continuity from the limit of the universal nonlinear LL, this impurity must carry the same quantum numbers as the fermion, and its overlap  with the fermion must be finite. The width of all subbands is small compared to the distance between $k$ and the Fermi points, so the spectra in all subbands can be linearized. The low-energy excitations near $\pm k_F$ can be modeled using a linear LL Hamiltonian with Fermi velocity $v$ and Luttinger parameter $K$. In analogy to the previous section, the mobile impurity Hamiltonian thus becomes for $|k| < k_F$,
\begin{align}\label{eq:phen_H0Hd}
 H_0 &= \frac{v}{2\pi} \int dx \left[ K (\nabla \theta)^2 + \frac{1}{K} (\nabla \phi)^2\right], \notag \\
 H_d &= \int dx\ d^{\dag}(x) \left[ \varepsilon_\th(k) - i v_d \nabla \right] d(x).
\end{align}
It is assumed throughout the paper that microscopic interactions decay faster than $\propto 1/x,$ so
that compressibility and sound velocity are finite.
The impurity velocity is given by $v_d = \partial \varepsilon_\th(k)/\partial k$.
In general, there are interactions between the impurity and right- and left-moving particles near the Fermi points,
\begin{align}\label{eq:phen_Hint}
 H_{int}
 &=
 \int dx \left[ V_R(k) \rho_R(x) + V_L(k) \rho_L(x) \right] d(x) d^\dag(x) \notag \\
 &= \int dx \left[ V_R(k) \nabla \frac{\theta - \phi}{2\pi} - V_L(k) \nabla \frac{\theta + \phi}{2\pi} \right] d d^\dag.
\end{align}
Unlike in the perturbative regime, Eq.~(\ref{eq:A_MI}),  where  $V_R(k)$ and $V_L(k)$ can be directly evaluated,  here they have to be treated as  momentum-dependent phenomenological parameters.
Similar to Sec.~\ref{sec:perturbative}, they determine the exponents of threshold singularities in all dynamic response functions.

The Hamiltonian (\ref{eq:phen_H0Hd})-(\ref{eq:phen_Hint}) was derived from simple phenomenological considerations, and provides a generalization of the LL theory beyond the low-energy limit. We note that effective quantum impurity Hamiltonians have been used before in the literature to describe the response properties when the effective impurity belongs to a band different from the one which forms the LL~\cite{sorella96,sorella98,castella93,furusaki99, carmelo99, balents00,mishchenko11, friedrich07, lamacraft09}. Here we extend their application to the case when both bands coincide.
Similarly to the case of a conventional LL, the Hamiltonian (\ref{eq:phen_H0Hd})-(\ref{eq:phen_Hint}) predicts not only the edge exponents in terms of $V_R(k)$ and $V_L(k),$ but also the structure of the finite-size theory (see Sec.~\ref{sec:finite}). For integrable models, some finite-size properties can be analytically extracted from the exact solutions~\cite{cheianov08,pereira08,pereira09,zvonarev09b,karimi11,shashi10,imambekov08}, and in combination with Eqs.~(\ref{eq:phen_H0Hd})-(\ref{eq:phen_Hint}) lead to nonperturbative predictions for the edge exponents. Various examples of this procedure will be reviewed in Sec.~\ref{sec:exact}. In the rest of this section, we will describe an alternative
procedure to phenomenologically relate the edge exponents to $\varepsilon_\th(k)$ only, which works for generic Galilean invariant systems~\cite{imambekov09b}.

Basically, the parameters $V_R(k)$ and $V_L(k)$ can be fixed by considering the shift of the edge $\varepsilon_\th(k)$ as a reaction to a uniform density variation and to a Galilean boost. These variations change $\rho_R$ and $\rho_L,$ and the resulting change in $ \varepsilon_\th(k)$ is determined by $V_R(k)$ and $V_L(k)$. First, consider the effect of a uniform variation of the system density $\delta \rho$. Because the edge of support is measured with respect to the chemical potential $\mu$, its position shifts by
\begin{align}\label{eq:phen_deltaE}
    \delta E = \left[ \frac{\partial \varepsilon_\th(k)}{\partial \rho} + \frac{\partial \mu}{\partial \rho} \right] \delta \rho
    = \left[ \frac{\partial \varepsilon_\th(k)}{\partial \rho} + \frac{\pi v}{K}\right] \delta \rho.
\end{align}
The second equality follows from the general result for the compressibility of a LL~\cite{giamarchi04}. Next, $\delta E$ is calculated using the mobile impurity Hamiltonian (\ref{eq:phen_H0Hd})-(\ref{eq:phen_Hint}), where a uniform density variation leads to a nonzero expectation value $\expcts{\nabla \phi}{} = -\pi \delta \rho$. We use this Hamiltonian to calculate the energy of a state containing an impurity at momentum $k$, and compare the energies with and without the density variation $\delta \rho$. A difference only emerges in $\expct{H_{int}}$ because it contains a term linear in $\expct{\nabla \phi}{}$. The total shift in the energy of the state reads
\begin{align}\label{eq:phen_VR_p_VL}
    \delta E = - \frac{V_R(k) + V_L(k)}{2} \delta \rho.
\end{align}
Equating the energy shifts (\ref{eq:phen_deltaE}) and (\ref{eq:phen_VR_p_VL}) leads to  $V_R(k) + V_L(k) = -2[\partial \varepsilon_\th(k)/\partial \rho + \pi v/K]$.

A second equation is needed to fix the difference $V_R(k) - V_L(k)$. For Galilean invariant systems, it can be derived by considering the shift of the edge position as a response to a boost with a small velocity $\delta u$~\cite{baym67}.
 Let us consider a hole at the edge of support in the moving frame which has momentum $k'$  and energy $\varepsilon'=\varepsilon_\th(k')$. Galilean invariance requires that in the rest frame it has momentum and energy
\begin{align}
  k = k'+m \delta u, \;\;
  \varepsilon = \varepsilon' + k' \delta u  + m\delta u^2/2.
\end{align}
Since the hole was at the edge in the moving frame and the system is Galilean invariant, this also corresponds to the edge
in the rest frame. Thus the shift of the threshold in the rest frame to linear order in $\delta u$ equals
\begin{align}\label{eq:phen_deltaE_prime}
    \delta E' &= \left[\varepsilon_\th(k-m\delta u) + (k-m\delta u) \delta u + m\delta u^2/2- \varepsilon_\th(k) \right] \notag \\
&\approx \left[ k - m \frac{\partial \varepsilon_\th(k)}{\partial k} \right] \delta u.
\end{align}
Again, we can now calculate the same energy using the mobile impurity Hamiltonian (\ref{eq:phen_H0Hd})-(\ref{eq:phen_Hint}). The boost changes the Fermi momenta of the right-moving and left-moving particles, $k_{F}^{R,L} = \pm k_F + m \delta u$, and thus leads to a difference between the densities of right- and left-movers. This, in turn, leads to a nonzero expectation value $\expcts{\nabla \theta}{} = m \delta u$. Substituting this into Eq.~(\ref{eq:phen_Hint}), we calculate the energy of a state with impurity at momentum  $k$ with the help of Eqs.~(\ref{eq:phen_H0Hd}) and (\ref{eq:phen_Hint}). Subtracting the corresponding energy for $\delta u = 0$, we find the energy shift due to the boost,
\begin{align}\label{eq:phen_VR_m_VL}
    \delta E' = - \frac{V_R(k) - V_L(k)}{2\pi} m \delta u.
\end{align}
Combining Eqs.~(\ref{eq:phen_deltaE_prime})-(\ref{eq:phen_VR_m_VL}) leads to  $V_R(k) - V_L(k)= 2 \pi [\partial \varepsilon_\th(k)/\partial k-k/m]$. Thus, both interaction potentials $V_{R,L}(k)$ can be expressed in terms of derivatives of the threshold $\varepsilon_\th(k)$ for arbitrary momenta $-k_F < k < k_F$.

In order to calculate the dynamic correlation functions, the interaction Hamiltonian $H_{int}$, see Eq. (\ref{eq:phen_Hint}), is removed like in Sec.~\ref{sec:perturbative} using a unitary transformation $U^\dag (H_0 + H_d + H_{int}) U$, where
\begin{align}\label{eq:U_phen}
 U = \exp\left\{ i \int dx \left(\frac{\delta_+(k)}{2\pi}  [\tilde \theta - \tilde  \phi] -\frac{\delta_-(k)}{2\pi}  [\tilde \theta + \tilde \phi]\right) d d^\dag \right\},
\end{align}
and $\tilde \theta$ and $\tilde \phi$ are defined in Eq.~(\ref{eq:rescaling}). Such a transformation removes $H_{int}$ if the momentum dependent phase shifts $\delta_+(k)$ and $\delta_-(k)$ are chosen such that
\begin{align}\label{eq:VRLtilde}
 \delta_-(k) = - \frac{\tilde V_L}{v_d+v}, \quad  \delta_+(k) = - \frac{\tilde V_{R}}{v_d-v},
\end{align}
where $\tilde V_{L(R)}$ are the couplings which are obtained after rescaling the bosonic fields using Eq.~(\ref{eq:rescaling}). They are related to parameters of $H_{int}$ by
\begin{align} \left(V_{L}-V_{R}\right)/\sqrt{K}=\tilde V_{L}-\tilde V_{R}
,\label{eq:Vviadelta1}\\
\left(V_{L}+V_{R}\right)\sqrt{K}=\tilde V_{L}+\tilde V_{R}. \label{eq:Vviadelta2}
\end{align}
Using the expressions for $V_R(k) \pm V_L(k),$ they can now be expressed in terms of derivatives of $\varepsilon_\th(k)$,
\begin{align}\label{eq:phen_phi}
 \frac{\delta_{\pm}(k)}{2\pi}&= \frac{1}{2 (\pm \frac{\partial \varepsilon_\th(k)}{\partial k} - v)}
 \bigg\{ \frac{1}{\sqrt{K}} \left[ \frac{k}{m} - \frac{\partial \varepsilon_\th(k)}{\partial k} \right] \notag \\
&\pm \sqrt{K} \left[ \frac{1}{\pi} \frac{\partial \varepsilon_\th(k)}{\partial \rho} + \frac{v}{K} \right] \bigg\}.
\end{align}
These phase shifts have the following symmetry property:
\begin{align}
\delta_{\pm} (k)= - \delta_{\mp}(-k)\;\; \mbox{for}  \;|k|<k_F.
\end{align}

The knowledge of the phase shifts allows the calculation of the edge singularities of $A(k,\varepsilon)$ at arbitrary momenta. In the interval $k \in [-k_F, k_F]$, the edge of support for negative (positive) energies is located at $\varepsilon = \varepsilon_\th(k)$ ($\varepsilon = -\varepsilon_\th(k)$). A periodic continuation of the functions $\pm \varepsilon_\th(k)$ yields the edge of support for arbitary $k$, as depicted in  Fig.~\ref{fig:AWeaklyInteracting}. Such thresholds, which can be constructed from $\varepsilon_\th(k)$ in the main region $k \in [-k_F, k_F]$ by using shifts and inversions, are called {\it shadow bands}. Hence, for $(2n-1) k_F < k < (2n + 1) k_F$ ($n \in \mathbb{Z}$), the lower and upper edges of support are at the energy $\varepsilon = \pm \varepsilon_\th(k_n)$, respectively, where
\begin{align} \label{eq: k_n_def}
k_n=k - 2 n k_F \in [-k_F, k_F].
\end{align}
Let us first discuss the hole sector, $\varepsilon \approx \varepsilon_\th(k_n) < 0$, at arbitrary $k$. The configuration responsible for the edge singularity can again be determined by requiring energy and momentum conservation. It contains an impurity hole on mass shell near momentum $k_n$, which carries almost the entire energy $\varepsilon$. The remaining momentum $2 n k_F$ is absorbed by creating $n$ low-energy holes near the right Fermi point and $n$ low-energy particles near the left Fermi point. Therefore, one projects the fermion operator as follows,
\begin{align}\label{eq:phen_Psi_proj1}
 \Psi(x) &\to e^{i k x} [\psi_L^\dag \psi_R]^n d \propto e^{i k x} \exp[-2 i n \phi(x)] d(x),
\end{align}
where we used the bosonization formula $\psi_{R,L} \propto \exp[-i ( \pm \phi - \theta)]$.
In this sector, the spectral function is the Fourier transform of $\expcts{\Psi^\dag(0,0) \Psi(x,t)}$ and its threshold singularities can be calculated using Eq.~(\ref{eq:phen_Psi_proj1}).

An analogous calculation is used in the particle sector, $\varepsilon \approx -\varepsilon_\th(k_n) > 0$ at arbitrary $k $. Here, the threshold configuration is a generalization of the one used in the perturbative calculation in Eq.~(\ref{eq:PT_Psik_project}). For $k_F < k < 3 k_F$, it contains an impurity hole at momentum $-k_{1}=2 k_F - k \in [-k_F, k_F]$ on mass shell, as well as two particles close to the right Fermi point. Similarly, for general $(2n - 1) k_F < k < (2n + 1) k_F$, the impurity carries the momentum
$-k_n,$ and there are $n+1$ particles at the right Fermi point and $n-1$ holes at the left Fermi point.
The corresponding projection now reads,
\begin{align}\label{eq:phen_Psi_proj2}
 \Psi^\dag(x) &\to e^{i k x} [\psi_R^\dag \psi_L]^{n-1} \psi_R^\dag \psi_R^\dag d \notag \\
 &\propto e^{i k x} \exp[ 2 i n \phi(x)] \exp[-2 i \theta(x)] d(x).
\end{align}
The spectral function can now be found by calculating the correlation function $\expct{\Psi(x,t) \Psi^\dag(0,0)}{}$ with the help of the diagonalized mobile-impurity Hamiltonian (\ref{eq:phen_H0Hd})-(\ref{eq:phen_Hint}). Combining the particle and hole sectors, the spectral function near the edges of support $\mp \varepsilon_\th(k_n)$ at arbitrary momentum $k$ reads, respectively,
\begin{align}\label{eq:phen_A}
 A(k,\varepsilon) \propto \theta[\varepsilon_\th(k_n) \pm \varepsilon] | \varepsilon_\th(k_n) \pm \varepsilon |^{-\mu_{n,\pm}(k)}.
\end{align}
Here the threshold exponents are given by \cite{imambekov09b}
\begin{align}
 \mu_{n,\pm}(k) = 1-\mu_{n,\pm,R} - \mu_{n,\pm,L}, \label{eq:munpm}
\end{align}
where $\mu_{n,\pm,R}$ and $\mu_{n,\pm,L}$ denote the contributions due to left and right Fermi points.
They are given by
\begin{align}
\mu_{n,-,\alpha} &= \left[  n \sqrt{K} - \delta_{\alpha}(k_n)/(2\pi) \right]^2,  \label{eq:mun-} \\
\mu_{n,+,\alpha} &= \left[ n \sqrt{K} + \alpha/\sqrt{K} + \delta_{\alpha}(-k_n)/(2\pi) \right]^2, \label{eq:mun+}
\end{align}
where $\alpha=R,L=+,-$. In conclusion, the spectral function at the edge of support still displays a sharp power-law even for momenta far away from the Fermi points. However, in contrast to the linear LL theory, the exponents become nonuniversal and momentum-dependent. In particular, they depend on the microscopic interactions not only through the Luttinger parameter $K,$ but also through the phase shifts $\delta_\pm(\pm k_n)$, which according to Eq.~(\ref{eq:phen_phi}) are determined by the shape of the edge $\varepsilon_\th(k)$.

Let us  now turn to the DSF $S(q,\omega)$. The key to the calculation of its edge singularities is finding the correct projection of the density operator $\rho(x) = \Psi^\dag(x) \Psi(x)$. It was shown in Eq.~(\ref{eq:LL_S}) that for $q \to 0$, $S(q,\omega)$ has a rectangular shape and the lower edge of support is at $-\tilde \xi(k_F - q) = v q - q^2/(2\tilde{m})$. The configuration which gives rise to this threshold consists of a hole on mass shell and a particle near the Fermi point. Therefore, by continuity the threshold for general momenta $0 < q < 2k_F$ is at $-\varepsilon_\th(k_F - q)$, and the density operator must be projected to $\rho(x) \propto \psi^\dag_R(x) d(x)$. Here, $\psi^\dag_R$ creates a particle near the right Fermi point and $d$ creates a hole at momentum $k_F - q$.

Let us generalize this argument to arbitrary momenta, $2 n k_F < q < 2(n+1) k_F$ ($n \in \mathbb{Z}$). Because $S(q,\omega)$ is symmetric, let us focus on $q \geq 0$. In analogy to the previous paragraph, the configuration responsible for the edge singularity at momentum $q$ now contains an impurity hole $d$ at momentum
\begin{equation}
 q_n = (2n + 1) k_F - q \in [-k_F, k_F].
\end{equation}
Moreover, it contains $n + 1$ particles near the right Fermi point and $n$ holes near the left Fermi point.
 Therefore, the density operator for momenta near $q$ is projected as
\begin{align}\label{phen_rho_project}
 \rho(x) &\to e^{i q x} \psi^\dag_R [\psi^\dag_R \psi_L]^n d \notag \\
&\propto e^{i q x} \exp[ i (2 n + 1) \phi(x)] \exp[-i \theta(x)] d(x),
\end{align}
and $S(q,\omega)$ can be found by calculating the correlation function $\expct{\rho(x,t) \rho(0,0)}{}$ using the projection (\ref{phen_rho_project}). The total energy of this configuration is close to $ -\varepsilon_\th(q_n)$. Hence, as a result one finds that
\begin{align}\label{eq:phen_S}
 S(q,\omega) \propto \theta[ \omega + \varepsilon_\th(q_n)] [\omega + \varepsilon_\th(q_n)]^{-\mu_n(q)}
\end{align}
with a momentum-dependent exponent
\begin{align} \label{eq:DSF_exponents}
 \mu_{n}(q) &= 1-\mu_{n,R} - \mu_{n,L}, \\
 \mu_{n,\alpha} &= \left[  \frac{(2n+1) \sqrt{K}}{2} + \frac{\alpha}{2\sqrt{K}} +\frac{\delta_{\alpha}(q_n)}{2\pi} \right]^2,\notag
\end{align}	
where $\alpha=R,L=+,-$, and $\mu_{n,R}$ and $\mu_{n,L}$ denote contributions due to left and right Fermi points.

For momenta close to $\pm k_F$ and interaction potentials decaying faster than $1/x^2$, the results of this section reproduce the universal phase shifts  of Sec.~\ref{sec:universal} at arbitrary interaction strength. Moreover, for $V(x) \propto 1/x^2$, the corresponding phase shifts confirm those found from the Bethe ansatz solution of the Calogero-Sutherland model, see Eq.~(\ref{eq:CS_phases}). For generic weak interaction potentials, the threshold position $\varepsilon_\th(k)$ can be calculated perturbatively and the phase shifts $\delta_\pm(k)$ can be derived from Eq.~(\ref{eq:phen_phi}). In this limit it is possible to recover the results of Sec.~\ref{sec:perturbative}.

The crucial step in the calculation of the exponents we outlined above is the
identification of the fermionic operator at the edges in terms of
the corresponding impurity operator, see Eqs.~(\ref{eq:phen_Psi_proj1}) and (\ref{eq:phen_Psi_proj2}).
A comparison with the solvable cases above shows that such
an identification indeed holds in the vicinity of the Fermi points for
any interaction strength, as well as at arbitrary momenta for weak
interactions. However, the identification of the state which corresponds to the edge of support may have to be modified
if the interactions affect the impurity spectrum too much, so that even in the region $|k|<k_F$ the impurity band doesn't
correspond to edge of support. The simplest scenario would be if at some $k$ the true edge of support
contains a single impurity with momentum close to $k$, as well as a low energy particle-hole pair at one of the branches.
We expect that it doesn't happen as long as
\begin{align}
\left|\frac{\partial\varepsilon_\th(k)}{\partial k}\right| <v \;\; \mbox{for} \; |k|<k_F. \label{eq:strong_ineq}
\end{align}
Equation (\ref{eq:strong_ineq}) guarantees that the phases in Eq.~(\ref{eq:phen_phi}) are continuous
functions of momentum, and the state which corresponds to the edge of
the spectral function in the basic region will contain a single impurity.
We note that the condition (\ref{eq:strong_ineq}) can be violated, \eg, for a microscopic
model which describes ultracold fermions with resonant interactions \cite{imambekov10}.

\subsection{Phenomenology of spin liquids}

\label{sec:spinchains}

Besides fermionic systems, spin chains constitute another noteworthy branch in the family of 1D quantum liquids. There is a plethora of experimentally accessible compounds which form antiferromagnetic Heisenberg chains and whose spin structure factor can be probed using neutron scattering.
In addition to neutron scattering, electron spin resonance (ESR) can be used to probe some of the results discussed here in the presence of Dzyaloshinskii-Moriya interactions~\cite{karimi11,povarov11}. In the present context, antiferromagnetic spin-$1/2$ systems are of particular importance because their low-energy degrees of freedom often fall into the universality class of a linear LL~\cite{giamarchi04}. The goal of this section is to extend the phenomenological considerations of Sec.~\ref{sec:phenomenology} towards response functions of 1D spin liquids. Beyond the low-energy limit, various microscopic models are used for the theoretical description of spin-$1/2$ chains. The phenomenological results of this section apply to a wide variety of model Hamiltonians of the form
\begin{align}\label{eq:spin_Hgen}
    H &= \sum_{n,n'} J_{n-n'} \left[ S^x_n S^x_{n'} + S^y_n S^y_{n'} + \Delta_{n-n'} S^z_n S^z_{n'} \right] \notag \\
    &- h \sum_{n} S^z_n,
\end{align}
where $\vec{S}_n = (S^x_n, S^y_n, S^z_n)$ are spin-$1/2$ operators located on the lattice sites $n$. The coupling between spins on different lattice sites is denoted by $J_{n}$. The exchange is anisotropic for $\Delta_n \neq 1$. An applied magnetic field in $z$-direction is denoted by $h$. Importantly, we assume throughout this section that the parameters $J_n$, $\Delta_n$ and $h$ are chosen such that the system is a liquid, \ie, the spectrum is gapless.

In order to introduce some general concepts of spin liquids on a simpler model, we start the discussion from the spin-$1/2$ XXZ model,
\begin{align}\label{eq:spin_HXXZ}
 H_{\rm XXZ} &= J \sum_{n} \left[ S^x_n S^x_{n+1} + S^y_n S^y_{n+1} + \Delta S^z_n S^z_{n+1} \right] \notag \\
 &- h \sum_{n} S^z_n.
\end{align}
The XXZ model can be solved exactly using the Bethe ansatz~\cite{korepin93,orbach58}
and this aspect will be discussed in more detail in Sec.~\ref{sec:XXZ}. At $\Delta=0$ the Hamiltonian (\ref{eq:spin_HXXZ}) is easily mapped onto a model of free spinless fermions, see Eq.~(\ref{eq:spin_XXZ_fermi}) below.
For $\Delta = 1$ the nearest-neighbor couplings are isotropic, and this case is referred to as XXX model.  If additionally $h=0,$ it  also  becomes $SU(2)$ invariant. For $|\Delta| > 1$ and $h=0$, the system is gapped and the ground state becomes ferromagnetic or antiferromagnetic, depending on the sign of $\Delta$.
On the other hand, the system remains gapless for $|\Delta| < 1$ and sufficiently small magnetic fields~\cite{haldane80,takahashi05,korepin93}. The low-energy sector in the corresponding parameter range can be modeled as an LL,
characterized by a Luttinger parameter $K \geq 1/2$. The limit $K=1/2$ is reached for the isotropic Heisenberg model ($\Delta = 1$, $h=0$), while the noninteracting case ($\Delta = 0$) leads to $K = 1$.
Certain properties of XXZ model which are specific to its integrability can be calculated using approaches which combine the phenomenology with the results of the Bethe ansatz~\cite{pereira06,pereira08,cheianov08,pereira09}, see Sec.~\ref{sec:XXZ}. In this section, however,  we will focus on its generic properties.

The XXZ model and its extensions we consider here preserve rotation symmetry in the $xy$ plane. The dynamic responses of this type of spin chain are therefore encoded in the transversal and longitudinal spin
structure factors, respectively,
\begin{align}\label{eq:spin_S_def_mp}
 S^{-+}(q,\omega) &= \sum_{n} e^{-i q n} \int dt e^{i \omega t} \expct{S^-_n(t) S^+_0(0)}{}, \\
 S^{zz}(q,\omega) &= \sum_{n} e^{-i q n} \int dt e^{i \omega t} \expct{S^z_n(t) S^z_0(0)}{},\label{eq:spin_S_def_zz}
\end{align}
where $S^\pm_n = S^x_n \pm i S^y_n$ and we assumed an infinite number of lattice sites. Due to the presence of the lattice, the quasimomentum $q$ is bounded, $|q| \leq \pi$.
Let us also mention that although we focus
on the spin structure factors, a similar phenomenology can also be used to find other correlation functions, \eg, the spin-exchange structure factor~\cite{klauser11} which determines the rate of resonant inelastic X-ray scattering~\cite{ament11}.

The spin operators can be mapped onto spinless fermions using a Jordan-Wigner transformation~\cite{giamarchi04},
\begin{align}\label{eq:spin_JW}
S^+_n &\to (-1)^n c^\dag_n e^{i \pi \phi_n},\;\;
S^-_n \to (-1)^n c_n e^{-i \pi \phi_n}, \notag \\
 S^z_n &\to c^\dag_n c_n - \frac{1}{2},
\end{align}
where $c^\dag_n$ ($c_n$) is a creation (annihilation) operator for a fermion on lattice site $n$ and the exponential of $\phi_n = \sum_{j = -\infty}^{n-1} c^\dag_j c_j$ denotes a Jordan-Wigner string. The latter is needed to ensure the proper anticommutation relation for fermions on different lattice sites, $\{ c^\dag_n, c_m \} = \delta_{nm}$. Because the operator $S^z_n$ has the simple representation (\ref{eq:spin_JW}) in terms of the fermionic density $\rho_n = c^\dag_n c_n$, $S^{zz}(q,\omega)$ is identical to the DSF of a system of spinless fermions if the density is measured with respect to the half-filled band.

The XXZ Hamiltonian has the following fermionic representation,
\begin{align}\label{eq:spin_XXZ_fermi}
 H_{\rm XXZ} &= -\frac{J}{2} \sum_n (c^\dag_n c_{n+1} + \text{h.c.}) - h \sum_n \left(\rho_n - \frac{1}{2} \right) \notag \\
&+ J \Delta \sum_n \left( \rho_n - \frac{1}{2} \right) \left( \rho_{n+1} - \frac{1}{2} \right).
\end{align}
In the fermionic language, a nonzero $\Delta$ corresponds to a short-range interaction potential. The magnetic field $h$ plays the role of the chemical potential.
The Jordan-Wigner transformation can also be applied to the more general Hamiltonian (\ref{eq:spin_Hgen}) and leads to higher-order, possibly non-local interaction terms in addition to Eq.~(\ref{eq:spin_XXZ_fermi}).

Due to this mapping on spinless fermions, many of the methods presented in the previous sections can be generalized to spin-$1/2$ chains with gapless spectra, but with certain peculiarities. First, one sees that unlike for Galilean invariant systems, the kinetic part in Eq.~(\ref{eq:spin_XXZ_fermi}) can have different signs of the nonlinearity. This results in a much wider variety of threshold behaviors. Second, for a general filling fraction ($h\neq 0$) the concept of an edge of support is, strictly speaking, not defined. This is a consequence of the presence of the lattice which results in ``foldings'' of the shadow bands of Sec.~\ref{sec:phenomenology} into the reduced zone scheme.
 However, the contributions to response functions from higher shadow bands are suppressed exponentially in their order. Indeed, in the language of fermions, the foldings come from umklapp processes which create additional particle-hole pairs and change the momentum of fermion system by multiples of the Brillouin zone period. If the Fermi wave vector is incommensurate with it, the creation of pairs is capable of reducing the energy of the ``deep'' hole to an arbitrarily low value~\cite{pereira09}. The lower the resulting energy value, the more particle-hole pairs need to be created. The creation of these additional particle-hole pairs makes the Anderson orthogonality catastrophe stronger and thus suppresses the corresponding contributions to the response functions.

In the most natural case of $h=0$, Eq.~(\ref{eq:spin_Hgen}) is $Z_2$-invariant, which corresponds to a particle-hole symmetric Hamiltonian in the fermionic representation~(\ref{eq:spin_XXZ_fermi}). As a consequence of particle-hole symmetry, the leading correction to spectrum is cubic in momenta. This new feature appears due to the presence of a lattice and has ramifications regarding the universal description of Sec.~\ref{sec:universal}. At the same time, the edge of support is still well-defined, since shadow bands fold onto each other~\cite{pereira09}; we will focus on the $h=0$ case in the bulk of this section, deferring to its end the consideration of the $h\neq 0$ case.

\subsubsection{XY model at zero magnetic field}
\label{sec:XY}

First, we will illustrate some of the generic properties of spin liquids and their field theoretical description  by considering in detail the isotropic $XY$ model which corresponds to $\Delta=0$ in the Hamiltonian~(\ref{eq:spin_HXXZ}). Its counterpart (\ref{eq:spin_XXZ_fermi}) describes free fermions. Their single-particle spectrum reads
\begin{equation}
\label{eq:xi-free}
\xi(k) = -J [\cos(k) - \cos(k_F)],\quad  k \in [-\pi, \pi],
\end{equation}
and at  $h=0$,  it is particle-hole symmetric, $k_F=\pi/2$.
In order to evaluate the longitudinal structure factor $S^{zz}_{\rm XY}(q,\omega)$ with the help of  quantum impurity model,  we need to find the appropriate projections of the operators $S^z_n$.
The lower threshold for $S^{zz}_{\rm XY}(q,\omega)$ is reached at the energy
\begin{align}
\omega_L(q) = - \xi(\pi/2 - q)=\xi (\pi/2+q).
\end{align}
The two corresponding configurations consist, respectively, of a deep hole with an accompanying low-energy particle, and a finite-energy particle accompanied by a low-energy hole (the low-energy particle and hole, respectively, are near the right Fermi point). Introducing the operator $d_1$ creating a hole with a momentum close to $\pi/2-q$, and the operator $d_2^\dagger$ creating a particle with momentum in the vicinity of $\pi/2+q$, we may write~\cite{pereira08}
\begin{align}\label{eq:spin_Sz_proj}
 S^z_n \to e^{iqn} \psi^\dag_R d_1+e^{i q n} d_2^\dag \psi_R.
\end{align}
Having two rather than one threshold configurations in the projection Eq.~(\ref{eq:spin_Sz_proj}) is a consequence of the particle-hole symmetry. Similar to the free-fermion density structure factor (\ref{eq:dsf-freefermions2}), the threshold frequency $\omega_L(q)$ reaches zero at $q=0$ and $q=2k_F=\pi$. The threshold energy as a function of $q$ is determined by the free-fermion spectrum (\ref{eq:xi-free}), and the corresponding exponent $\mu_z^-=0$.

\begin{figure}
\includegraphics[width=8cm]{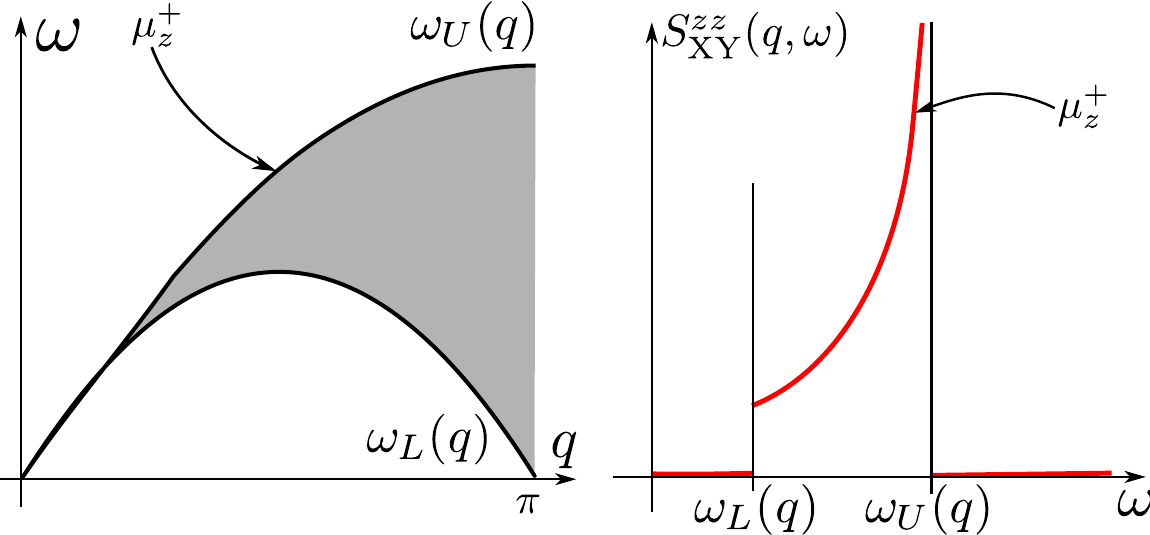}
\caption{\label{fig:SpinSzzXY} (Color online) Longitudinal spin structure factor $S^{zz}_{\rm XY}(q,\omega)$ for the isotropic XY model, see Eq.~(\ref{eq:spin_Szz_XY}). For fixed $q < \pi$, the function has a  step-function threshold at the lower edge of support $\omega_L(q)$ and diverges as an inverse square-root at the upper edge of support $\omega_U(q)$.}
\end{figure}

The nonzero values of $S^{zz}_{\rm XY}(q,\omega)$ are confined to a finite energy window \cite{mueller79,mueller81}. The upper threshold at $h=0$ is generated by particle-hole pairs with particle and hole momentum in the vicinity of $(\pi+q)/2$ and $(\pi-q)/2$, respectively. The projection of $S^z_n$ onto these states reads \cite{pereira09}
\begin{align}\label{eq:spin_Sz_proj_ph}
 S^z_n \to e^{iqn} d_1^\dag d_2\,,
\end{align}
and the threshold energy is given by
\begin{align}
\omega_U(q) = \xi(\pi/2 + q/2) - \xi(\pi/2-q/2).
\end{align}
 The structure factor is proportional to the convolution of the particle and hole spectral functions, resulting in the exponent $\mu_z^+ = 1/2$, independently of $q$. A direct calculation of the structure factor yields~\cite{mueller79,mueller81}
\begin{align}\label{eq:spin_Szz_XY}
 S^{zz}_{\rm XY}(q,\omega) = \frac{\theta[ \omega - \omega_L(q)] \theta[\omega_U(q) - \omega]}{[\omega_U^2(q) - \omega^2]^{1/2}}
\end{align}
The threshold energies and the edges of support of $S^{zz}_{\rm XY}(q,\omega)$ are depicted in Fig.~\ref{fig:SpinSzzXY}.

The evaluation of the transversal structure factor $S^{-+}_{\rm XY}(q,\omega)$, even at $\Delta = 0$, is more complicated due to the presence of the string operators in the fermionic representation of $S^\pm_n$, see Eq.~(\ref{eq:spin_JW}). The Jordan-Wigner transformation shifts the locations of zero modes of $S^{-+}_{\rm XY}(q,\omega)$ to $q=\pi \pm 2nk_F$, as in the conventional LL theory~\cite{giamarchi04} (note that zero mode at $q=\pi$ is present at any value of $h$). The thresholds of $S^{-+}_{\rm XY}(q,\omega)$ are determined by the same configurations that define the thresholds of the spectral function $A(k,\omega)$ for spinless fermions at momenta $k=\pi - q + k_F.$
For simplicity, we will refer here to the vicinity of $q = \pi$, \ie, $0< k-k_F\ll k_F$. Note that in the case of $h=0$, which we are still interested in, the threshold exponents obtained below are valid for the entire interval $|q|\leq\pi$.

\begin{figure}
\includegraphics[width=8cm]{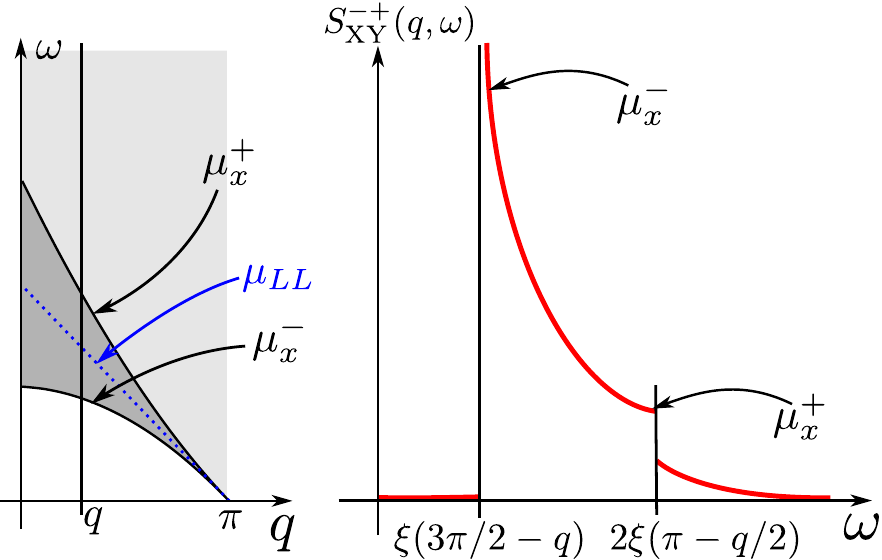}
\caption{\label{fig:SpinSxy} (Color online) Transversal structure factor $S^{-+}_{\rm XY}(q,\omega)$ for the isotropic XY model for $q \approx \pi$ at $h=0$. A divergent power-law with exponent $\mu^-_x =1/2$ is found at  the lower edge $\xi(3\pi/2-q)=-\xi(q-\pi/2).$  Note that the linearized Luttinger liquid (LL) theory predicts only one singularity with a different exponent $\mu_{\rm LL}=3/4.$ At higher energy, $\xi (\pi-q/2)-\xi(q/2)=2\xi (\pi-q/2),$ the transversal structure factor has a step-like feature, $\mu^+_x = 0$. }
\end{figure}

The mapping onto spinless fermions allows us to use the results of Sec.~\ref{sec:universal} for the threshold configurations. The singularity in $S^{-+}_{\rm XY}(q,\omega)$ at energy $\omega =\xi(3\pi/2-q)=-\xi(q-\pi/2)$ is the counterpart of the ``mass shell'' singularity in the spectral function of free fermions, see Sec.~\ref{sec:perturbative_spectral}. The particle-like configuration contributing to the projection of spin operators takes the form
\begin{align}\label{eq:spin_Sp_proj1}
 S^+_n \to e^{i \pi n} e^{-i k n} d^\dag \exp\left[ i \pi \int^x dy \rho(y) \right],
\end{align}
where $d^\dag$ creates an impurity particle at momentum $k_d =3\pi/2-q$. As $\Delta=0$, there is no interaction between the impurity and excitations near the Fermi level, and a straightforward calculation yields the threshold exponent $\mu_x^-=1/2$. Remarkably, even at $\Delta=0$ the nonlinear LL theory yields a result for $S^{-+}_{\rm XY}(q,\omega)$ which is different from the result $\mu_{\rm LL}=3/4$ obtained from an analysis of a linearized fermionic spectrum~\cite{giamarchi04}. The qualitative behavior of $S^{-+}_{\rm XY}$ is illustrated in Fig.~\ref{fig:SpinSxy}.

Turning now to the peculiarity in $S^{-+}_{\rm XY}(q,\omega)$ at the upper threshold $\xi (\pi-q/2)-\xi(q/2)=2\xi (\pi-q/2)$, we notice that the configuration representing $S_n^+$ and consisting of a minimal number of excitations should contain a particle-hole pair at momenta $k=\pi/2\pm (\pi- q)/2$ and a low-energy particle near the Fermi level,
\begin{align}\label{eq:spin_Sp_proj-u}
 S^+_n \to e^{i \pi n} e^{-i k n} d_1^\dag d_2\psi_R^\dagger \exp\left[ i \pi \int^x dy \rho(y) \right].
\end{align}
The time evolutions of the two impurities' creation and annihilation operators are independent of each other and of the evolution of the rest of the product, which includes the creation operator of a particle near the Fermi point and the string operator. The time evolution of the impurities is controlled by the corresponding single-particle Hamiltonians. The third component of the product can be treated by the standard  bosonization. This way, the spin correlation function in real space-time factorizes into a product of  three functions. Taking the proper long-time asymptotes and performing a Fourier transform, one finds $\mu_x^+=0.$
 Note that because of the presence of the string operator in Eq.~(\ref{eq:spin_Sp_proj-u}), the structure factor $S^{-+}_{\rm XY}(q,\omega)$ remains finite after the drop at $\omega=2\xi (\pi-q/2)$, see the sketch in Fig.~\ref{fig:SpinSxy}.
The closed-form analytical expression for the correlation function $\expct{S^-_n(t) S^+_0(0)}$ has been obtained recently~\cite{zvonarev09c} as a Fredholm determinant, and in principle contains the full information
about the properties of the spectral function $ S^{-+}_{\rm XY}(q,\omega).$

\subsubsection{Generic spin chains at zero magnetic field}
\label{sec:chains_generic}

Moving away from the XY model to the general Hamiltonian (\ref{eq:spin_Hgen}) introduces interactions between the  Jordan-Wigner fermions.
The location of zero modes of the spin liquid ($q=0$ and $q=\pi$ at $h=0$) is preserved, but the interactions affect the detailed shape of the thresholds $\omega_L(q)$ and $\omega_U(q)$, and generically lead to smearing of
$\omega_U(q)$ away from the low-energy regions.
The particle-hole symmetry inherent to the case of zero field
brings several new qualitative features which are absent for models in the continuum.
As a consequence of particle-hole symmetry, the quadratic term in the band curvature vanishes ($\tilde{m} = \infty$) and the leading spectrum nonlinearity at momenta close to $\pm k_F= \pm \pi/2$  becomes cubic. This makes a direct application of the universal results of Sec.~\ref{sec:universal} to the investigation of the singularities at $\omega=\omega_L(q)$ impossible. One still may use perturbation theory  of Sec.~\ref{sec:perturbative}  if the interaction constants $\Delta_n$ are small,
and, \eg, for the XXZ model one obtains $\mu_z^-\approx 2 \Delta/\pi$~\cite{pereira08}.
If the phase shifts are known non-perturbatively, then
the exponents for $S^{zz}(q,\omega)$ and $S^{-+}(q,\omega)$  can be evaluated using Eqs.~(\ref{eq:DSF_exponents}) and (\ref{eq:bosonic_exponents}), respectively.
 For the case of isotropic exchange ($\Delta_n = 1$),
the $SU(2)$ invariance of the equal-time correlations dictates~\cite{giamarchi04} $K=1/2$ irrespective of $J_n.$
The universal values of the threshold exponents can be established similarly with the help of $SU(2)$ symmetry in conjunction with the phenomenology.

The quasiparticle configurations with energies near the upper threshold $\omega_U(q),$
which is well defined in the low-energy region ($q\ll \pi$),
involve two mobile impurities.
Generically these two interact with each other, but as long as $h=0$ and particle-hole symmetry for
fermions is preserved, there is no interaction of the
two-impurity configuration
with the low-energy excitations~\cite{pereira08}. That brings universality to the exponents $\mu_z^+$ and $\mu_x^+$. The response function $S^{zz}(q,\omega)$ is associated with the creation of a particle ($d_1^\dagger$) and a hole ($d_2$) having respective momenta $\pi/2+p\pm q/2$, where $p \ll q$. Here, $q$ and $p$ are the total and relative momentum, respectively. The relative velocity of particle and hole motion vanishes at $p=0$, so the particle-hole interaction effect is strong even at infinitesimal interaction. The DSF near the threshold is proportional to the probability density $w(q,p)$ to create a particle and a hole at the same point, and to the joint density of states,
\begin{align} \label{eq:SqwU}
S(q,\omega)\propto\int dp\;\delta[\varepsilon(q,p)-\omega]w(q,p).
\end{align}
Here, $\varepsilon(q,p) = \omega_U(q)+E_r(p)$ is the energy of the pair near the threshold. The energy of the relative motion $E_r(p)=p^2/(2M)$ depends on $q$ via $1/M\propto d^2\omega_U/dq^2\propto q$, while $w(q,p)\propto |\psi_{p}(x=0)|^2$ is found from the solution $\psi_p(x)=\cos(p|x|+\vartheta)$ of the Schr\"odinger equation for the relative motion in the pair, ${\cal H}_{r}\psi_p(x)=E_r(p)\psi_p(x).$ Here
\begin{align}
{\cal H}_{r}=- \frac{1}{2M} \frac{\partial^2}{\partial x^2} +U_{12} \delta(x),
\end{align}
and $U_{12}\propto q^2$ is the effective interaction between the particle and hole (note that $M < 0$). The solution at small $|E_r(p)|\ll |M|U_{12}^2 \propto q^3$ yields $\psi_{p}(x=0)\propto \sqrt{E_r(p)/(|M|U_{12}^2)}.$
Therefore, using Eq.~(\ref{eq:SqwU}) at $ 0<\omega_U(q)-\omega \ll |\omega_U(q)- v q |,$ one obtains
\begin{align}\label{eq:spin_Sz_sqrt}
 S^{zz}(q,\omega) &\propto q^{-7/2} \theta[\omega_U(q) - \omega] \sqrt{\omega_U(q) - \omega} \notag \\
 &+ \text{regular terms}.
\end{align}
The momentum independent threshold exponent $\mu_z^+ = -1/2$ differs from its value at $\Delta_n=0$: interactions cause a discontinuous change in the edge exponent~\cite{pereira08}.

To describe $S^{-+}(q,\omega)$ near the upper threshold,
we may use projection of spin operators of Eq.~(\ref{eq:spin_Sp_proj-u}). The real space-time correlation function factorizes into  a product of two functions. The first one describes the evolution of the particle-hole pair, identical to that appearing in the $S^{zz}(q,\omega)$ correlation function. The second function in the product comes from the evolution of the string operator multiplied by the creation operator of a fermion near the $k_F=\pi/2$ point; that function depends on $K$. The resulting convolution yields $\mu_x^+=-3/2+1/(2K).$

As we may see from the derivation of Eq.~(\ref{eq:spin_Sz_sqrt}), the asymptotic behavior of  $S^{zz}(q,\omega)$ at small positive $\omega_U(q)-\omega$ is independent of the sign of the particle-hole interaction. However, at $U_{12}>0$ (we assume that $M<0$, as it is in the XXZ model), a bound particle-hole state is formed. It would show up, for instance, as an additional peak in $S^{zz}(q,\omega)$ at some value $\omega > \omega_U(q).$
Generically the energy of such bound state is finite for all momenta (as is illustrated by the XXZ model~\cite{pereira09}), and can be of the order of bandwidth for strong interactions.
The possible bound state peak  as well as  the upper threshold $\omega_U(q)$  (away from special points $q=0$ and $q=\pm\pi$) appear on the background of the spectral continuum, and will be generically smeared out.

Considering the upper threshold, we find $\mu_x^+=\mu_z^+$ at $SU(2)$ invariance,
$K=1/2$. This is not a coincidence but rather a consequence of $SU(2)$ symmetry which enforces $S^{xx}(q,\omega) = S^{yy}(q,\omega) = S^{zz}(q,\omega)$. Now we will use that constraint on the correlation functions in order to obtain their threshold behavior at $\omega=\omega_L(q)$. Together with the constraint, phenomenology alone
is sufficient to establish momentum-independent $\mu_z^-=1/2$ as the universal value independent of the  microscopic details.
 Using the mobile impurity Hamiltonian (\ref{eq:phen_H0Hd})-(\ref{eq:phen_Hint}) and the projections similar to Eq.~(\ref{eq:spin_Sz_proj}) and (\ref{eq:spin_Sp_proj1}),
 the edge exponents of both $S^{-+}(q,\omega)$ and $S^{zz}(q,\omega)$ can be calculated independently as functions of the phase shifts $\delta_\pm(q)$. Moreover, since the system is half-filled, umklapp scattering is allowed and can produce terms of the form $[\psi_L^\dag \psi_L^\dag \psi_R \psi_R]^m$ ($m \in \mathbb{Z}$) in the projections of $S^z_n$ and $S^+_n$. The edge exponents of the longitudinal and transversal spin structure factors, $\mu^{-}_{z,m}$ and $\mu^{-}_{x,m}$, become functions of $m$. As a consequence, the structure factors at the edge are characterized by sums of the form,
\begin{align}
 S^{zz}(q,\omega) \propto \sum_m |\omega - \omega_L(q)|^{-\mu^-_{z,m}}, \notag \\
 S^{-+}(q,\omega) \propto \sum_n |\omega - \omega_L(q)|^{-\mu^-_{x,n}}.
\end{align}
A calculation similar to Eqs.~(\ref{eq:munpm})-(\ref{eq:mun+}) combined with $K = 1/2$ leads to
\begin{align}
 \mu^-_{z,m} &= 1 - \frac{1}{2} \left( \sqrt{2} + \frac{\delta_+ - \delta_-}{2\pi} \right)^2  \\
&- \frac{1}{2} \left(- \frac{4 m - 1}{\sqrt{2}} + \frac{\delta_+ + \delta_-}{2\pi} \right)^2 \notag \\
 \mu^-_{x,n} &= 1 - \frac{1}{2} \left( \frac{\delta_+ - \delta_-}{2\pi} \right)^2
- \frac{1}{2} \left(- \frac{4 n + 1}{\sqrt{2}} + \frac{\delta_+ + \delta_-}{2\pi} \right)^2. \notag
\end{align}
As a consequence of $SU(2)$ invariance
the entire series for $\mu^-_{z,m}$ and $\mu^{-}_{x,n}$ have to coincide at each $q.$ This requirement
leads to  the  momentum-independent
phase shifts $\delta_\pm=\mp \pi/\sqrt{2}$
and exponents~\cite{imambekov09b},
\begin{align}\label{eq:spin_S_XXX}
 S^{zz}(q,\omega) =\tfrac{1}{2} S^{-+}(q,\omega) \propto |\omega - \omega_L(q)|^{-1/2}.
\end{align}
It coincides with the result of linear LL theory for $S^{-+}(q,\omega)$ at $q \to \pi$, $\mu_{\rm LL} = 1 - 1/(4K) = 1/2$~\cite{giamarchi04}. Note that similar to equal-time correlators~\cite{affleck89,singh89}, Eq.~(\ref{eq:spin_S_XXX}) will also have logarithmic corrections due to existence of marginally relevant terms in the Luttinger Hamiltonian.

\subsubsection{Generic spin chains at finite magnetic field}
\label{sec:chains_finiteh}

A finite magnetic field ($h\neq 0$) violates the $Z_2$ invariance of the spin liquid and destroys the particle-hole symmetry of the corresponding Hamiltonian in the fermionic variables. It shifts the Fermi points $\pm k_F$ away from $\pm \pi/2$;
their new positions $\pm k_F=\pm \pi (\langle S^z \rangle +1/2),$  are fully determined by the induced magnetization $\langle S^z\rangle.$
The positions of two of the low-energy regions, $q=0$ for $S^{zz}(q,\omega)$ and $q=\pi$ for $S^{-+}(q,\omega)$, do not shift.
The other two regions are shifted from $q=\pi$ to $q=\pm 2 k_F$ in $S^z(q,\omega),$ and from $q=0$ to $q=\pm (\pi-2k_F)$ in $S^{-+}(q,\omega)$~\cite{giamarchi04}.
As we already mentioned earlier in this section, folding of the shadow bands at $h\neq 0$ makes the thresholds in the response functions at arbitrary $q$ ill-defined. However, sharp threshold lines in the $(q,\omega)$-plane remain intact in the vicinities of the $\omega=0$ singularities of the response functions. Moreover, now the expansion in $k-k_F$ of the elementary excitations spectrum does contain the quadratic term, $1/\tilde{m}\neq 0$, so one may use the universal nonlinear LL theory, see Sec.~\ref{sec:universal}.
This theory is valid as long as the interaction potential $J_{n} \Delta_{n}$ decays faster than $1/n^2$.
For $k \to \pm k_F$, the phase shifts $\delta_\pm(k)$ reach the universal values (\ref{eq:LL_phases}).

The threshold behavior of $S^{zz}(q,\omega)$ at $q \to 0$ is identical to Eq.~(\ref{eq:LL_S}),
\begin{equation}
\label{eq:szz-exp}
S^{zz}(q,\omega) =
\frac{K |\tilde m|}{|q|} \theta\left(\frac{q^2}{2|\tilde m|} - |\omega - v q| \right),
\end{equation}
and the exponents $\mu_z^\pm=0$ are independent of the interactions. Here $v$ and $K$ are the characteristics of the linear LL, and $1/\tilde{m}=(v/K^{1/2})\partial v/\partial h+v^2/(2K^{3/2})\partial K/\partial h$  is phenomenologically related to these quantities~\cite{pereira06}.
Similarly, for $q \to \pi$ \cite{imambekov09a}
\begin{align}
\label{eq:spm-exp}
 S^{-+}(q,\omega) \propto \left| \omega - \left(v |q - \pi| \pm \frac{(q - \pi)^2}{2 \tilde m} \right) \right|^{\mp \frac{1}{\sqrt{K}} + \frac{1}{2K}}.
\end{align}
Note that the exponents $\mu_x^{\pm}=\pm(1/\sqrt{K})-1/(2K)$ differ from the linear LL prediction $\mu_{\rm LL} = 1 - 1/(4K)$\cite{giamarchi04}.

We also note that the region of validity of Eqs.~(\ref{eq:szz-exp})-(\ref{eq:spm-exp}) in $q$ becomes narrow ($\propto h$), as the magnetic field decreases. The interesting question of whether there is a universal description of the crossover from the results of Sec.~\ref{sec:chains_generic} to the universal description at nonzero magnetic fields remains to be addressed. The behavior of the response functions in the $(q,\omega)$ plane away from the zero-frequency special points is generally model-dependent.

\subsection{Bosonic systems with or without spin}

\label{sec:bosonic}

The interest in interacting 1D bosonic systems, see the recent review by~\textcite{cazalilla11}, has soared in past years mostly thanks to an increasing variety of experiments. The 1D bosons can be realized using Josephson junction arrays~\cite{fazio01}, helium in nanopores~\cite{wada01,delmaestro10,delmaestro11}, as an effective description of magnons in magnetic insulators~\cite{giamarchi08}, and most prominently in systems of ultracold atomic gases. In the latter case, one quite naturally often ends up with exactly solvable models, and these will be reviewed in detail in Sec.~\ref{sec:exact}. Here we will focus on a general phenomenological description of bosonic systems, both spinless and spinful.

Let us start with the discussion of the spinless case. In the absence of interactions, 1D bosonic systems are very different from fermionic ones. The former have a quadratic spectrum of excitations, while the excitation spectrum of the latter is linear. However, even for infinitesimally small repulsive interactions, bosons acquire a finite compressibility and their spectrum also becomes linear. The description of such bosons within the linear LL theory is not that different from the description of fermions~\cite{efetov75,haldane81b,giamarchi04}. The slowly varying component of the bosonic field is given by
$\Psi(x) \sim e^{i\theta(x)}.$
The difference between this equation
and, \eg, the bosonized expression for the right-moving fermionic field (\ref{eq:perturb_bosonize}) is the factor $\exp{\left[i(\pi \rho x -\phi)\right]}$,
which is nothing but a bosonized version of the Jordan-Wigner string operator $\exp{[i\pi\int_{-\infty}^{x}\Psi^{\dagger}(y) \Psi(y)dy]}$ discussed in Sec.~\ref{sec:spinchains}. As a consequence of that, the edge of support of the bosonic spectral function, which we denote by $A(q,\varepsilon)$, is shifted by $k_F$ compared to fermionic case.
For $\varepsilon>0$, it coincides with that of the DSF, and for historical reasons we shall denote it by $\varepsilon_2(q)>0$ in the region $0<q<2k_F$. For weakly interacting bosons, $\varepsilon_2(q)$ corresponds to excitation spectrum of dark solitons, see Sec.~\ref{sec:Lieb-Liniger}.

The Jordan-Wigner string does not appear in the bosonized form of the density operator. Therefore, the low-energy projection of the density operator does not depend on statistics of the particles (bosons or fermions). If one considers the DSF, it is not possible to distinguish spinless bosonic systems from spinless fermionic systems. Hence, the phenomenological relations of Sec.~\ref{sec:phenomenology} determine the DSF exponents (\ref{eq:DSF_exponents}) as well as the phenomenological phase shifts (\ref{eq:phen_phi}), assuming that the effective threshold position is identified as
\begin{align}
\varepsilon_{\rm th}(k=k_F-q)=-\varepsilon_2(q)<0 \;\;\mbox{for}\;\; 0<q<2k_F.
\end{align}
The same phase shifts and effective Hamiltonian also determine
the edge exponents of the spectral function, when combined with proper operator identifications similar to Eq.~(\ref{eq:spin_Sp_proj1}) for spin chains. Similarly to the equal-time field correlator, only contributions arising from the Jordan-Wigner string from the vicinities of the Fermi points have to be treated differently compared to the fermionic case.
Near the thresholds in the regions $2nk_F<q<2(n+1)k_F$, the exponents for $\varepsilon \gtrless 0$ are given by, \cite{imambekov09b}
\begin{align} \label{eq:bosonic_exponents}
 \mu^{b}_{n,\pm}(q) &= 1-\mu^b_{n,\pm,R} - \mu^b_{n,\pm,L},\\
 \mu^b_{n,-,\alpha} &= \left[ (2n+1)\sqrt{K}/2 + \delta_{\alpha}(q_n)/(2\pi) \right]^2, \notag \\
\mu^b_{n,+,\alpha} &= \left[ (2n+1)\sqrt{K}/2 + \alpha/\sqrt{K} - \delta_{\alpha}(-q_n)/(2\pi) \right]^2, \notag
\end{align}
where $\alpha=R,L=+,-,$ and $q_n= (2n+1)k_F-q$. The notation is illustrated in Fig.~\ref{fig:LiebLinigerFig}. Since the operators $S^\pm$ carry the same Jordan-Wigner string as the boson creation-annihilation operators, Eqs.~(\ref{eq:bosonic_exponents}) can also be used to express the exponents of $S^{-+}(q,\omega)$ in terms of the phase shifts.

Next, let us consider a system of interacting bosons which possess (iso)spin, and we will focus on the discussion of the $SU(2)$ symmetric case. Such systems in quasi 1D  configurations have already been realized with ultracold atomic gases~\cite{mcguirk02,higbie05, sadler06}. In addition to the DSF and the spectral function, we will be also interested in the longitudinal and transverse spin structure factors, $S^{zz}(q,\omega)$ and $S^{\pm \mp}(q,\omega)$ respectively. These are
defined by
\begin{align} \label{eq:SSFs}
  S^{\pm \mp}(q,\omega) &= \int dx dt e^{i(\omega t-qx)} \expct{S^\pm(x,t) S^\mp(0,0)}, \notag \\
  S^{zz}(q,\omega) &= \int dx dt e^{i(\omega t-qx)} \expct{S^z(x,t) S^z(0,0)},
\end{align}
where $S^\pm(x) = S^x(x) \pm i S^y(x)$. In terms of the physical particles, the spin density $\vec{S}(x) = (S^x(x), S^y(x), S^z(x))$ reads
\begin{align}
 \vec{S}(x) = \sum_{\sigma,\sigma'} \Psi^\dag_\sigma(x) \vec{S}_{\sigma\sigma'} \Psi_{\sigma'}(x),
\end{align}
where $\vec{S}_{\sigma\sigma'}$ denotes the vector of spin matrices (half of the Pauli matrices for spin-$1/2$).

Bose statistics requires that the total $N$-particle wave function, which consists of a spatial part and a spin part, must be symmetric under exchange of the positions and spins of any two particles. \textcite{eisenberg02} found that the unique spatial component of the ground state wave function is symmetric, so the spin wave function has to be symmetric as well. As a consequence, the ground state of the spinful Bose system is ferromagnetic (we will assume that the magnetization is pointing in $+z$ direction). This is in striking contrast to the case of spinful fermions, where the ground state is usually a spin singlet \cite{lieb62}. Due to the spin-polarized ground state, the bosonic system responds very differently to external perturbations which conserve the total spin and perturbations which change it. For simplicity, in what follows we will discuss only the case of (iso)spin$-1/2.$

The longitudinal spin structure factor $S^{zz}(q,\omega)$ does not change the total spin and is simply proportional to the DSF $S(q,\omega)$ for all $(q,\omega)$. The transverse spin structure factor $S^{-+}(q,\omega)$ vanishes because the system is already in a state with largest possible $S^z$, and the action of the operator $S^+(0)=\Psi^\dag_\uparrow(0) \Psi_\downarrow(0)$ destroys it. On the other hand, $S^{+-}(q,\omega)$ involves the action of the operators $S^-(0)=\Psi^\dag_\downarrow(0) \Psi_\uparrow(0)$ and $S^+(x) = [S^-(x)]^\dag$ that flip the spin of a single boson. The excitation of lowest energy which is created when $S^-(0)$ acts on the ferromagnetic ground state is a magnon. In an $SU(2)$ invariant system, the magnon spectrum towards small momenta is quadratic, $\omega_m(q) = q^2/(2 m^*)$, where $m^*$ denotes the effective magnon mass.
Due to the quadratic spectrum, the threshold energy at small $q$ of the transversal structure factor is always lower than that of $S^{zz}(q,\omega)$ and $S(q,\omega)$. Note that the quadratic form of the magnon spectrum makes the calculation of the transversal response function based on the linear LL theory impossible~\cite{zvonarev09a,zvonarev07a,
matveev08,akhanjee07,kamenev09}.

A magnon can be thought of a spin-down impurity moving in a liquid of spin-up particles. Unlike the mobile impurities  introduced in previous sections as an effective description,
these magnons can be separately experimentally addressed, as has been demonstrated recently in experiments with ultracold gases~\cite{palzer09,catani11}. An effective Hamiltonian
similar to Eqs.~(\ref{eq:phen_H0Hd})-(\ref{eq:phen_Hint}) can be used to describe the singularities in the spin structure factors, if one introduces an operator $d^\dag(x)$ creating a bosonic spin-down impurity with momentum near $q$ and energy near $\omega_m(q)$. Then the projection of the operator $S^-(x)$ onto the three-subband model reads
\begin{align}\label{eq:boson_proj_sp}
 S^-(x) \propto e^{-i q x} d^\dag(x) e^{i \theta(x)}.
\end{align}
The phenomenological approach of Sec.~\ref{sec:phenomenology} can also be directly generalized, and leads again to explicit predictions for the transverse spin structure exponent
at the magnon spectrum $\mu_m(q)$ which were presented by~\textcite{kamenev09}.
The general expressions simplify somewhat in the limit $|q|\ll m^* v$ due to the quadratic spectrum, and in order $O(q^4)$ are given by
\begin{align}
\mu_m(q) =1 -\frac{K q^2}{2(\pi \rho)^2}\left[1+\left(\frac{q}{m^* v}\right)^2(3+ 4\sigma + \sigma^2)\right], \notag
\end{align}
where $\sigma=-\rho/(2m^*)\partial m^*/\partial \rho$.

\subsection{Fermionic systems with spin}
\label{sec:spinful}

In this section, some of the concepts introduced in the previous
sections will be applied for the exploration of the properties of
spin-$1/2$ fermions in 1D. Compared to spinless fermions, a
complication arises due to the spin-charge separation encountered in
interacting 1D systems: a generic density-density
interaction couples fermions of opposite spins and lifts the
degeneracy between charge modes (excitations symmetric in spin-up and spin-down) and spin modes (excitations antisymmetric in spin-up and spin-down) present in the noninteracting system.
Within the linear LL theory,
the eigenstates of the interacting system are then
linear combinations of spin-up and spin-down excitations and can be
interpreted as density waves which carry either only spin or only
charge. These two types of excitations propagate at different
velocities,
$v_s$ and $v_c$, respectively.

The phenomenon of spin-charge separation is the hallmark of the linear spinful LL theory~\cite{dzyaloshinskii74}: the fermionic Green's function
is a product of two parts which describe excitations propagating with velocities $\pm v_s$ and $\pm v_c$, respectively.
The fermionic spectral function which can be obtained from the Green's function by Fourier transformation
thus has power-law singularities at the spin and charge modes \cite{voit93a,voit93b,schonhammer92}.
The spin and charge density structure
factors, on the other hand, are delta-functions reflecting the
linear spectra of the two types of bosonic modes.

The existence of two distinct excitation energies for a
given momentum has already been observed in a number of experiments
with solid state systems~\cite{auslaender02,auslaender05,kim06,jompol09}.
Moreover, experiments using 1D ultracold fermions~\cite{liao10}
could allow the observation of spin-charge separation in real space~\cite{recati03,kollath06}.

In this section, we will review recent theoretical progress \cite{schmidt10a,schmidt10b,pereira10} in
understanding the notion of spin-charge separation in 1D models beyond
the linear LL paradigm.
This question has received some attention over the
years.
The existence of a certain form of spin-charge separation at all energy scales is implicitly built into the structure of
the exact solutions of spinful integrable models
which will be discussed in more detail in Sec.~\ref{sec:exact}.
For instance, \textcite{ogata90} demonstrated that the eigenstates
of the strongly interacting 1D Hubbard model (see Sec.~\ref{sec:XXZ}) factorize into charge parts
and spin parts. The former are identical to the eigenmodes of free
spinless fermions, whereas the latter are solutions of the XXZ model.
In an extension of this approach called pseudofermion dynamical theory~\cite{carmelo99,carmelo04,carmelo05,carmelo06}, some of the results of the present section have been envisioned before for the integrable 1D Hubbard model, but its field-theoretical basis and applicability for general nonintegrable models remain unclear.

The approach of \textcite{ogata90} can be applied
even without an underlying lattice or integrability, because
strong repulsion
makes the charge mode stiffer while softening the spin excitations.
The enhanced rigidity of the charge spatial structure allows
one~\cite{matveev04,matveev04b,matveev07,matveev07b} to invoke the notion of a \emph{Wigner
crystal} \cite{wigner34}, in which fermions arrange on a 1D lattice, as a starting point for the consideration of the
dynamics, notwithstanding the absence of long-range crystalline
order in 1D. The dispersion of the spin excitations is suppressed
in the amplitude of the repulsion
potential. In the limit of strong repulsion
it is tempting to dispense with the spin dynamics altogether,
and to assume that arbitrarily oriented spins are attached to the sites of
the Wigner crystal and that its excitations are sound waves. Such a model
was dubbed \emph{spin-incoherent} LL. The bandwidth of the spin excitations
is zero in the spin-incoherent LL,
and at any temperature
the real-space electron Green's function decays exponentially in the spatial
coordinate. This reflects the absence of any order in the spin system in the absence of exchange interaction between the spins
~\cite{cheianov04,cheianov04b,fiete04,fiete06,fiete09}. We refer the reader to an excellent review of spin-incoherent LLs for details~\cite{fiete07}.

In contrast to the notion of the spin-incoherent LL, here we concentrate on the generic case of comparable (but different) velocities of spin and charge excitations. As we shall see, the
concept of spin-charge separation survives, albeit in a modified form,
beyond the low-energy limit and the assumption of a linear spectrum.
In the following, we shall focus on the case of repulsive
interactions. In this case, the system remains gapless in both charge
and spin sectors, and charge excitations propagate faster than spin
excitations ($v_c > v_s$). The parameter $K_c$ characterizing the
linear LL of charge excitations is in the range $0 < K_c
< 1$. For simplicity, we will only consider systems without Zeeman splitting,
although a small magnetic field might be beneficial to detect spin-charge separation in experiments~\cite{rabello02,grigera04}.
The resulting $SU(2)$ invariance generically leads to the spin LL parameter $K_s=1$.

The investigation of a nonlinear spectrum in the bosonic language runs
into similar problems as in the spinless case
(Sec.~\ref{sec:universal}). A nonlinearity leads to interactions
between the bosonic modes. To lowest order, the corresponding
self-energy diverges at energies $\omega = v_{c,s} k$. It is still
possible to
investigate certain properties using the bosonic approach~\cite{brazovskii93,brazovskii94,vekua09}, \eg,
the dynamic response functions sufficiently far
away from the thresholds \cite{pereira10,teber07}, but closer to the
threshold perturbation theory fails. An efficient resummation
scheme which eliminates these divergencies has not been developed yet.
Therefore, it is easier to generalize the fermionic quasiparticle methods used in
the previous sections to spinful systems. The path to a fermionic
description is similar to the one followed in Sec.~\ref{sec:universal}
for spinless fermions. For a linear fermion spectrum, the
Hamiltonian becomes a sum of independent charge and spin parts,
$H_{\rm LL} = H_c + H_s$. The charge term $H_c$ has the form of the
linear LL Hamiltonian (\ref{eq:HLL}) and it is
characterized by the charge velocity $v_c$ and the Luttinger
parameter $K_c$. Its eigenmodes are bosonic charge density waves.
The spin part $H_s$ also contains a quadratic term of the form
(\ref{eq:HLL}), characterized now by a different Luttinger parameter
$K_s$ and the spin velocity $v_s$. In addition, however, $H_s$
generally contains a sine-Gordon term, which describes spin-flip
scattering between the physical fermions. For repulsive interactions
it is marginally irrelevant and vanishes in the limit of small
bandwidth \cite{solyom79}. Then, the eigenmodes of $H_s$ are bosonic
spin density waves.

Within the linear LL theory, the physical fermions can be expressed is terms of left- and right-movers using $\Psi_{\sigma}(x) = e^{i k_F x} \Psi_{R\sigma}(x) + e^{-i k_F x} \Psi_{L\sigma}(x)$ where $\sigma = \uparrow,\downarrow$ denotes the spin. The bosonization rules for the fermionic operators are given by~\cite{giamarchi04}
\begin{align}\label{eq: bos_iden}
 \Psi_{\alpha\sigma}(x) \propto  e^{-i/\sqrt{2} [ \alpha \phi_c(x) - \theta_c(x) + \alpha \sigma \phi_s(x) -\sigma \theta_s(x) ]},
\end{align}
where $\alpha = R,L = +,-$.
Because $H_c$ and $H_s$ are formally identical
to linear LL Hamiltonians, it is possible to express
them in terms of free quasiparticles by generalizing
Eq.~(\ref{eq:LL_HLL}),
\begin{align}\label{eq:sLL_H}
 H_c &= - \sum_{\alpha} i \alpha v_c \int dx : \tPsi_{\alpha c}^\dag(x) \nabla \tPsi_{\alpha c}(x) :, \notag \\
 H_s &= - \sum_{\alpha} i \alpha v_s \int dx : \tPsi_{\alpha s}^\dag(x) \nabla \tPsi_{\alpha s}(x) :,
\end{align}
where $\alpha = R,L = +,-$. The fermionic quasiparticles
$\tPsi_{\alpha\nu}(x)$ ($\nu = c,s$) describe low-energy charge and
spin excitations, which we shall refer to as holons and
spinons, respectively. As in the spinless case, see
Eq.~(\ref{eq:LL_PsiR}), the physical fermions $\Psi_{\alpha\sigma}(x)$
can be expressed in terms of the quasiparticles by using the
bosonization  identity  (\ref{eq: bos_iden}),
then rescaling the bosonic fields and finally
refermionizing them. For the right-movers, the result reads
\begin{align}\label{eq:sLL_spin_psi_tpsi}
 \Psi_{R\uparrow}(x)
&\propto
 \tPsi_{R c}(x) F_{R c}(x) \tPsi_{R s}(x)
F_{R s}(x), \notag \\
 \Psi_{R\downarrow}(x)
&\propto
 \tPsi_{R c}(x) F_{R c}(x) \tPsi^{\dag}_{R s}(x)
F^\dag_{R s}(x).
\end{align}
The string operators are given by
\begin{align}\label{eq:sLL_spin_string}
  F_{R\nu}(x) = \exp\left\{ -i \int_{-\infty}^x dy \left[
      \delta_{+\nu} \trho_{R \nu}(y) + \delta_{-\nu}
      \trho_{L\nu}(y) \right] \right\},
\end{align}
where $\trho_{\alpha\nu}(x) =
\tPsi^\dag_{\alpha\nu}(x)\tPsi_{\alpha\nu}(x)$ ($\alpha = R,L$, $\nu = c, s$) denote
the right- and left-moving quasiparticle densities, respectively.
Note that Eq.~(\ref{eq:sLL_spin_psi_tpsi}) contains charge and spin string operators, although we assume only $K_c\neq 1$. The
appearance of the string operators in Eq.~(\ref{eq:sLL_spin_psi_tpsi}) is an inevitable consequence of the attempted
``splitting'' of a spinful fermion into two spinless fermionic excitations. The phase shifts are determined by $K_c$,
\begin{align}\label{eq:sLL_phases}
  \frac{\delta_{+c}}{2\pi} &= 1 - \sqrt{\frac{1}{8 K_c}} - \sqrt{\frac{K_c}{8}}, \notag \\
  \frac{\delta_{-c}}{2\pi} &= \sqrt{\frac{1}{8 K_c}} - \sqrt{\frac{K_c}{8}} ,\notag \\
  \frac{\delta_{+s}}{2\pi} &= 1 - \frac{1}{\sqrt{2}}, \;
  \frac{\delta_{-s}}{2\pi} = 0.
\end{align}
Note that the Klein factors, which are required to ensure the correct
fermionic anticommutation relations, were neglected in
Eq.~(\ref{eq:sLL_spin_psi_tpsi}). This is justified as long as it is
used to calculate expectation values of operators which conserve
charge as well as spin. This is indeed the case for all dynamic
response functions.

According to Eq.~(\ref{eq:sLL_spin_psi_tpsi}), the creation of a
physical spin-up (spin-down) particle leads to the formation of a
holon and the creation (annihilation) of a spinon. The string
operators $F_{R\nu}(x)$ reflect the shake-up of the spinon and holon
Fermi seas due to the addition of the physical fermion.
The phase shifts (\ref{eq:sLL_phases}) characterize the strength of
the shake-up. Keeping the spectrum of fermions linear as in Eq.~(\ref{eq:sLL_H}), and using
Eqs.~(\ref{eq:sLL_spin_psi_tpsi})-(\ref{eq:sLL_phases}), one may readily reproduce
the known results for the spectral function of fermions in a
spinful linear LL~\cite{schmidt10b}. The spectral function $A(k,\varepsilon)$ for fixed
$k > k_F$ now displays \emph{two} peaks: divergent power-law
singularities are located at the
energies of the spinons and the
holons, $\varepsilon = v_s(k - k_F)$ and $\varepsilon = v_c(k -
k_F)$. Within the linear LL theory, the exponents at both thresholds are determined by the phase
shifts $\delta_{\pm \nu}$ and thus by the Luttinger parameter $K_c$
\cite{schonhammer92,voit93a,voit93b,meden99}.

A nonzero curvature of the physical fermion spectrum generally leads
to interactions between the spinons and holons. Moreover, it also
bends the spectra of both the spinons and the holons. For $k \to k_F$,
the leading correction to the holon spectrum is quadratic,
\begin{align}\label{eq:sLL_xic}
 \tilde \xi_c(k) \approx v_c( k - k_F) + \frac{(k - k_F)^2}{2\tilde m}.
\end{align}
The effective mass $\tilde m$ can be determined phenomenologically and the
result resembles the spinless case (\ref{eq:LL_meff}),
\begin{align}\label{eq:sLL_meff}
  \frac{1}{\tilde m} = \frac{v_c}{\sqrt{2} K_c} \frac{\partial}{\partial\mu} (v_c \sqrt{K_c}).
\end{align}
The shape of the spinon spectrum $ \tilde  \xi_s(q)$, on the other hand,
becomes rather different. From Eq.~(\ref{eq:sLL_spin_psi_tpsi}), it
can be inferred that the spin-up/spin-down symmetry of the physical
fermions entails particle-hole symmetry of the spinons. Therefore,
there is no quadratic term in $\tilde \xi_s(q)$. Instead, for $q \to 0$ the
leading nonlinearity is cubic,
\begin{align}\label{eq:sLL_xis}
  \tilde  \xi_s(q) \approx v_s q - \gamma q^3 \quad \mbox{for } |q|\ll k_F,
\end{align}
with $\gamma > 0$. The spinon spectrum near the Fermi points is therefore
similar to the low-energy spectrum of the XXX model
in zero magnetic field
encountered
in Sec.~\ref{sec:spinchains}. In both cases, the shape is a direct
consequence of $SU(2)$ symmetry.

For repulsive interactions
one has $v_s < v_c$, so for a given momentum $k \approx \pm k_F$ exciting a spinon requires
less energy than exciting a holon.
Since the
thresholds in all dynamic correlation functions should be continuous, they
coincide with a shifted spinon mass shell $\tilde \xi_s(q)$ for arbitrary
momenta.
The precise shape of $\tilde \xi_s(q)$ away from Fermi points
depends on the microscopic details of the interaction and is generally
unknown.

The power-law singularities characterizing the dynamic responses
beyond the linearized theory
can be determined by generalizing the mobile-impurity Hamiltonian
(\ref{eq:LL_imp}) to the spinful case.
Within the linearized theory, all
threshold singularities in the dynamic response functions are
characterized by configurations where a spinon carries almost the entire available energy.
Hence, the quantum numbers of the effective impurity should coincide with that of a spinon near the Fermi points.
By continuity, this remains true also for momenta away from Fermi points.
For a spinon impurity
at momentum $q_d = k - k_F$ where $k \in [-k_F, k_F]$, we use the
projection $\tPsi_{Rs}(x) \to e^{i q_d x} \td(x) + \tpsi_{Rs}(x)$. The
spinons and holons near the Fermi points, as well as the impurity at
momentum $q_d$ are then described by
\begin{align}\label{eq:sLL_imp}
 H_0
&= - \sum_{\alpha} i \alpha v_c \int dx \left[ : \tpsi_{\alpha c}^\dag(x) \nabla \tpsi_{\alpha c}(x) : \right] \notag \\
&- \sum_{\alpha} i \alpha v_s \int dx \left[ : \tpsi_{\alpha s}^\dag(x) \nabla \tpsi_{\alpha s}(x) :  \right], \notag \\
 H_d &= \int dx \td^\dag_s(x) \left[ \tilde \xi_s(q) - i v_d \nabla \right] \td_s(x),
\end{align}
where $\alpha = R, L = +, -$ and $v_d = \partial \tilde  \xi_s(q)/\partial q.$ The interaction between
the impurity and the low-energy spinon and holon degrees of freedom
leads to
\begin{align}\label{eq:sLL_Hint}
 H_{int} &= \int dx \sum_{\alpha=R,L} \sum_{\nu = c,s} \tilde{V}_{\alpha\nu}(k) \trho_{\alpha\nu}(x) \td_s^\dag(x) \td_s(x).
\end{align}
The term $H_{int}$ can be removed by using a unitary transformation
similar to Eq.~(\ref{eq:U_phen}).

The impurity causes phase shifts of the quasiparticles near the Fermi
points,
\begin{align}\label{eq:sLL_Deltadelta}
 \Delta\delta_{\alpha\nu}(k) = \frac{\tilde{V}_{\alpha\nu}(k)}{v_d - \alpha v_\nu},
\end{align}
where $\alpha = R,L = +,-$ and $\nu = c,s$. Let us first investigate
these phase shifts in the limit $k \to k_F$. For interactions between
the physical fermions which decay faster than $\propto 1/x^2$, the interaction
potentials $\tilde{V}_{\alpha\nu}(k)$ fulfill
$\tilde{V}_{\alpha\nu}(k) \propto (k - k_F)^2$. The denominators in
Eq.~(\ref{eq:sLL_Deltadelta}) are determined by the cubic shape of the
spinon spectrum (\ref{eq:sLL_xis}). In the limit $k \to k_F$, one
finds $v_d \pm v_c \to \text{const.}$ and $v_d + v_s \to
\text{const.}$, so the phase shifts $\Delta\delta_{\pm c}$ and $\Delta
\delta_{-s}$ vanish. In contrast, $v_d - v_s \propto (k - k_F)^2$, so
the phase shift $\Delta \delta_{+s}$ remains nonzero even for $k \to
k_F$.

This is a striking contrast to the spinless case, where all phase shifts vanish in the limit $k \to k_F$, see Eq.~(\ref{eq:LL_imp}). The latter is a consequence of the quadratic spectrum of the spinless fermions near $k_F$  \cite{imambekov09a}. The spinon spectrum (\ref{eq:sLL_xis}), on the other hand, is cubic due to $SU(2)$-symmetry. Thankfully, the phase shifts $\Delta \delta_{\pm s}(k)$ can be determined for arbitrary momenta by exploiting the $SU(2)$ symmetry.

In order to determine these phase shifts, we use the same argument
that led to the spin structure factor of the XXX model in
Eq.~(\ref{eq:spin_S_XXX}): for an $SU(2)$ symmetric system, the
threshold exponents of the two components of the spin structure factors
$S^{-+}(q,\omega)$ and $S^{zz}(q,\omega)$, defined in Eq.~(\ref{eq:SSFs}), have to coincide at
arbitrary momenta $q$.

As previously, the first step in calculating these exponents consists in
identifying the threshold configuration and projecting the operators
$S^\pm(x)$ and $S^z(x)$ accordingly. Then, the mobile impurity
Hamiltonian (\ref{eq:sLL_imp})-(\ref{eq:sLL_Hint}) can be used to calculate the threshold
exponents as functions of $\Delta \delta_{\alpha\nu}(k)$. The
requirement of identical exponents for $S^{-+}(q,\omega)$ and
$S^{zz}(q,\omega)$ unambiguously entails that for arbitrary momentum
$k$ \cite{schmidt10a,pereira10}
\begin{align}\label{eq:sLL_spin_phases}
 \delta_{\pm s} + \Delta \delta_{\pm s}(k) = 0.
\end{align}

Having fixed the phase shifts $\Delta\delta_{\pm s}$ by
Eq.~(\ref{eq:sLL_spin_phases}), only the values of $\Delta \delta_{\pm
  c}(k)$ are left to be determined.
For integrable models, $\Delta \delta_{\pm c}$ can be extracted
exactly from the finite-size corrections of the Bethe ansatz spectrum~\cite{carmelo08, essler10}
similar to the procedure discussed in Sec.~\ref{sec:finite}.
 For generic Galilean invariant models, a
generalization of the phenomenological approach of
Sec.~\ref{sec:phenomenology} is possible. It relates $\Delta
\delta_{\pm c}$
in the interval $|k|<k_F$ to the shape of the edge of support $\epsilon_s(k)=\tilde \xi_s(k-k_F)<0$
\cite{schmidt10a}:
\begin{align}\label{eq:sLL_phen}
 \frac{\Delta \delta_{\pm c}(k)}{2\pi} = \pm \frac{\frac{k - k_F}{m \sqrt{K_c}}
\pm \sqrt{K_c} \left( \frac{2}{\pi} \frac{\partial \epsilon_s(k)}{\partial \rho_c}
+ \frac{\partial \epsilon_s(k)}{\partial k} \right)}{2 \sqrt{2} \left( \frac{\partial \epsilon_s(k)}{\partial k} \mp \frac{k_F}{m K_c}
 \right)},
\end{align}
where $m$ is the mass of the physical fermions. In the limit $k \to
\pm k_F$, the low-energy expansion of the spinon spectrum $\tilde \xi_s(q)$
leads back to $\Delta \delta_{\pm c} = 0$. This is a consequence of the conventional spin-charge separation of the linear LL theory. The
equations (\ref{eq:sLL_spin_phases}) and (\ref{eq:sLL_phen}) fix all
parameters in the mobile impurity Hamiltonian (\ref{eq:sLL_imp}) and
are thus the key for the calculation of edge exponents in the various
dynamic response functions.

\begin{figure}
\includegraphics[width=8cm]{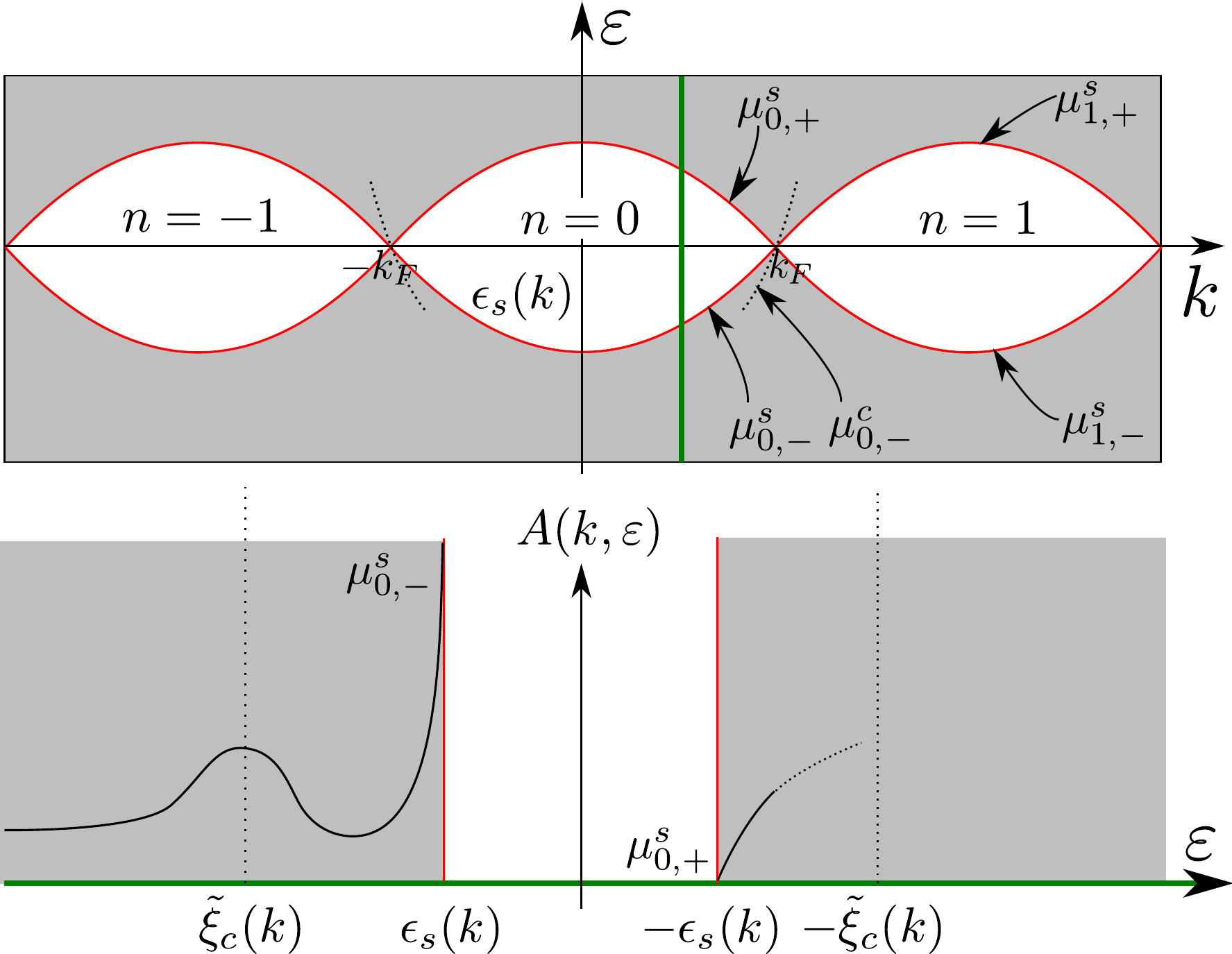}
\caption{\label{fig:Aspinful} (Color online) Spectral function $A(k,\varepsilon)$ for spinful fermions ($|k| < k_F$), see Eq.~(\ref{eq:sLL_A}). For repulsive interactions, the edge of support is determined by the spinon spectrum  $\epsilon_s(k).$ The power-law singularity at the holon mass shell $\tilde{\xi}_c(k)$ is generally broadened away from Fermi points.}
\end{figure}

Let us start with the spectral function. For the calculation of the
threshold behavior of $A(k,\varepsilon)$ for $0 < k < k_F$, one needs to consider the
response of the system to the addition of a physical spinful
fermionic hole
with a given momentum $k$ and the threshold energy $\epsilon_s(k)<0.$ The incoming hole creates a spinon on the mass shell,
which absorbs the entire energy, as well as a holon at a Fermi point.
That spinon is protected from decay by energy and momentum
conservation. If the energy of the incoming hole is slightly below
$\epsilon_s(k)$, the excess energy
is used for the creation of
low-energy particle-hole pairs in the holon sector. However, due to
Eq.~(\ref{eq:sLL_spin_phases}) it cannot be used for the creation of
additional spinons! Since the velocity of the spinon impurity is
smaller than the holon velocity, $v_d = \partial \epsilon_s(k)/ \partial k
< v_c$, the spinon becomes spatially separated from the holons. While
this effect is reminiscent of the conventional spin-charge separation
of the linear LL theory, it should be pointed out that this new form
of spin-charge separation
survives for  arbitrary momenta only for energies close to the threshold.
Finding the threshold exponents of the spectral function $A(k,\varepsilon)$ at $k > 0$ requires a projection of the fermionic operator $\Psi_{R\sigma}(x)$ in Eq.~(\ref{eq:sLL_spin_psi_tpsi}). Due to $SU(2)$ symmetry, the result does not depend on the choice of $\sigma = \uparrow,\downarrow$. For $0 < k < 2 k_F$, one can therefore project
\begin{align}
 \Psi_{R\uparrow}(x) \to e^{i k x} \tpsi_{Rc} F_{Rc} \td_s F_{Rs}.
\end{align}
For general $(2n - 1)k_F < k < (2n + 1)k_F$ ($n \in \mathbb{Z}$), the excess momentum can be used to create additional particle-hole  interbranch pairs similar to Sec.~\ref{sec:phenomenology}. The mobile-impurity Hamiltonian now allows a calculation of the spectral function and the result is,
\begin{align}\label{eq:sLL_A}
 A(k,\varepsilon) \propto \theta[\epsilon_s(k_n) \pm \varepsilon] |\epsilon_s(k_n) \pm \varepsilon|^{-\mu_{n,\pm}^s(k)},
\end{align}
where $k_n = k - 2 n k_F  \in [-k_F, k_F]$. The location of the edges and the respective exponents are sketched in Fig.~\ref{fig:Aspinful}. The momentum-dependent exponents are given by
\begin{align}
\label{eq:sLL_mu_A}
& \mu_{n,\pm}^s(k) =   1 - \left( -\frac{(2n + 1)
\sqrt{K_c}}{\sqrt{8}} -\frac{1}{\sqrt{8K_c}} + \frac{\Delta \delta_{+c}}{2\pi} \right)^2 \notag \\
&-\left( -\frac{(2n + 1)
\sqrt{K_c}}{\sqrt{8}} +\frac{1}{\sqrt{8K_c}} + \frac{\Delta \delta_{-c}}{2\pi} \right)^2 - m^2_\pm,
\end{align}
where $\Delta \delta_{\pm c} = \Delta \delta_{\pm c}(k_n)$ and $m_\pm = (n + 1/2 \pm 1/2) \text{ mod } 2$.

Farther away from the threshold, a second peak emerges at energies which correspond to the holon mass shell, $\varepsilon \approx \tilde \xi_c(k)$. Near this energy, the incoming physical particle triggers the formation of a holon on its mass shell as well as a low-energy spinon. Away from the Fermi momentum, interactions between holons and spinons generally lead to a nonzero decay rate for holons and thus a smearing of this peak. This will be discussed in more detail in Sec.~\ref{sec:QP_Relax}.

Let us conclude the discussion of the spectral function by a closer look at momenta $k \approx k_F$. In that limit, $\Delta \delta_{\pm c} = 0$ and the exponent (\ref{eq:sLL_mu_A}) coincides with the prediction of the linear LL theory, $\mu_{0,-}^s = 1 - 1/(4K_c) - K_c/4$ \cite{giamarchi04}.
At $k \to k_F$, the peak at the holon mass shell becomes sharp. However, the exponent  $1-\left[\delta_{-,c}/(2\pi)\right]^2-\left[\delta_{+,s}/(2\pi)-1\right]^2-\left[\delta_{-,s}/(2\pi)\right]^2$ found in the vicinity of the holon mass shell, $|\varepsilon - \xi_c(k)| \ll (k - k_F)^2/(2\tilde{m})$, is different from the LL prediction, just as in the case of spinless fermions. At larger detunings from the threshold, the power-law behavior with the LL value of the exponent is restored~\cite{schmidt10b}. Unlike in the case of spinless fermions, here the crossover function between the two asymptotes of $A(k,\varepsilon)$ is not known. The spectral function $A(k,\varepsilon)$ and notations for exponents are illustrated schematically in Fig.~\ref{fig:Aspinful}.

Next, we turn to the discussion of the spin structure factors. Due to $SU(2)$ symmetry, they are related by $S^{zz}(q,\omega) = \frac{1}{2} S^{-+}(q,\omega)$. The calculation of $S^{zz}(q,\omega)$ requires a projection of the spin density operator $S^z(x)$. The configuration of lowest energy for the momentum $0 < q < 2k_F$ contains a single particle-hole pair in the spinon sector,
\begin{align}\label{eq:sLL_proj_Sz}
 S^z(x) \to e^{i q x} \tpsi_{Rs}^\dag F_{Rs} d_s F_{Rs}.
\end{align}
For general momenta $2n k_F < q < 2 (n + 1) k_F$, the projection contains $n$ additional low-energy particle-hole pairs which carry the momentum $2 n k_F$. Using the mobile impurity Hamiltonian, it can be shown that
\begin{align}\label{eq:sLL_Szz}
 S^{zz}(q,\omega) \propto \theta[ \omega - |\epsilon_s(q_d)|] [ \omega - |\epsilon_s(q_d)|]^{-\mu_n^{\rm DSF}(q)}
\end{align}
where $q_d = (2n + 1)k_F - q$ and the exponents are
\begin{align}\label{eq:sLL_mu_DSF}
 \mu_{n}^{\rm DSF}(q) &= \frac{1}{2} - \left( \frac{n
\sqrt{K_c}}{\sqrt{2}} + \frac{\Delta \delta_{+c}}{2\pi}\right)^2
-\left(\frac{n
\sqrt{K_c}}{\sqrt{2}}+ \frac{\Delta \delta_{-c}}{2\pi}\right)^2.
\end{align}

The phase shifts are taken at momentum $q_d$, \ie, $\Delta \delta_{\pm c} = \Delta \delta_{\pm c}(q_d)$. Note that at $q\to 0$ the scattering phase shifts $\Delta\delta_{\pm c}\to 0$. Therefore, the exponent of the spin structure factor approaches the universal value $1/2$, coinciding with the respective spin liquid exponent of the spin-$1/2$ XXX chain, see Sec.~\ref{sec:spinchains}. Similar to the spin chain model, the region of frequencies $|\omega-|\epsilon_s(q_d)||$ where Eq.~(\ref{eq:sLL_mu_DSF}) is applicable, shrinks as $q^3$. The latter parameter defines the width of the peak in the spin structure factor. Its detailed structure has not been investigated yet. In the linear LL, it is replaced by a $\delta$-function at $\omega=v_s q$.

The quadratic dispersion of the holon spectrum leads to $\Delta\delta_{\pm c}\to 0$ for $k \to \pm k_F$ and results in a rectangular-shaped peak in the charge DSF $S(q,\omega)$ similar to the case of spinless fermions, see Eq.~(\ref{eq:perturb_S1}). For $\omega \approx v_c q$ and $|q| \ll k_F$, the result up to order $q^2$ reads
\begin{align}\label{eq:sLL_S}
 S(q,\omega) = \frac{2 \tilde m K_c}{|q|} \theta\left( \frac{q^2}{2 \tilde m} - |\omega - v_c q|\right)
\end{align}
with an effective mass $\tilde m$. Because the width of the peak $\delta \omega(q) = q^2/\tilde m$ is proportional to $q^2$ whereas its height scales as $1/|q|$, the limit $q \to 0$ reproduces the linear LL result, $S(q,\omega) = 2 K_c |q| \delta(\omega - v_c q)$. The DSF (\ref{eq:sLL_S}) already satisfies the $f$-sum rule \cite{nozieres97}. Hence, additional features may exist with weights at most of order $(q/k_F)^3$.

A second peak in $S(q,\omega)$ occurs at energies close to the spinon mass shell. Indeed, the coupling between spinons and holons has the remarkable consequence that the lower edge of support of the \emph{charge} DSF now coincides with the shifted \emph{spinon} spectrum. The weight of this additional peak can be estimated by using perturbation theory in the spin-charge coupling amplitudes $\kappa_\pm$, which can be also defined phenomenologically \cite{pereira10}, see Eq.~(\ref{zeta_phen}).

For $\omega \approx v_s q$,
the total weight in the vicinity of $\omega \approx v_s q$ equals  $K_c (\alpha_- + \alpha_+)^2q^3/12,$
where $\alpha_\pm = \kappa_{\pm}/(v_c \pm v_s)$. At small $q$, the perturbation theory correctly predicts the peak with weight
$\propto q^3$ at the spinon mass shell. However, it is unable to predict the precise shape of the peak. An analysis using a mobile impurity Hamiltonian reveals again that at the lower threshold $S(q,\omega)$ has a power-law singularity \cite{pereira10}. The singularity remains intact for arbitrary momenta \cite{schmidt10a}. The calculation of the exponent requires a projection of the charge density operator $\rho(x).$ Using Eq.~(\ref{eq:sLL_spin_psi_tpsi}), one finds that the configuration with least energy for a momentum $0 < q < 2k_F$ reads
$ \rho(x) \to e^{i q x} \tpsi_{Rs}^\dag F_{Rs} d_s F_{Rs}$
and has the energy $\approx |\epsilon_s(k_F -q)|.$ Interestingly, this configuration is identical to the threshold configuration for the spin structure factor (\ref{eq:sLL_proj_Sz}). Therefore, near the edge of support $S(q,\omega) \propto S^{zz}(q,\omega).$
In particular, the threshold exponent for $S(q,\omega)$ is given by Eq.~(\ref{eq:sLL_mu_DSF}).

\subsection{Finite-size and finite-temperature effects}
\label{sec:finite}

One of the successes of the linear LL theory is its ability to easily predict finite-size and finite-temperature effects. This can be achieved because the Gaussian Hamiltonian of the LL is conformally invariant~\cite{gogolin98},
which results in a
universality of the finite-size corrections,
such as in the $\propto 1/L$ correction to the ground state energy or the $\propto T$ correction to specific heat at low temperatures~\cite{affleck86,blote86}.
It can be used as a powerful tool when combined with exact solutions, since the energies in the latter case can be evaluated up to $1/L$ corrections:
such an approach was used to evaluate the Luttinger parameters of
the Lieb-Liniger, the XXZ~\cite{bogoliubov87} and the 1D Hubbard models~\cite{frahm90}.
Conformal invariance also fixes the time dependence and the finite-size effects in correlation functions.

For example, for spinless fermions at $\rho |x\pm vt| \gg 1,$
one has~\cite{cazalilla04}
\begin{align}\label{eq:xt_LL}
 & \expct{\Psi^\dag(0,0) \Psi(x,t)}_{\rm LL}\sim  \\
& \sum_n\frac{\rho e^{i(2n+1)k_Fx}}{2i(-1)^{n}}\frac{C_n}{\left(i
\rho(vt+x)+0\right)^{\mu_{L}}\left(i
\rho(vt-x)+0\right)^{\mu_{R}}},
\notag
\end{align}
where $C_n$ are
``nonuniversal'' prefactors, and
\begin{align}
\mu_{R(L)}=(2n+1)^2K/4\pm (2n+1)/2+1/4K\geq0.\nonumber
\end{align}
For a finite system with periodic boundary conditions on a circle of length $L,$ conformal invariance dictates~\cite{cazalilla04,shashi11} that the $n$th term in Eq.~(\ref{eq:xt_LL}) gets modified to
\begin{align}
\frac{\rho e^{i(2n+1)k_Fx}C_n}{2i(-1)^{n}} \prod_{L,R}\left(\frac{\pi
e^{i\pi(vt\pm x)/L}}{i\rho L\sin{\frac{\pi(vt\pm x)}{L}}+0}\right)^{\mu_{L(R)}}. \label{eq:Cazalilla}
\end{align}

The goal of this section is to promote the phenomenological theory based on impurity Hamiltonians to the same status.
It will not only provide new predictions, but will also  serve as a calculational tool
to extract information from exactly solvable models (see Sec.~\ref{sec:exact}) and interpret the results
of numerical simulations.

Let us start with the discussion of finite-size effects.
For concreteness, we will focus on spinless fermions and the vicinity of the edge of support.
The finite-size spectrum of $\propto1/L$ corrections is that of a shifted
Gaussian conformal theory~\cite{tsukamoto98},
and can be determined using conventional techniques. The correction to the position of the edge in standard  notations~\cite{korepin93} is given by~\cite{pereira08,pereira09}
\begin{align}
\Delta E=\frac{2\pi v}{L}\left[ \frac1{4K}\left(\Delta N-n_{imp}\right)^2+K(D-d_{imp})^2\right], \label{eq:energyL}
\end{align}
where $\Delta N$ and $D$ are quantum numbers specifying the excitations, while $n_{imp}$ and $d_{imp}$
are related to phenomenological phases in Eq.~(\ref{eq:phen_phi}) via
\begin{align}
n_{imp}=\frac{\delta_-(k)-\delta_+(k)}{2\pi/\sqrt{K}},\;
d_{imp}=-\frac{\delta_+(k)+\delta_-(k)}{4\sqrt{K}\pi}. \label{eq:nddelta}
\end{align}

The quantum numbers $\Delta N$ and $D$ are related to the numbers $N_{R}$ and $N_L$
of fermions created at each Fermi point, which define the exponents as discussed in Sec.~\ref{sec:phenomenology}. For fermionic models, they
are given by $\Delta N=N_R+N_L$ and $D=(N_R-N_L)/2.$ Thus, Eq.~(\ref{eq:energyL}) allows one to calculate the
phase-shifts $\delta_{\pm}(k)$ by, \eg, numerical tracking of the $\propto 1/L$ corrections to energies.

We will now consider the finite-size $\propto 1/L$ structure of
low-lying levels at fixed $k$, as well as the scalings of various  matrix elements~\cite{shashi11}.
Using a resolution of the identity in the expectation value $\langle \Psi^{\dagger}(0,0)\Psi(x,t)\rangle$,
we get
\begin{align}
\langle \Psi^{\dagger}(0,0)\Psi(x,t)\rangle=\sum_{s'}e^{-i (k_{s'}x-\varepsilon_{s'}t)}|\langle s', N-1|\Psi|N\rangle|^2, \label{eq:Lehmann_finite_size}
\end{align}
where $|s', N-1\rangle$ denote eigenstates with $N-1$ particles. In a finite-size system, the formfactors in this equation have to be matched with the field-theoretical predictions.

To understand the procedure, let us first consider the scaling of the formfactors between low-energy states for 1D fermions~\cite{bogoliubov87}, which can be described by the conventional LL theory.
We can now expand terms in Eq.~(\ref{eq:Cazalilla}) using a Fourier series as
\begin{align}
\left(\frac{\pi
e^{i\pi(vt\pm x)/L}}{iL\sin{\frac{\pi(vt\pm x)}{L}}+0}\right)^{\mu}\!\!&=\sum_{n_\mp\geq
0}  C(n_\mp,\mu)\frac{e^{ 2i\pi
n_{\mp}\frac{vt\pm x}{L}}}{(L/2\pi)^{\mu}},\notag \\
\label{eq:finite_size}
C(n_\pm,
\mu)&=\frac{\Gamma(\mu+n_\pm)}{\Gamma(\mu)\Gamma(n_\pm+1)}.
\end{align}
In this equation, the summation only over $n_{+}$ is implied for right branch contribution, and the summation only over $n_-$ is implied for left branch contribution. Plugging Eq.~(\ref{eq:finite_size}) into Eq.~(\ref{eq:Cazalilla}), and comparing the result to Eq.~(\ref{eq:Lehmann_finite_size}), one can clearly identify contributions from excitations at the right (left)
Fermi branches with energies
$2\pi vn_{\pm}/L>0,$
respectively.
To accommodate the additional momentum $-(2n+1)k_F,$ one needs to put $n+1$ holes at the right Fermi point and $n$ particles at the left Fermi point. The contributions from $n_+=n_-=0$ give the scalings of the ``parent'' formfactors~\cite{bogoliubov87}
\begin{align}
\left|\langle n, N-1|\Psi|N\rangle\right|^2 \approx \frac{ C_n \rho_0}{2(-1)^{n}}  \left(\frac{2\pi}{\rho_0 L}\right)^{\frac{(2n+1)^2 K^2+1}{2K}},\label{eq:parent}
\end{align}
where $|n,N-1\rangle$ is the lowest energy state of $N-1$ fermions with momentum $-(2n+1)k_F.$
The nontrivial scaling of this formfactor with $L$ is a consequence of the criticality
of the LL. The studies of scalings of the
formfactors serve as a tool to evaluate the nonuniversal prefactors $C_n$~\cite{shashi10,shashi11}, which are usually not known except for a few cases~\cite{popov80,vaidya79,jimbo80,gangardt04,gangardt01,astrakharchik06,lukyanov03}.
Within $\propto 1/L$
accuracy, for $n_{\pm}\geq 2$ the excited states of $N-1$ particles are
degenerate, while
the degeneracy within each ``multiplet'' is lifted by $\propto 1/L^2$
corrections due to the nonlinear spectrum. The universal Hamiltonian of Sec.~\ref{sec:universal} predicts the
distribution of the spectral weight within each ``multiplet'', as has been shown by~\textcite{shashi11}.

Let us now apply a similar logic to the finite-size behavior of the response functions near the edge of support,
and for concreteness we will focus on the spectral function $A(k,\varepsilon)$ for $|k|<k_F$ and $\varepsilon<0$.
In addition to the LL, we now also need to
take into account the finite-size quantization of the
momentum of the impurity moving with velocity
$|v_d|<v.$ For an infinite-size system, $A(k,\varepsilon)$ in the vicinity of the edge $\varepsilon_\th(k)<0$ (see notations in Sec.~\ref{sec:phenomenology}) can be written as
\begin{align}
A(k,\varepsilon)=A_{0,-}(k)\int dx dt
e^{-i\delta \varepsilon t}D(x,t)L(x,t)R(x,t),\label{eq:Aint}
\end{align}
where $\delta\varepsilon=\varepsilon-\varepsilon_\th(k),
D(x,t)=\delta(x-v_d t)$ is the impurity correlator,
$L(R)(x,t)=(i(vt \pm x)+0)^{-\mu_{0,-,L(R)}},$ and
$A_{0,-}(k)$
is a ``nonuniversal'' prefactor,
see Eqs.~(\ref{eq:phen_A}) and (\ref{eq:munpm}). After $(x,t)$ integration, Eq.~(\ref{eq:Aint}) results in
\begin{align}
A(k,\varepsilon)= \frac{2\pi \theta(-\delta\varepsilon ) A_{0,-}(k)
|\delta\varepsilon|^{-\mu_{0,-}} }{\Gamma(1-\mu_{0,-})(v+v_d)^{\mu_{0,-,L}}(v-v_d)^{\mu_{0,-,R}}}.\label{eq:A2def}
\end{align}
In finite-size systems, $L(x,t)$ and $R(x,t)$ get modified, see Eq.~(\ref{eq:finite_size}). Similarly, the change of $D(x,t)$ to  $\sum_{n_D} e^{2i\pi
n_D(x-v_dt)/L}$ corresponds to the quantization of the impurity momentum.
Since we are considering fixed $k,$ the total momentum of the excitations on the left and right branches should be equal to the inverse of the shift of the momentum of the impurity $2\pi
n_D/L$, which implies  $n_D=n_--n_+.$
Combining these terms, we get for $A(k,\varepsilon)$
\begin{align}
 &\sum_{n_\pm \geq 0}
 \delta\left(\delta\varepsilon-\Delta E + \frac{2\pi n_+}{L}(v-v_d)+\frac{2\pi
 n_-}{L}(v+v_d)\right) \times \notag \\
&A_{0,-}(k) \frac{(2\pi)^{\mu_{0,-,R}+\mu_{0,-,L}+1}}{L^{\mu_{0,-,R}+\mu_{0,-,L}}} C(n_+,\mu_{0,-,R}) C(n_-,\mu_{0,-,L}).
 \label{eq:fsize}
\end{align}
At $\propto 1/L$ accuracy, the finite-size structure of the response function is
given by the sum of two generically incommensurate frequency ``ladders'', with the relative spectral weights
in each ``multiplet'' controlled by the phase shifts $\delta_{\pm}(k).$  Equation~(\ref{eq:fsize}) allows for an analytical or numerical evaluation of $A_{0,-}(k)$ based on the scaling of the single formfactor with $n_+=n_-=0$ as
\begin{align}
\left|\langle k;N-1|\Psi|N\rangle\right|^2\approx \frac{A_{0,-}(k)}{L}\left(\frac{2\pi}{L}\right)^{ \mu_{0,-,R} +
\mu_{0,-,L}},  \notag
\end{align}
where $|k;N-1\rangle$ in the eigenstate of $N-1$ particles corresponding to the edge of support at $|k|<k_F.$
The structure described by Eqs.~(\ref{eq:parent}) and (\ref{eq:fsize}) can be explicitly
confirmed for certain integrable models~\cite{kitanine09a,kitanine09b,kitanine11,shashi10} using known expressions for
the finite-size formfactors. This provides a stringent microscopic check of the phenomenological impurity Hamiltonians and allows to analytically
calculate various ``nonuniversal'' prefactors for these models. In addition, it allows to calculate various prefactors perturbatively~\cite{shashi11}.

Let us now comment on the effects of finite temperatures, which are quite different for nonlinear LLs compared to linear ones. In the latter, conformal invariance allows one to calculate~\cite{giamarchi04} most of the finite-temperature effects by simple substitutions such as $v/L \to i T$.
This cannot be done for nonlinear LLs.
The full analysis of the finite-temperature effects,
especially in kinetic problems (see Sec.~\ref{sec:kinetics}), remains an open problem.
Let us make some general remarks here
focusing on the response functions of spinless fermions.

For the interacting case, a finite temperature smears the sharp edges of support of the spectral function. However, far away from the Fermi points and at $|\varepsilon_\th(k)|\gg T$ (we use $k_B=1$ throughout the text), there is a large interval of energies where the effect of temperature can be captured by substituting the Fermi point correlators by their finite-temperature versions~\cite{karimi11}, \ie, in Eq.~(\ref{eq:Aint})
\begin{align}
L_T(R_T)(x,t) \to \left(\frac{2\pi T/v}{\sin{\left[2\pi i T(t\pm x/v)+0\right]}}\right)^{-\mu_{0,-,L(R)}}. \notag
\end{align}
At the same time, the impurity correlator $D(x,t)$ can be kept as a delta-function,
because the correction to the impurity ``occupation number'' is  exponentially suppressed.
These substitutions result in universal functions characterizing the smearing of the edge singularities
\begin{align}
A_T(k,\varepsilon)=A_{0,-}(k)\int dt
e^{-i\delta \varepsilon t}L_T(v_d t,t)R_T(v_d t,t). \notag
\end{align}
Note that these functions can be evaluated numerically and generically have a strongly non-Lorentzian shape. The temperature mostly affects the response functions at energies of the order of $\sim T$ from the edges of support.

In the vicinities of the Fermi points, one can use the universal Hamiltonian of Sec.~\ref{sec:universal}
to take a finite temperature into account.
A na\"ive extension of the above argument would imply a smearing of the nonlinear LL physics for the DSF at temperatures on the order of $\sim q^2/(\tilde{m}).$ We note, however, that this is not the case, as can be illustrated by the DSF of noninteracting fermions. The latter can be straightforwardly evaluated
as in Sec.~\ref{sec:perturbative_DSF},  and in the limit $q\ll k_F$ reads
\begin{align} \label{eq:S0_temp}
S_0(q,\omega)=\frac{m}{q}\prod_{\pm}n_F\left[v_F\frac{\pm m(\omega-v_F q)- q^2/2}{q}\right],
\end{align}
where the Fermi-Dirac distribution function $n_F(\epsilon)=1/[\exp{(\epsilon/T)}+1]$ replaces the step functions in Eq.~(\ref{eq:dsf-freefermions2}).
One sees that the zero-temperature result~(\ref{eq:dsf-freefermions2}) survives up to temperatures on the order of $\sim v_F q.$
Similarly, the universal Hamiltonian of Sec.~\ref{sec:universal} implies that at finite temperatures Eq.~(\ref{eq:LL_S}) is replaced by
\begin{align}\label{eq:S_temp}
S(q,\omega)=\frac{K \tilde  m}{q}\prod_{\pm}n_F\left[v\frac{\pm
\tilde m(\omega-v q)- q^2/2}{q}\right],
\end{align}
so that the DSF is barely different from its $T=0$ form as long as the temperature is small compared to
$vq.$ The mechanism of the DSF smearing by finite temperature expressed in Eq.~(\ref{eq:S0_temp}) is the same as in Eq.~(\ref{eq:S_temp}). At $q\ll k_F$, the DSF involves only contributions from the right-moving quasiparticle and quasihole with momenta of the order $\sim q$ around the Fermi point. A finite temperature smears their velocities
by $\delta v\sim T/(\tilde m v).$ The corresponding energy variation $q \delta v \sim qT/(\tilde m v)$ is small compared to $q^2/\tilde m$
as long as $T \ll vq.$

There is a substantial difference between the domains of applicability of Eqs.~(\ref{eq:S0_temp}) and Eq.~(\ref{eq:S_temp}) though. In the former, the limits $q\to 0$ and $T\to 0$ may be taken in any order. The latter is valid only in the limit when, in addition to keeping $\tilde m(\omega-v q)/q^2$ constant at $q\to 0$ as in Eqs.~(\ref{eq:LL_S}) and (\ref{eq:LL_A}), $T/(qv)$ is kept constant. As a result, Eq.~(\ref{eq:S_temp}) does not hold at fixed $T$ and $q \to 0,$ which will be important in Sec.~\ref{sec:Drude}.

Unlike the DSF, the spectral function in the universal limit is a convolution of contributions from both left and
right Fermi points. The kinematic considerations of Sec.~\ref{sec:perturbative} and Sec.~\ref{sec:universal} imply that the quasiparticles at the left branch have energies of the order $\sim (k-k_F)^2/\tilde{m},$, so the finite-temperature smearing of the nonlinear effects in the spectral function is a two-step process~\cite{ma11}.

First, at temperatures $\sim (k-k_F)^2/\tilde{m}$, the contribution from the left branch
gets smeared out. The contribution from the right branch gets affected significantly only at temperatures on the order of $\sim v(k-k_F).$ It should be noted that the spectral function of chiral fermions at the edges of quantum Hall states~\cite{granger09,altimiras09,lesueur10,altimiras10,paradiso11,chang03,jolad10,lunde10,heyl10,neuenhahn09} should be more robust to finite temperatures due to the absence of the counter-propagating  branch.

\subsection{Real-space correlation functions}
\label{sec:realspace}

In the previous sections, we showed that the power-law singularities of the spectral function $A(k,\varepsilon)$ near the edge of support can be described with the help of mobile-impurity Hamiltonians for 1D fermionic, bosonic and spin systems. We will show in this section that the existence of these threshold singularities has important implications for the space-time correlation functions in the limit of large $x$ and $t$. In particular, we will elucidate the connection between the threshold singularities and the breakdown of conformal invariance, focusing on the case of spinless fermions.

Using a Lehmann spectral representation~\cite{abrikosov63}, various space-time Green's functions can be obtained by Fourier transforming the spectral function. For example,
\begin{align}\label{real_Glesser}
\expct{\Psi^\dag(0,0) \Psi(x,t)} =
 \frac{1}{2\pi} \int dk \int_{-\infty}^0 d\varepsilon e^{i k x} e^{-i \varepsilon t} A(k,\varepsilon).
\end{align}
In a similar way, $\expct{\Psi(x,t) \Psi^\dag(0,0)}$ can be expressed via the Fourier transform of $A(k,\epsilon > 0)$ in the particle sector.

As is well known in the theory of Fourier transformations~\cite{bleistein86}, the non-analyticities of $A(k,\epsilon)$ control the long space-time behavior of its Fourier transforms. The spectral function is non-analytic in the vicinities of the Fermi points and their $2nk_F$ images. We showed in the previous sections that the regions where $A(k,\varepsilon)$ deviates significantly from the predictions of the linear LL theory become narrow in the limit $\varepsilon \to 0$. As a consequence, the effects of the spectrum nonlinearity are suppressed near $\varepsilon \approx 0$ when integrating over $k$.
Neglecting the nonlinear effects in the vicinities of the Fermi points produces the well-known space-time power-law behavior of the correlation functions in Eq.(\ref{eq:xt_LL}) at $\rho |x\pm vt| \gg 1$. This result is manifestly conformal invariant due to the conformal invariance of the LL Hamiltonian (\ref{eq:HLL}). If one considers the limit
\begin{align} \label{eq:vdlimit}
t\to \infty \;\;\mbox{and}\;\;v_d=x/t\;\; \mbox{fixed},
\end{align}
Eq.~(\ref{eq:xt_LL}) results in the same set of power-law tails in $t$ irrespective of the ratio $v_d/v.$

\begin{figure}
\includegraphics[width=8cm]{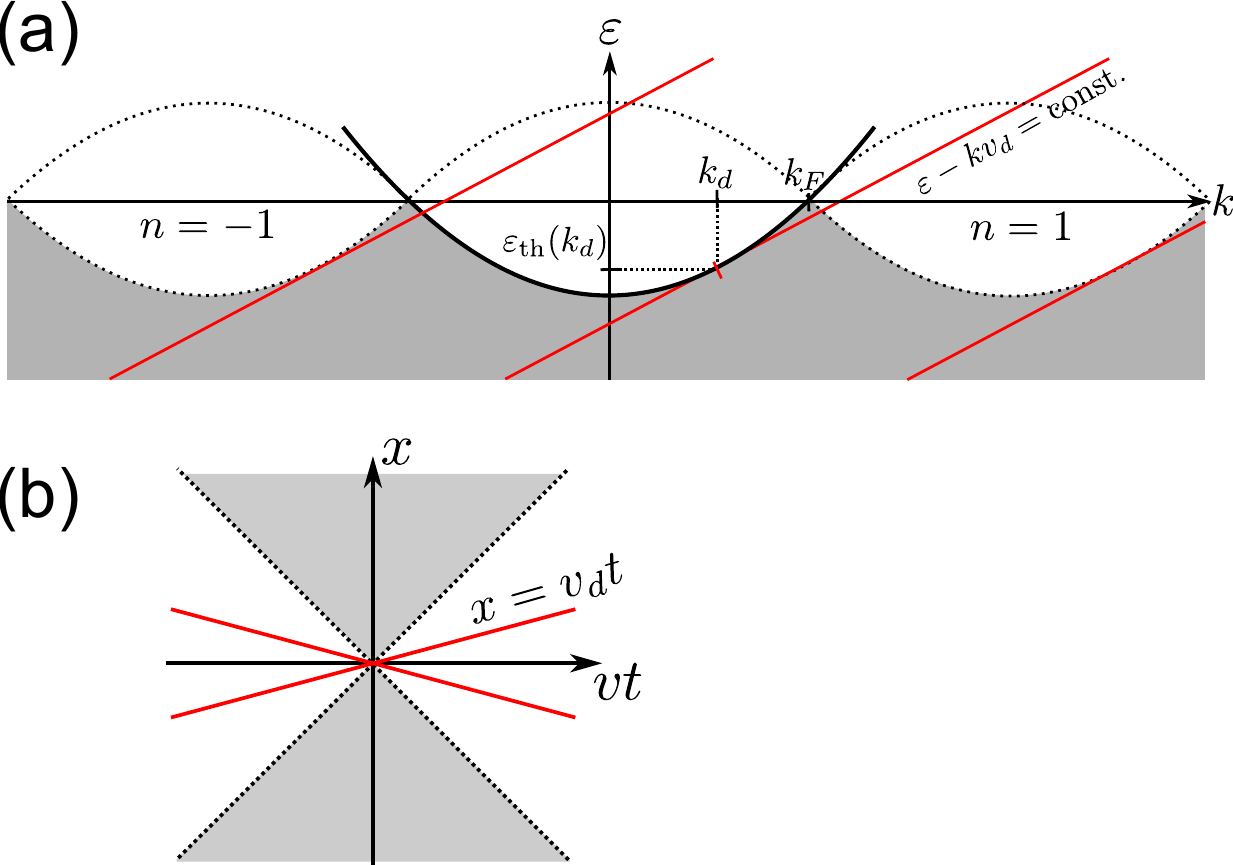}
\caption{
\label{fig:SaddlePoint}
 (Color online) (a) Calculation of the real-space correlation functions. The integration range for the calculation of $\expct{\Psi^\dag(0,0) \Psi(x,t)}$ from $A(k,\varepsilon)$ is shaded. For large $t$ and fixed $x/t=v_d$, the contributions to the integral come from a region around $\varepsilon = 0$, and from points at
which the lines $\varepsilon - k v_d = {\rm const.}$ touch the edge of support. (b) Breakdown of conformal invariance in $\expct{\Psi^\dag(0,0) \Psi(x =v_d t,t)}$. In the shaded areas ($|v_d|>v$) only the linear Luttinger liquid (LL) power laws survive, see Eq.~(\ref{eq:xt_LL}). In the white areas, for $|v_d|<v$, new power laws appear in addition, see Eq.~(\ref{eq:xt_main}).
}
\end{figure}

We will show now that in the limit (\ref{eq:vdlimit}), the threshold singularities in $A(k,\varepsilon)$
may generate a new set of power-law tails in $t.$  For generic interacting spinless fermions, such power laws appear only for $|v_d|<v$ and, moreover, the exponents depend on the ratio $v_d/v.$ The origin of these power laws is very similar to that of van Hove singularities in the density of states of noninteracting systems~\cite{landau80} and is illustrated in Fig.  \ref{fig:SaddlePoint}a. To study the limit (\ref{eq:vdlimit}), it is convenient first to make a rotation of coordinates in the $(x,t)$ and $(k,\varepsilon)$ planes to
\begin{align}\label{eq:rotation}
  x' &= \frac{x-v_d t}{2}, &t' &=\frac{x+v_d t}{2v_d},\notag \\
  k' &= \varepsilon/v_d +k, &\varepsilon'&=\varepsilon-k v_d.
\end{align}
In these coordinates, one has $kx-\varepsilon t =k' x' - \varepsilon' t',$ and we are interested in the limit $x'=0$, $t'=t \to \infty.$ Therefore, to study this limit one needs first to keep $\varepsilon'$ constant (tangential lines in Fig.~\ref{fig:SaddlePoint}a) and integrate over $k'.$ The non-analyticities near the Fermi points and their $2nk_F$ images produce the conventional LL power laws (\ref{eq:xt_LL}). In addition, after the integral over $k'$ is performed, for $|v_d|<v$, the existence of sharp edges of support produces new non-analyticities in $\varepsilon'$
from the vicinities of the touching points, see Fig.~\ref{fig:SaddlePoint}a. The origin of these non-analyticities is very similar to that of van Hove singularities,
but unlike the latter they also depend on the threshold exponents of the spectral function. The touching condition implies that $v_d$ is nothing but the impurity velocity. Each ``shadow band'' produces a separate power law and an evaluation of the Fourier transform (\ref{real_Glesser}) results in
\begin{align}\label{eq:xt_main}
& \expct{\Psi^\dag(0,0) \Psi(x = v_dt,t)} = \expct{\Psi^\dag(0,0) \Psi(v_dt,t)}_{\rm LL}   \\
&+ \sqrt{\frac{m_d}{-2 i \pi}}\sum_n \frac{e^{-i \pi \mu_{n,-}/2} A_{n,-}(k_d) e^{-i \epsilon(k_d) t + i (k_d + 2 n k_F)x} }{t^{1/2} (v t+x)^{\mu_{n,-,L}}(v t-x)^{\mu_{n,-,R}}} \notag
\end{align}
where the correlation function of the linear LL $\expct{\Psi^\dag(0,0) \Psi(v_dt,t)}_{\rm LL}$ is defined in Eq.~(\ref{eq:xt_LL}) and the momentum $k_d$ is defined by the touching condition
$ v_d = \frac{\partial \varepsilon_{\th}(k)}{\partial k}\bigg|_{k = k_d}.$
The effective mass $m_d$ is given by $1/m_d  = \partial v_d/\partial k|_{k=k_d}$. The exponents $\mu_{n,-,R(L)}$ are defined by Eq.~(\ref{eq:mun-}) and $A_{n,-}(k_d)$ are the non-universal prefactors from Sec.~\ref{sec:finite}.

We note that for sufficiently weak interactions the new ``nonlinear'' tails in $t$ decay {\it slower} than the linear LL tails for all $|v_d|<v$. Indeed, for noninteracting fermions and $v_d=0,$ the contributions from the Fermi points decay
as $\propto 1/t,$ whereas the contribution from the bottom of the band decays only as $\propto 1/\sqrt{t}$ due to a conventional van Hove singularity there \cite{gutman08}. For the integrable Lieb-Liniger model, expansions similar to our Eq. (\ref{eq:xt_main}) have been obtained from purely microscopic considerations~\cite{kozlowski11a,kozlowski11b}
and match the exact results for the exponents calculated in Sec.~\ref{sec:Lieb-Liniger}. Similar results have been obtained for a gas of 1D Lieb-Liniger anyons~\cite{patu09}.

In Sec.~\ref{sec:universal}, we have established that in the vicinity of the Fermi points, the response functions are universal within the nonlinear LL theory. The results of this section then imply that for $|v_d\pm v|\ll v,$ the time dependence of the field correlator, and its crossover
between linear and nonlinear regimes is universal as well. Using the universal expressions (\ref{eq:LL_phases}) for the phase shifts in this limit and Eqs.~(\ref{eq:xt_LL}) and (\ref{eq:xt_main}),
one can establish that
$\mu_{L}+\mu_{R}> \frac12 +\mu_{0,-,R}+\mu_{0,-,L}.$
Therefore, the nonlinear results always decays slower in time than the linear LL result.
Using a more careful analysis of the prefactors~\cite{imambekov09a},
one can also establish the scaling of the crossover time $t_c$ between the two regimes as
$\rho v t_c \sim \left(\frac{\tilde{m} (v-v_d)}{k_F}\right)^{\frac{4-6\sqrt{K}+4K}{2-5\sqrt{K}+2K}}.$
For weakly interacting fermions, this condition reduces to  $\rho v_F t_c \sim v^2_F/(v_F - v_d)^2.$

Let us now comment on the space and time behavior of the transverse spin correlation function of $SU(2)$ invariant spinful bosons. In Sec.~\ref{sec:bosonic}, it has been established that for small momenta $S^{+-}(q,\omega) \propto (\omega-q^2/(2m^*))^{1-Kq^2/(2\pi^2 \rho^2)}.$
Of particular interest is the regime when repulsive interactions between spin-up and spin-down particles are strong.
Then the spin-down impurity cannot exchange positions with other particles and effectively becomes trapped~\cite{zvonarev07a}.
In the thermodynamic limit, the magnon mass $m^*$ diverges and the bandwidth of spin excitations becomes very narrow.
This is very reminiscent of the
narrow-band spinon excitations of strongly-interacting $s=1/2$ fermions discussed by~\textcite{matveev04,matveev07,matveev07b}, and mentioned in Sec.~\ref{sec:spinful}.
In both cases, the large difference between the bandwidths of spin and charge excitations results in the existence of a new regime where a certain new universal behavior of correlations takes place.
Performing the inverse of the Fourier transformation (\ref{eq:SSFs}) within the saddle point approximation, one obtains~\cite{zvonarev07a}
$
\expct{S^{+}(x,t) S^{-}(0,0)}\propto
\left[\frac{K}{2(\pi\rho)^2} \ln\left(E_F t\right)+\frac{it}{2m^*}\right]^{-1/2}
\times\exp\left\{\frac{i m^* x^2}{2t-2 i K m^*/(\pi\rho)^2 \ln(E_F t)}\right\},
$
where $E_F\sim (\pi \rho)^2/(2m)$ is introduced to provide a short-time cutoff. For generic interactions, the logarithmic term can be ignored and one obtains the scaling $x^2 \propto t/m^*$
characteristic of a single-particle wave packet spreading. However, in the limit of infinitely strong interactions $m^*$ diverges and one obtains a logarithmic scaling $(\pi \rho x)^2 \sim K\ln(E_F t).$ For large but finite $m^*$ a logarithmic scaling law is applicable in an intermediate time interval, the length of which grows with the increase of $m^*.$

\section{EXACTLY SOLVABLE MODELS}
\label{sec:exact}

In the previous section, we concentrated on the properties of generic 1D quantum systems
beyond the linear LL description.
However, in 1D there exists a class of
exactly solvable (or integrable) models, for which energy spectra and thermodynamical properties can be calculated exactly using the Bethe ansatz. The calculation of their
correlation functions, on the other hand, is a much more complicated task~\cite{korepin93},
because it requires not only the knowledge of the exact energies, but also of the matrix elements (formfactors).
In addition, one needs to be able to sum over all excited stated in the Lehmann representation
in order to obtain answers in the thermodynamic limit.
The exact expressions for the formfactors are usually known only in finite-size systems~\cite{slavnov89,slavnov90,ha96,kojima97,kitanine99},
and so far only few-spinon approximations~\cite{bougourzi96,karbach97,bougourzi98} or fully numerical summations over formfactors~\cite{caux05a,caux05b,caux06,caux07,gritsev10,kohno10} have been implemented for certain gapless models.

In this section, we review
some recent results for the correlation functions of integrable models
which have been obtained by combining exact results with field theories beyond the linear LL theory.
We will show that a plethora of new results can be derived using this approach.
More importantly,
exactly solvable models also provide nontrivial checks of the phenomenology and provide additional
verification of the effective impurity models. It should be noted that, historically, the analysis of exactly solvable models~\cite{haldane80,haldane81b} played an
important role in the justification of the linear LL theory. Now
these models are proving their worth
to extensions of the LL theory as well.

The rest of this section is organized as follows. In Sec.~\ref{sec:CS} we review some recent progress for the Calogero-Sutherland model~\cite{calogero69,calogero71,sutherland71,sutherland04}. In Sec.~\ref{sec:Lieb-Liniger} we discuss the Lieb-Liniger model and demonstrate several approaches~\cite{imambekov08,pereira08,cheianov08} to extract parameters of the effective impurity Hamiltonians. In Sec.~\ref{sec:Yang-Gaudin} we consider
Yang-Gaudin models~\cite{yang67,gaudin67,gaudin83} which describe spinful multi-component systems interacting via a contact potential. In Sec.~\ref{sec:XXZ}, we discuss 1D lattice models, such as the spin-1/2 XXZ model (which is equivalent to spinless fermions on a lattice), and 1D Hubbard model~\cite{lieb68,essler05}.

\subsection{Calogero-Sutherland  model}

\label{sec:CS}

In this subsection, we will discuss
the Calogero-Sutherland (CS) model~\cite{calogero69,calogero71,sutherland71,sutherland04},
the most well studied model with inverse square interactions~\cite{haldane88, shastry88, kuramoto91,kuramoto09}. Such models are special among exactly solvable models, because their ground state wavefunctions
can often be expressed as Jastrow-type products, and their dynamical correlation functions can be derived
in a closed form as multiple integrals
because of special properties of their excitations~\cite{haldane91,haldane92,haldane93,ha94a,ha94b,ha95,lesage95,ha96,talstra94,kato98,takashi99,arikawa99,yamamoto00,arikawa01,arikawa04,arikawa06, pustilnik06b}.
The special quantum mechanical properties of a system with inverse-square ($\propto 1/r^2$) interaction potential can already be seen at the classical level, because
a liquid of such particles admits a description using integrable hydrodynamics~\cite{stone08,kulkarni09,abanov09,abanov05}.
For concreteness, here we will focus on the DSF of the CS model following \textcite{pustilnik06b,khodas07a},
and illustrate connections to the phenomenology discussed in Sec.~\ref{sec:phenomenology}.
Most of the related results on other inverse-square models are reviewed in a recent book by \textcite{kuramoto09}.

The CS Hamiltonian is given by
\begin{align}
H_{CS}=-\frac{1}{2m}\sum_{i=1}^N\frac{\partial^2}{\partial x_i^2}
+ \sum_{i<j}V(x_i-x_j). \label{eq:CS}
\end{align}
Here $m$ is the mass, and the interaction potential is
\begin{equation}
V(x) = \frac{\lambda(\lambda - 1)\!/m}
{(L/\pi)^2\sin^2\!\bigl(\pi x/L\bigr)}
= \frac{\lambda(\lambda - 1)}
{m x^2}\quad (\text{for } L \to \infty) \notag
\end{equation}
where $\lambda>1/2$ is a dimensionless interaction strength defining the LL parameter as $K=1/\lambda$.

\begin{figure}
\includegraphics[width=8cm]{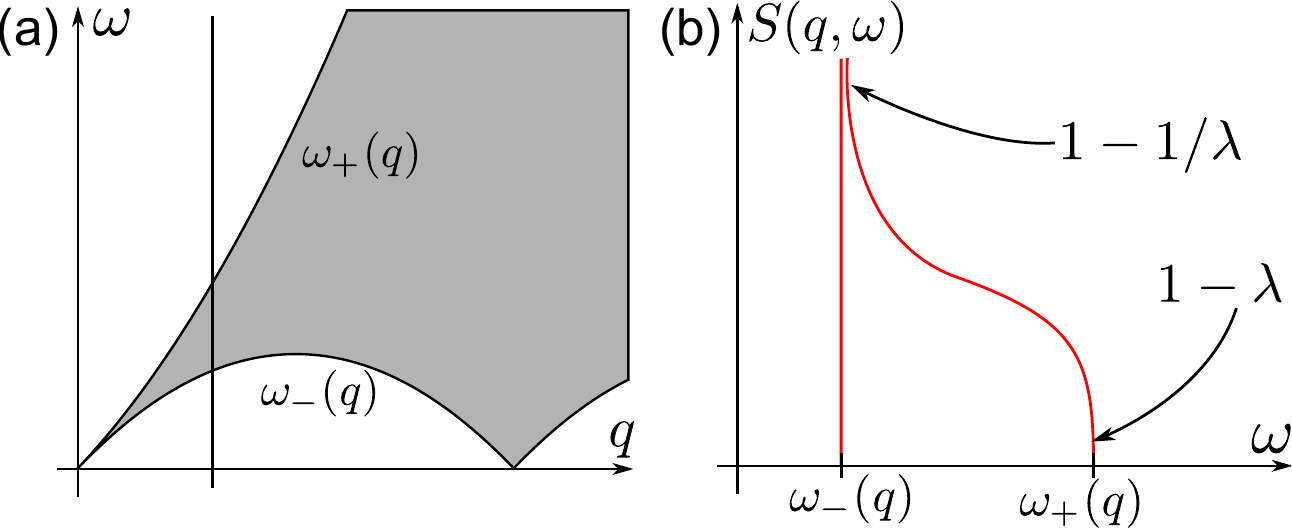}
\caption{\label{fig:alapustilnik} (Color online)
(a) The DSF $S(q,\omega)$ of the Calogero-Sutherland (CS) model differs from zero only in a
finite interval of frequencies $\omega_-<\omega<\omega_+.$ At the boundaries of
this interval, $S(q,\omega)$ exhibits power-law singularities, see Eq.~(\ref{eq:CSexp})
(b) Dependence of $S(q,\omega)$ on $\omega$ at a fixed $q<2k_F$ and for
repulsive interactions, $\lambda>1$.}
\end{figure}

Excitations can be simply classified using the language of quasiparticles and quasiholes
with respect to a filled ``Fermi sea''. The long range
of the interactions results in a rather peculiar excitation spectrum. For quasiparticles with velocities $v_+$ larger in absolute value than the sound velocity $v=\pi \lambda \rho/m,$ the spectrum is given by
$m(v_+^2-v^2)/2$. In contrast, for quasiholes with velocities $|v_-|<v$, the spectrum is given by $m\lambda(v^2-v_-^2)/2.$ The discontinuity of the effective mass near the Fermi points
can be expected from perturbation theory, since the Fourier transform of the $1/r^2$ potential is nonanalytic, $V_k\propto |k|.$ Because of that, the universal Hamiltonian of Sec.~\ref{sec:universal} is not applicable to the CS model.

A special feature of this model is that for rational $\lambda=r/s$ (where $r$ and $s$ are coprime), the operator $\ \rho^{\dagger}_{q>0}$ acting on the ground state creates only right-moving excitations: $s$ quasiparticles and $r$ quasiholes~\cite{ha94b,ha95,ha96}. This is not a generic property of integrable systems and has a profound effect on the dynamic response functions.
In particular, since no left-moving excitations are created, the DSF $S(q,\omega)$ is nonzero only in a finite interval of energies, $\omega_-(q)<\omega<\omega_+(q),$
see Fig.~\ref{fig:alapustilnik}. The existence of an upper threshold is not expected for generic 1D systems, see Sec.~\ref{sec:perturbative_DSF}. The upper threshold corresponds to a configuration where the entire energy is given to a single quasiparticle, while for $q<2k_F$ the lower threshold corresponds to all energy being given to a single quasihole. One can show that for $q>0$,
\begin{align}
\omega_+(q)&=vq+q^2/(2m),\label{eq:CSomegap}\\
\omega_-(q)&=vq-\lambda q^2/(2m) \;\mbox{for}\; q<2k_F.\label{eq:CSomegam}
\end{align}
The DSF can be written as
\begin{align}
S(q,\omega)=q^2\int \prod_{i,j} dv_{+,i} dv_{-,j}\,F_{s,r}\,
\delta\bigl(q-P\bigr)\,
\delta\bigl(\omega-E\bigr), \label{eq:CSSqw}
\end{align}
where $P$ and $E$ are the total momentum and energy of the excitations, respectively,
and the expression for the formfactor $F_{s,r}$ is given by~\cite{ha94b,ha95,ha96,haldane93b}
\begin{align}
\propto
\frac{
\prod_{i<i'}|v_{+,i}-v_{+,i'}|^{2\lambda}\!
\prod_{j<j'}| v_{-,j}- v_{-,j'}|^{2/\lambda}
}
{
\prod_{i,j}(v_{+,i}- v_{-,j})^2
\bigl(v_{+,i}^2 - v^2\bigl)^{1-\lambda}
\bigl(v^2-v_{-,j}^{2}\bigl)^{1-1/\lambda}
} \,. \notag
\end{align}
The analysis of the multidimensional integral in Eq.~(\ref{eq:CSSqw})
performed by \textcite{pustilnik06b} then yields a power-law behavior in the allowed regions as
\begin{align}\label{eq:CSexp}
\frac{S(q,\omega)}{m/q} \propto
\left|\frac{\,\omega_+-\omega_-}{\omega-\omega_\pm}\right|^{1-\lambda^{\pm 1}}
\mbox{for} \;
|\omega-\omega_\pm|\ll \omega_+-\omega_-.
\end{align}
The prefactors of the DSF have also been evaluated  following Sec.~\ref{sec:finite}~\cite{shashi10}.

The analysis of the spectral function can be performed similarly~\cite{khodas07a},
and for $|k|<k_F$, $\varepsilon<0$ results in the power-law behavior
\begin{align}
A(k,\epsilon)\propto \theta(\varepsilon_{\rm th}(k)-\varepsilon)(\varepsilon_{\rm th}(k)-\varepsilon)^{1-(\lambda-1)^2/(2\lambda)}.
\end{align}
These exponents at generic edges of support $\pm\varepsilon_{\rm th}(k)=\mp\omega_-(k_F-k)$ can be simply recovered from the
phenomenological considerations of Sec.~\ref{sec:phenomenology},
with Eq.~(\ref{eq:phen_phi}) resulting in momentum-independent phase shifts
\begin{align}
\frac{\delta^{CS}_-}{2\pi}=-\frac{\delta^{CS}_+}{2\pi}=\frac12\left(\frac{\lambda-1}{\sqrt{\lambda}}\right)=\frac12\left(\frac{K-1}{\sqrt{K}}\right).
\label{eq:CS_phases}
\end{align}
We see that near the right Fermi point, the phase shift $\delta^{CS}_-$ follows the prediction of Eq.~(\ref{eq:LL_phases}) in Sec.~\ref{sec:universal}, while $\delta^{CS}_+$ differs from the universal prediction, as expected due to the slow decay of the inverse-square potential.

\subsection{Lieb-Liniger model}

\label{sec:Lieb-Liniger}

Arguably the simplest exactly solvable model, the seminal Lieb-Liniger~\cite{lieb63a,lieb63b,korepin93} model of 1D bosons interacting via a contact potential,
played an important role in the development of both Bethe ansatz ideas~\cite{korepin93} as well as the LL description~\cite{efetov75,haldane81b}.
Although the studies of this model were mostly an academic exercise for more than 40 years, its experimental realization in ultracold atomic gases~\cite{kinoshita04,torla04, kinoshita05,kinoshita06,amerongen08,haller09}
now allows for a parameter-free comparison of theoretical predictions with measurements.
Generally, ultracold 1D Bose gases are realized by loading a Bose-Einstein condensate into a deep 2D optical lattice in the $y-z$ plane formed by perpendicular laser beams, or using atom chips. The tight transversal confinement inhibits the occupation of higher transverse modes and provides a clean realization of the Lieb-Liniger model~\cite{olshanii98}.
These experiments stimulated significant interest in the long-standing problem of
calculating its correlation functions~\cite{korepin93}, which can be measured using
interference~\cite{polkovnikov06,imambekov07,hofferberth07,imambekov08b,hofferberth08,donner07}, photoassociation~\cite{kinoshita05},
analysis of particle losses~\cite{torla04,haller11},
density fluctuation statistics~\cite{jacqmin11,armijo10}, time-of-flight correlation
statistics~\cite{hodgman11,imambekov09_pra,manz10}, scanning electron microscopy~\cite{guarrera11,guarrera12},
or Bragg and photoemission spectroscopy~\cite{stamper-kurn99,papp08, clement09,fabbri09,ernst10,stewart08,gaebler10,dao07,veeravalli08}. Recently many new theoretical results
were obtained in this direction~\cite{khodas07b,khodas08,imambekov08,shashi10,caux06,caux07,calabrese07, gangardt03a,kheruntsyan03,gangardt03b,sykes08,deuar09,kormos09,kormos10,kormos11,pozsgay11,cheianov06a,cheianov06b,kitanine09a,kozlowski11a,kozlowski11b,cazalilla11, golovach09, cherny09},
but a fully analytical calculation of the correlation
functions is still lacking. In this subsection, we will review recent progress for the Lieb-Liniger model based on combining the phenomenology beyond the LL theory with the Bethe ansatz.

The exactly solvable Lieb-Liniger model is
given by
\begin{align}
H_{\rm LiLi}=-\frac{1}{2m}\sum_{j=1}^{N} \frac{\partial^2}{\partial
x_j^2} + 2c\sum_{1\leq j<k\leq N} \delta(x_j - x_k),
\label{eq:LiebLiniger}
\end{align}
where $c>0$ is the interaction constant and $m$ is the particle mass. In the thermodynamic
limit, the ground state is fully characterized by the
dimensionless parameter
\begin{align}
\gamma = 2 m c/\rho,
\end{align}
where $\rho=N/L$ is the
density. The regime of weak interactions corresponds to $\gamma\ll 1$,
while strong repulsion, \ie, the Tonks-Girardeau limit~\cite{girardeau60} corresponds
to $\gamma \gg 1.$ The LL parameter equals $K=v_F/v,$
where $v$ is the sound velocity and $v_F=\pi \rho/m$ is the Fermi
velocity of a noninteracting Fermi gas of density $\rho.$ The parameter
$K$ is uniquely defined by $\gamma,$ with $K\approx
\pi\gamma^{-1/2}$ for $\gamma\ll 1$ and $K\approx 1+4/\gamma$ for
$\gamma\gg 1$~\cite{cazalilla04}.

Let us briefly review the solution of the Lieb-Liniger model to
introduce the notation. We will mostly follow the conventions of
\textcite{korepin93}. The ground state quasimomenta $\nu_j$ ($1\leq j \leq
N$) are given by the solutions of nonlinear Bethe equations
\begin{align}
L \nu_j + \sum_{k=1}^N \theta (\nu_j-\nu_k)=2\pi n_j, \label{eq:Betheeqs}
\end{align}
where $\theta(\nu)=2 \arctan{\frac{\nu}{2mc}}$ is the two-particle
phase shift and the ground state quantum numbers are $n_j=j-1 -(N-1)/2$.
In the
thermodynamic limit, this system gives rise to the linear integral
equation
\begin{align} \label{eq:rhonu}
 \rho(\nu)-\frac{1}{2\pi} \int_{-Q}^{Q} K(\nu, \eta) \rho(\eta)d \eta= \frac{1}{2\pi}.
\end{align}
Here, $\rho(\nu)=\lim_{L\to\infty} 1/(L(\nu_{k+1}- \nu_k))$ is the
density of roots, $K(\nu, \eta)= 4mc/[(2mc)^2+(\nu - \eta)^2]$, and
$Q$ ($-Q$) is the highest (lowest) filled quasimomentum; $Q$ is
defined
by the normalization condition $
\rho=\int_{-Q}^{Q}\rho( \nu) d\nu$.

Particle-like excitations (Lieb-I mode) can be constructed by
adding an extra quasimomentum $|\lambda|>Q$, while hole-like
excitations (Lieb-II mode) are obtained by removing a quasimomentum
$|\lambda|<Q$.
Such excitations change the total number of particles,
and it is customary in the literature~\cite{korepin93} to change the boundary conditions for wavefunction from
periodic to antiperiodic when the number of particles changes by $\pm 1$.
Since all quasimomenta $\nu_j$ are coupled to each
other by Eq.~(\ref{eq:Betheeqs}), the addition of an extra particle or hole will shift all quasimomenta.
A convenient way to take this change into account is to introduce a
shift function
\begin{align}
F(\nu|\lambda)= \pm (\nu_j-\tilde
\nu_j)/(\nu_{j+1}-\nu_{j}), \label{eq:Fdef}
\end{align}
where $\tilde \nu_j$ are the new solutions with antiperiodic boundary conditions, and the upper (lower) sign corresponds to
an extra particle (hole). In the thermodynamic limit,
$F(\nu|\lambda)$ satisfies an integral equation
\begin{align}
F(\nu|\lambda) -
\frac{1}{2\pi}\int_{-Q}^{Q} K(\nu, \eta) F(\eta|\lambda)d \eta=
\frac{\theta(\nu - \lambda)}{2\pi} \label{eq:Feq}.
\end{align}
Due to the antiperiodic boundary conditions of $\tilde
\nu_j$, the shift function $F(\nu|\lambda)$ in fact
corresponds to the fermionic Cheon-Shigehara model~\cite{cheon98,cheon99},
which is dual to the Lieb-Liniger model. However, all results for the Lieb-Liniger model can be
formulated using
Jordan-Wigner strings and $F(\nu|\lambda),$ so we will use the fermionic language for consistency with Sec.~\ref{sec:phenomenology}.

\begin{figure}
\includegraphics[width=8cm]{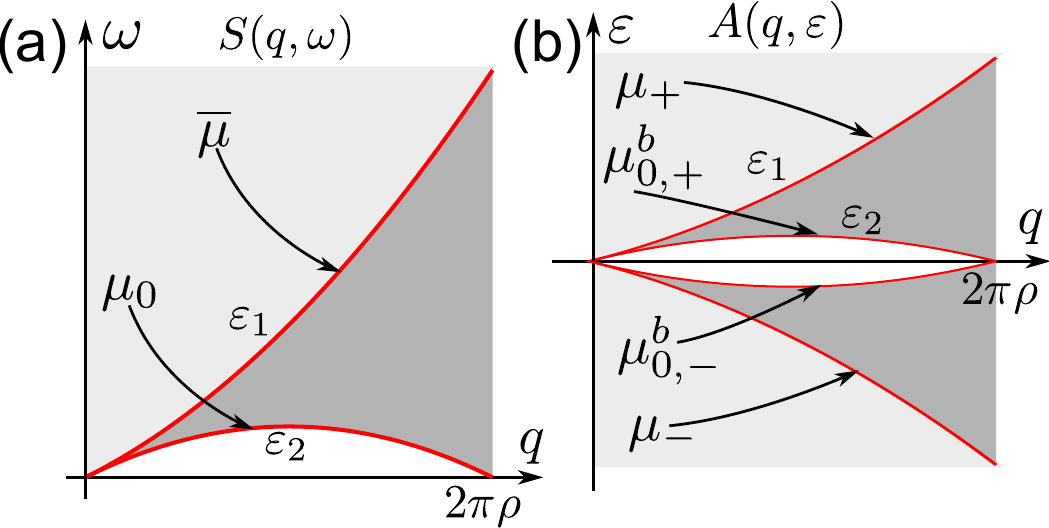}
\caption{\label{fig:LiebLinigerFig} (Color online) (a) Dynamic structure factor (DSF)
$S(q,\omega)$ and (b) spectral function $A(q,\varepsilon)$ for the Lieb-Liniger model. Shaded
areas indicate the regions where the functions are nonvanishing. Lieb's
particle mode $\varepsilon_1(q)$ and hole excitation mode
$\varepsilon_2(q)$ are indicated.
}
\end{figure}

As will be shown below, the shift functions $F(\pm Q| \lambda)$ play a crucial
role in the calculation of the edge singularities, so we will investigate them in more detail.
One can analytically derive the limiting behavior
\begin{align}
 F(Q| Q)&=1-\frac{1}{2\sqrt{K}}-\frac{\sqrt{K}}{2},\label{eq:F1exact} \\
F(-Q| Q)&=\frac{1}{2\sqrt{K}}-\frac{\sqrt{K}}{2},\label{eq:F2exact}
\end{align}
and $F(\pm Q| -Q)=-F(\mp Q|Q)$. Moreover, one can show,
\begin{align}
F(\pm Q| \lambda)\approx-\frac{\sqrt{K}}{2}+ \frac{2mc\sqrt{K}}{\pi \lambda}  \quad \text{for } Q, mc
\ll\lambda. \label{eq:F0exact}
\end{align}
Equations~(\ref{eq:F1exact})-(\ref{eq:F2exact}) have been derived by \textcite{korepin98}, and Eq.~(\ref{eq:F0exact}) follows from an expansion of the right hand side of Eq.~(\ref{eq:Feq}) combined with
\begin{align}
\rho(\pm Q)=\sqrt{K}/2\pi,
\end{align}
see, \eg, Eqs.~(I.9.20)-(I.9.22) in~\textcite{korepin93}.

The shift function can be used to calculate
the exact energies of Lieb's particle $(\varepsilon_1>0)$ and hole $(\varepsilon_2>0)$
excitations as a function of the momentum
$q(\lambda)$. They are given by
$\varepsilon_{1,2}(q)=\pm \epsilon(\lambda),$ with $\epsilon(\lambda)$
defined by
\begin{align} \label{eq:elambda}
\epsilon(\lambda)-\frac1{2\pi}\int_{-Q}^{Q} K(\lambda, \eta)\epsilon(\eta) d \eta=
\lambda^2/(2m) - \mu,
\end{align}
where $\mu$ is the chemical potential and $\epsilon(\pm Q)=0.$
The momentum corresponding to a quasimomentum $\lambda$ is given by
\begin{align} \label{eq:qlambda}
q(\lambda)=\pm \left( \lambda - \pi \rho+  \int_{-Q}^{Q}
\theta(\lambda - \nu) \rho(\nu) d \nu\right).
\end{align}
Here the upper (lower) sign corresponds to a particle (hole)
excitation with $\lambda>Q$ ($|\lambda|<Q$), and $q(Q)=0$, $q(-Q+0)=2\pi \rho=2k_F$.
Equations~(\ref{eq:elambda}) and (\ref{eq:qlambda}), together with the normalization condition mentioned earlier for $\rho(\nu)$ provide the full set of equations to determine the form of $\varepsilon_{1,2}(q)$, see Fig.~\ref{fig:LiebLinigerFig}.
Lieb's particle and hole modes can be simply understood in the limits of weak and strong interactions.

In the limit of weak interactions ($\gamma \ll 1$) 1D bosons form a quasicondensate characterized by slow algebraic decay of real-space correlation functions at long distances \cite{petrov00,mora03,popov80}. On a semiclassical level, its state can be described by a macroscopic wave function $\Psi(x,t)$ which is a solution of the 1D Gross-Pitaevskii equation~\cite{pitaevskii03},
\begin{align}\label{eq:boson_GP}
 i \partial_t \Psi(x,t) + \frac{1}{2m} \partial_x^2 \Psi(x,t) + 2c( \rho - |\Psi(x,t)|^2) \Psi(x,t) = 0.
\end{align}
One class of wave-like solutions of Eq.~(\ref{eq:boson_GP}) describes Bogoliubov quasiparticles with a dispersion relation $\varepsilon_1(q) = v q \sqrt{1 + (q/2m v)^2}$, where $v = \sqrt{\gamma} \rho /m$ is the sound velocity for the weakly interacting gas. A second class of solutions of Eq.~(\ref{eq:boson_GP}) are dark solitons~\cite{kulish76,khodas08} which have energies $\varepsilon_2(q) < \varepsilon_1(q)$. They correspond to localized perturbations of the quasicondensate density and travel at velocities $v_s(q) = \partial \varepsilon_2(q)/\partial q < v$. The spectrum is defined implicitly by
\begin{align}\label{eq:boson_soliton_spectrum}
 \varepsilon_2 &= \frac{4 \rho v}{3} \sin^3\left(\frac{\theta_s}{2}\right),\quad
 q = \rho [ \theta_s - \sin(\theta_s)],
\end{align}
where the parameter $\theta_s \in [0, 2 \pi]$ is related to the ratio of the soliton velocity $v_s(q)$ and the sound velocity $v$ by $\cos (\theta_s/2) = v_s/v$.

In the Tonks-Girardeau limit \cite{girardeau60} of strong interactions ($\gamma \gg 1$), a large contact repulsion enforces the equivalent of a Pauli exclusion principle, so the system becomes analogous to weakly interacting fermions. For $0 < q < 2 \pi \rho$, the Lieb-II mode approaches $\varepsilon_2(q) = v q - q^2/(2m)$. The Lieb-I mode approaches $\varepsilon_1(q) = v q + q^2/(2m)$ for any $q.$ The exact solution described above smoothly interpolates between the limits of strong and weak interactions. The small-$q$ behavior of the spectra preserves the form $\varepsilon_{1,2}(q) = v q \pm q^2/(2 \tilde{m})$ at any $\gamma$, with $v$ and $\tilde{m}$ being functions of $\gamma$. We find $\tilde{m} = 4 \pi^{1/2} m/(3 \gamma^{1/4})$ at $\gamma \ll 1$
from Eq.~(\ref{eq:LL_meff}). Matching the above asymptotes with the spectra of the corresponding wavelike solutions of Eq.~(\ref{eq:boson_GP}) determines the region of applicability of the asymptotes, $q \lesssim \rho \gamma^{3/4}$ \cite{khodas08}.

Let us now describe the response functions,
and for concreteness we will focus on the vicinity of
$\varepsilon_2(q)$. As has been explained in Sec.~\ref{sec:phenomenology},
the phase shifts $\delta_\pm(k)$ define the exponents of the dynamic response functions and
can be extracted from the Bethe ansatz by various techniques.
The first approach, which does not rely on integrability, is based on Eq.~(\ref{eq:phen_phi}), where $\varepsilon_{\rm th}(k=k_F-q)=-\varepsilon_2(q)<0.$
Another technique is based on the calculation of the finite-size ($\propto 1/L$) energy shifts defined by
Eqs.~(\ref{eq:energyL}) and (\ref{eq:nddelta}). This was pioneered by \textcite{pereira08,pereira09} and results in
(see Eqs.~(\ref{eq:energyL})-(\ref{eq:nddelta}) for definitions)
\begin{align}
n_{imp} &=\int_{-Q}^{Q}\rho_{imp}(\nu) d\nu,\\
2d_{imp}&=\int_{-\infty}^{-Q}\rho_{imp}(\nu) d\nu- \int_{Q}^{\infty}\rho_{imp}(\nu) d\nu,
\end{align}
where $\rho_{imp}(\nu)$ is defined by
\begin{align}
\rho_{imp}(\nu)-\frac1{2\pi}\int_{-Q}^{Q} K(\nu, \eta)\rho_{imp}(\eta) d \eta = -K(\nu,\lambda). \label{eq:phasesimp}
\end{align}
Finally, \textcite{cheianov08,imambekov08} showed that
\begin{align}
\delta_{\pm}(k)=2\pi F(\pm Q|\lambda).\label{eq:dpmviaF}
\end{align}
This result is based on an identification of the operator $U$ in Eq.~(\ref{eq:U_phen}) with a boundary condition changing
operator~\cite{schotte69,affleck94} of the LL particles whenever they pass the mobile impurity $d.$ The LL describes the low energy particles near Fermi points $\pm Q$, and thus the phase shifts $\delta_{\pm}(k)$ are proportional to the shifts of the quasimomenta of particles near these points.
Since the shift functions $F(\pm Q|\lambda)$ are proportional to the
shifts of the quasimomenta due to the presence of a hole, they are proportional to the phase shifts $\delta_{\pm}(k).$
The proportionality coefficient in Eq.~(\ref{eq:dpmviaF}) is fixed by requiring that the excitation of a particle near $\pm Q$
to the next allowed energy state corresponds to a phase shift $\pm2\pi.$

It is not at all obvious that three approaches described above should lead to the same phase shifts $\delta_{\pm}(k)$. It was shown analytically by \textcite{pereira09} that
the predictions of the latter two coincide, and it can be checked numerically that Eqs.~(\ref{eq:phen_phi})
result in the same phase shifts. Thus the coincidence of the three different predictions
of the phenomenological Hamiltonian given by Eqs.~(\ref{eq:phen_H0Hd})-(\ref{eq:phen_Hint}) provides an unambiguous microscopic confirmation
for its validity for the Lieb-Liniger model. Since the effective field theory near the edge of support did not rely on integrability,
it is natural to assume that such an approach holds universally for a large class of microscopic non-integrable models as well.
Equations~(\ref{eq:F1exact})-(\ref{eq:F2exact}) together with (\ref{eq:dpmviaF}) are nothing but the universal phase shifts given by
Eq.~(\ref{eq:LL_phases}). Here the universal phase shifts were derived in a purely microscopic fashion, which provides an independent
check of renormalization group (RG) arguments of Sec.~\ref{sec:universal}.

The phase shifts (\ref{eq:dpmviaF}) evaluated at $|\lambda|<Q$ provide nonperturbative expressions for the exponents of the DSF $S(q,\omega)$ and the spectral function $A(k,\varepsilon)$ at their edges of support, see Eq.~(\ref{eq:bosonic_exponents}). It should be noted however, that the response functions of integrable systems might also have protected singularities (or non-analyticities) within a continuum. For the Lieb-Liniger model, the most prominent of these occur at $\varepsilon = \pm \varepsilon_{1}(q)$. In addition, various weaker ``shadow'' singularities occur at $\varepsilon = \pm \varepsilon_{1}(\pm q- 2nk_F)$. The exponents can be calculated as in Sec.~\ref{sec:phenomenology} by introducing the effective mobile impurity with $v_d= \partial \varepsilon_{1}(q)/\partial q>v$, and they are given by (see Fig.~\ref{fig:LiebLinigerFig} for notations)
\begin{align}
\overline{\mu}&=1-\mu_{0,R}- \mu_{0,L}, \notag \\
\mu_{\pm} &= 1 - \mu^b_{0,\mp,R} - \mu^b_{0,\mp,L},
\end{align}
where $\mu_{0,R(L)},$ and $\mu^b_{0,\mp,R(L)}$ are defined by Eqs.~(\ref{eq:DSF_exponents}) and (\ref{eq:bosonic_exponents}), and one needs to use the phase shifts (\ref{eq:dpmviaF}) evaluated at $\lambda>Q.$

For singularities which occur within a continuum, one can also calculate the ``shoulder ratios'' of the weights right above and below the singular line.
For instance, for the DSF one obtains~\cite{pereira09}
\begin{align} \label{eq:shoulders}
 \lim_{\delta \varepsilon \to 0} \frac{S[q,\varepsilon_{1}(q) + \delta \varepsilon]}{S[q,\varepsilon_{1}(q) - \delta \varepsilon]}
= \frac{\sin( \pi \mu_{0,L})}{\sin(\pi \mu_{0,R} )}.
\end{align}
If the exponent $\overline{\mu}$ is negative (\ie, there is a non-analyticity and not a singularity), then  Eq.~(\ref{eq:shoulders}) describes only the non-analytic parts.
Similar to Eq.~(\ref{eq:LL_shoulder}), the shoulder ratios are determined only by the phase shifts.
The perturbative result (\ref{eq:perturb_nu}) can be also interpreted in terms of perturbative phase shifts (\ref{eq:deltas_pert}).
In addition to the phenomenologically determined shoulder ratios, for the Lieb-Liniger model one can combine the known expressions for the finite size formfactors~\cite{slavnov89, slavnov90}
with the field theoretical predictions of Sec.~\ref{sec:finite} to obtain predictions for the non-universal prefactors of the edge singularities~\cite{shashi10}.
Such a finite-size analysis also provides a microscopic justification for the existence of the singularities within a continuous spectrum for the Lieb-Liniger model.

\subsection{Yang-Gaudin models}
\label{sec:Yang-Gaudin}

It has been established by~\textcite{yang67} and~\textcite{gaudin67} that
the Hamiltonian (\ref{eq:LiebLiniger}) of the previous section remains exactly solvable using the so-called nested Bethe ansatz, even if one does not require the wave function to be
symmetric with respect to permutations of $x_i$ and $x_j.$ One can impose either symmetry or antisymmetry with respect to permutations of
certain subsets of $x_i.$ This leads to a family of exactly solvable models for 1D multi-component systems, such as spin-$1/2$~\cite{yang67,gaudin67}
and $SU(N)$~\cite{sutherland68} fermions, as well as Bose-Bose~\cite{li03,fuchs05,guan07} and Bose-Fermi~\cite{lai71,imambekov06a,imambekov06b,frahm05,batchelor05} mixtures.
In this section, we will review recent exact results for repulsive (iso)spin-$1/2$ bosonic and fermionic Yang-Gaudin models, illustrating connections with the universal phenomenological description of Sec.~\ref{sec:bosonic} and Sec.~\ref{sec:spinful}, respectively.

The general approach is based on the calculation of finite-size corrections to the edge state energies, and their interpretation in terms of phase shifts. This procedure is aided by the known finite-size structure of
the effective impurity theories described in Sec.~\ref{sec:finite}.

\begin{figure}
\includegraphics[width=8 cm]{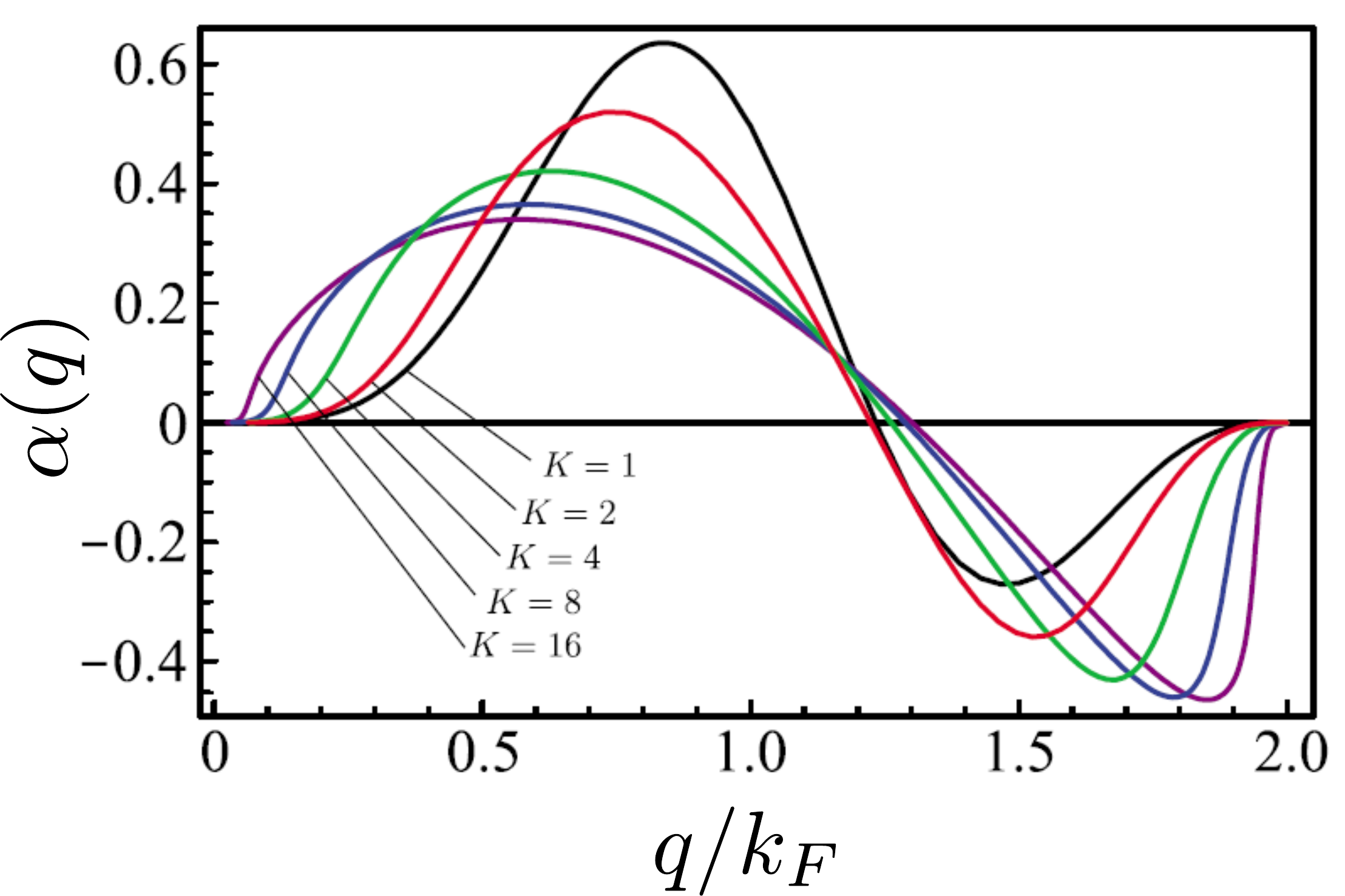}
\caption{(Color online) The function $\alpha(q)$  defining the transverse spin structure exponent for  isospin-$1/2$ bosonic Yang-Gaudin model, see Eq.~(\ref{eq:mumalpha}), is plotted for different
values of the dimensionless coupling constant $\gamma.$ The values of the
Luttinger parameter $K$ are indicated for each curve and correspond in
increasing order to $\gamma= \infty,\; 1.65, \; 0.56, \; 0.238 $ and $0.109$
respectively. Adapted from \textcite{zvonarev09b}.
}
\label{fig:alphak}
\end{figure}

Let us start from the discussion of bosons~\cite{zvonarev09b}, which have a ferromagnetic ground state~\cite{eisenberg02}. As in Sec.~\ref{sec:bosonic}, we choose the magnetization to be pointing in $+z$ direction. The microscopic wave functions in the ferromagnetic sector coincide with those of the Lieb-Liniger model, so all dynamic response functions which do not involve the spin-down state, such as $S(q,\omega)$, $S^{zz}(q,\omega)$, and the spectral function for spin-up particles $A_{\uparrow}(q,\varepsilon)$, coincide with those of the Lieb-Liniger model. The response functions which involve one spin-down state, \eg, $S^{+-}(q,\omega)$, will have singularities at the magnon spectrum $\omega_m(q).$ The states which contain one magnon are characterized by a set of quasimomenta $\{\nu_1,\ldots,\nu_N,\lambda\}$ which satisfy the set of equations~\cite{gaudin83}
\begin{equation}
L\nu_j + \sum_{k=1}^N \theta(\nu_j-\nu_k)= 2\pi n_j+\theta(2\nu_j-2\lambda)+\pi. \label{eq:YG_bosons}
\end{equation}
The total momentum $P$ and the energy $E$ are
\begin{align}
P=\sum_{j=1}^N \nu_j,\;\; E= \frac1{2m}\sum_{j=1}^N \nu_j^2. \label{eq:PE}
\end{align}
For $q\ll \pi \rho$, the magnon spectrum can be expanded as $\omega_m(q)\approx q^2/(2m^*),$ and the expression
for $m/m^*$ as a function of $\gamma$ can be obtained analytically from the exact solution~\cite{fuchs05}. It has the asymptotic behavior $1-2\sqrt{\gamma}/(3\pi)$ for $\gamma\ll 1$, and $2\pi^2/(3\gamma)$ for $\gamma\gg 1$. An analysis of the finite-size corrections to the energy of the magnon allows one to derive equations similar to Eqs.~(\ref{eq:Feq}) and (\ref{eq:dpmviaF}) which define the phase shifts for arbitrary interactions and momenta, see \textcite{zvonarev09b} for more details.
Similar to the Lieb-Liniger model, the phase shifts evaluated from the finite-size corrections coincide numerically with the phenomenological predictions~\cite{kamenev09}. In Fig.~\ref{fig:alphak}, we present the exact results of \textcite{zvonarev09b} for the transverse spin structure exponent $\mu_m$ after reparametrization
\begin{equation}
\mu_m(q)= 1 - \frac{K}2 \left(\frac{q}{k_F}\right)^2 - \frac{(K-1)^2}{K}\alpha(q), \label{eq:mumalpha}
\end{equation}
which is chosen such that $\alpha(q)$ vanishes at $q=0,2k_F.$

Considerably more complicated is the case of spin-$1/2$ fermions~\cite{essler10}, since for repulsive interactions the ground state is a singlet~\cite{lieb62} and both spin and charge Fermi
surfaces are present, as was discussed in Sec.~\ref{sec:spinful}. The existence of two Fermi points is implicitly built into the structure of the
Bethe ansatz solution, because instead of a single set of quasimomenta $\nu_i$ as in Eq.~(\ref{eq:Betheeqs}), one needs to introduce the spin quasimomenta $\Lambda_j$
which live in an auxiliary spin space. In a finite-size system, periodic boundary conditions lead to~\cite{yang67,lee11}
\begin{align} \label{eq:YGeq1}
L \nu_j &=2\pi I_j -\sum_{\alpha=1}^{N_\downarrow} \theta (2\nu_j-2\Lambda_\alpha), \\
\label{eq:YGeq2}
\sum_{j=1}^{N} \theta (2\Lambda_\alpha-2\nu_j) &=2\pi J_\alpha -\sum_{\beta=1}^{N_\downarrow} \theta (\Lambda_\alpha-\Lambda_\beta).
\end{align}
where in the first equation $j=1, ..., N,$ and in the second equation $\alpha=1, ..., N_\downarrow,$ while $I_j, J_\alpha$ are integer or half-integer depending on the parities of $N, N_{\uparrow}.$
The energies and momenta of the eigenstates are given by Eq.~(\ref{eq:PE}).

The ground state is characterized by two filled ``Fermi seas'', one for the quasimomenta $\nu_i,$ and another for the spin quasimomenta $\Lambda_\alpha.$ Similarly to the Lieb-Liniger model, excitations can be constructed by creating holes in these distributions. At zero magnetic field, the edge of support for the spectral function at $|k|<k_F$ corresponds to a spinon excitation, where a hole is created in the spinon Fermi sea while a holon is created at the Fermi surface, in complete accordance with the field-theoretical description of Sec.~\ref{sec:spinful}.
Equations~(\ref{eq:PE}) and (\ref{eq:YGeq1})-(\ref{eq:YGeq2}) contain the full information about the excitation spectrum of both holons and spinons. 
In addition, finite-size corrections to their energies 
can be analyzed similar to Sec.~\ref{sec:Lieb-Liniger}.
When combined with the extension of Sec.~\ref{sec:finite}, they
lead to explicit predictions~\cite{essler10} for the phase shifts in terms of microscopic parameters, and the obtained results coincide numerically with the phenomenological predictions
(\ref{eq:sLL_spin_phases}) and (\ref{eq:sLL_phen}).
This coincidence provides a nontrivial non-perturbative check of the RG arguments of Sec.~\ref{sec:spinful},  justifying the effective Hamiltonians of impurities with fractional quantum numbers.

Finally, let us mention that recent experimental progress with alkaline earth ultracold atoms~\cite{fukuhara07,desalvo10,taie10,takasu03,kraft09,stellmer09,deescobar09}
which naturally posses a higher symmetry of interactions~\cite{cazalilla09,gorshkov10} calls for extensions of the present approach to $SU(N)$ invariant Sutherland-type models~\cite{sutherland68}.

\subsection{Lattice models: XXZ, spinless fermions and 1D Hubbard model}

\label{sec:XXZ}
As has been discussed in Sec.~\ref{sec:spinchains}, the presence of a lattice leads to much wider possibilities
for threshold behaviors. In this section, we will review some recent results obtained by combining field-theoretical approaches with the exact solutions of XXZ~\cite{pereira08,cheianov08,karimi11},
spinless fermion~\cite{pereira09} and 1D Hubbard models~\cite{essler10}. The main modification for a generic non-integrable model at arbitrary filling is that, strictly speaking, the edges of support disappear due to
the presence of the lattice, as has been discussed in Sec.~\ref{sec:spinchains}. For integrable systems, this might not necessarily lead to a smearing of the singularities. Nevertheless, to avoid this
possible complication, we will discuss here models at commensurate fillings, such as half-filling.

The XXZ model is given by Eq.~(\ref{eq:spin_HXXZ}),
and its basic properties were discussed in Sec.~\ref{sec:spinchains}.
Using a Jordan-Wigner transformation it maps onto a Hamiltonian of spinless fermions with nearest-neighbor interactions, see Eq.~(\ref{eq:spin_XXZ_fermi}). The structures of the exact solutions of both models are the same and, \eg, $S^{zz}(q,\omega)$ of the XXZ model coincides with the DSF of the fermionic model, while $S^{+-}(q,\omega)$ and the spectral function differ
due to the Jordan-Wigner string. For concreteness, here we will focus on the XXZ model and refer the reader to Ref.~\cite{pereira09}, where spinless fermions have been discussed in detail.

Similarly to the Lieb-Liniger model, the XXZ wavefunction is written as a combination of plane waves~\cite{orbach58,korepin93}.  It is convenient to characterize them in terms of rapidities $\lambda$ which are
related to the bare two-particle phase shift $\theta(\lambda=\lambda_1-\lambda_2)$ and the bare momentum $p_0(\lambda)$ as
$
\theta= i \ln\!\left[\frac{\sinh(2i\eta+\lambda)}{\sinh(2i \eta-\lambda)}\right],\,
p_0 = i\ln\!
\left[\frac{\cosh(\lambda-i\eta)}{\cosh(\lambda+i\eta)}\right],
$
where $\eta$ conveniently parameterizes the interaction via $\Delta=-\cos{2\eta}.$
The solutions of the Bethe equations in terms of the rapidities can be imaginary, which generally
leads to a number of complications,
such as the existence of bound states discussed in Sec.~\ref{sec:spinchains}.
Inside the gapless regime, however, the ground state is constructed similar to the Lieb-Liniger model out
of real solutions which occupy a ``Fermi sea'' $(-\Lambda,\Lambda).$
Spin wave-like excitations also can be constructed by creating holes and adding particles with real rapidities on the top of the filled ``Fermi sea". The density of ground state roots $\rho(\lambda)$ and the shift function $F(\nu|\lambda)$ satisfy the equations
\begin{align} \label{eq:rho}
\rho(\nu)-\frac{1}{2\pi} \int_{-\Lambda}^{\Lambda} K(\nu, \mu) \rho(\mu)d \eta&= \frac{1}{2\pi}\frac{dp_0(\nu)}{d\nu}.\\
 F(\nu|\lambda) -
\frac{1}{2\pi}\int_{-\Lambda}^{\Lambda} K(\nu, \mu) F(\mu|\lambda)d \mu&=
\frac{\theta(\nu - \lambda)}{2\pi}, \label{eq:Fxxz}
\end{align}
where $K(\nu,\mu)=d\theta(\nu-\mu)/d\nu,$ and the normalization condition for $N$ spin-down particles on a lattice of size $M$ reads $\int_{-\Lambda}^{\Lambda} \rho(\lambda)d\lambda=N/M.$ Away from half-filling, the $\Lambda$ following from this equation is finite and the leading nonlinearity of the
spin wave spectrum is quadratic; the results of Sec.~\ref{sec:chains_finiteh} are applicable.
The peculiarity of the half-filled case ($M=2N$), expected from the considerations of Sec.~\ref{sec:spinchains}, manifests itself in the exact solution as $\Lambda\to \infty.$
In this case, all integral equations can be solved analytically by Fourier transformation, which leads, \eg, to an analytical expressions for
the Luttinger parameter $K=\left(2-2\arccos{\Delta}/\pi \right)^{-1}$ and the edge of support
$\omega_L(q) = v \sin(q)= \frac{\pi \sqrt{1-\Delta^2} }{2\arccos{\Delta}}\sin(q)$. As expected, the leading nonlinearity of spectrum is cubic.

The central objects which determine the exponents of the response functions are the phase shifts $\delta_{\pm}(k),$ which similarly to Eq.~(\ref{eq:dpmviaF}) are given by
$\delta_{\pm}(k)=2\pi F(\pm \Lambda|\lambda).$ However, one needs to take the limit $\Lambda\to \infty,$ and there is an ambiguity in the way this limit should be approached. This has lead to conflicting predictions by \textcite{pereira08} and \textcite{cheianov08}. The ambiguity was resolved by \textcite{imambekov09b} in favor of the former, based on a comparison with
the universal results of Sec.~\ref{sec:universal} and the $SU(2)$ symmetry arguments of Sec.~\ref{sec:spinchains} for $\Delta=1.$ The resulting phase shifts are momentum-independent, and they are given by
$\delta_-/(2\pi) = -\delta_+/(2\pi) = 1/(2\sqrt{K}) - \sqrt{K}/2$. The exponents of $S^{zz}(q,\omega)$ and $S^{+-}(q,\omega)$ can now be explicitly evaluated using Eqs.~(\ref{eq:DSF_exponents}) and (\ref{eq:bosonic_exponents}), and are given by~\cite{pereira09,karimi11}
$\mu^{-}_z=1-K, \;\quad \mu_x^{-}= 2-\frac1{2K} - K.$
These exponents are momentum-independent and interpolate between the results of the XY and the XXX models of Sec.~\ref{sec:spinchains}.

The results for the phase shifts and exponents away from half-filling can be obtained by numerically solving Eq.~(\ref{eq:Fxxz}). The behavior of the response function near the energy of the bound state can be also analyzed: such a bound state merges with the spinon excitation at finite momentum, and the field-theoretical description of the singularity changes at this point~\cite{pereira09,karimi11}.

Finally, let us also briefly comment on the application to the 1D Hubbard model~\cite{lieb68,essler05}, which describes spinful fermions on a lattice. The Lieb-Wu system of equations which determines the energies and momenta of the eigenstates is similar to Eqs.~(\ref{eq:YGeq1})-(\ref{eq:YGeq2}), but with $\sin{\nu_i}$ substituting $\nu_i$ inside the phase shifts. The finite-size corrections to the energies of holon and spinon excitations have been calculated~\cite{essler10}, and were used in conjunction with the field-theoretical results of Sec.~\ref{sec:spinful} to obtain predictions for some of the threshold  exponents. For certain values of the parameters they coincide numerically with the results of \textcite{carmelo04,carmelo06,carmelo08} obtained using completely different methods.
We note, however, that the detailed analysis of the 1D Hubbard model is rather complicated due to the rich kinematics, and some of the first steps in this direction have been made recently~\cite{pereira12}.

\section{KINETICS OF AND TRANSPORT IN A NONLINEAR LUTTINGER LIQUID}

\label{sec:kinetics}

In this Section, we will review some elementary processes of relaxation, as well as kinetic and transport phenomena emerging in a nonlinear LL. In statistical mechanics, one assumes that a generic macroscopic system, even if isolated from the rest of the world, will eventually reach a local thermal equilibrium. The density matrix of a finite-size part of such a system will reach the Gibbs distribution as long as that part comprises many particles. The parameters of the equilibrium distribution are fixed by additive conserved quantities (particle number, energy, momentum). Normally, we expect the approach to equilibrium to be controlled by a spectrum of relaxation rates found from an appropriate kinetic equation \cite{huang87}. However, there are prominent counterexamples to that common wisdom. The approach to thermal equilibrium of a system of interacting particles, in any dimension, may be hindered by disorder, resulting in a ``many-body localization''. This possibility was raised by \textcite{anderson58}, analyzed in the contexts of disordered solid-state conductors~\cite{basko06} and atomic cold gases~\cite{aleiner10} recently, and currently receives a considerable attention, see \textcite{pal10} and references therein. Closer to the subject of this review, the abundance of integrals of motion in a disorder-free system is also deemed to prevent equilibration \cite{polkovnikov11}. Such a possibility is foreseen in quantum integrable 1D  systems \cite{sutherland04}. Quite remarkably, an experimental investigation of the latter subtle roadblock to equilibration became possible in the context of cold gases \cite{kinoshita06}.  A restricted phase space for scattering events suppresses relaxation processes even in a generic (non-integrable) 1D system. Recently, some peculiar features of the electron equilibration were found in experiments with quantum wires formed within a GaAs heterostructure \cite{barak10} and carbon nanotubes~\cite{chen09}.

Related but not identical to the equilibration problem is the question about singularities in the dependence of the response functions on momentum and frequency. As we saw in Sec.~\ref{sec:Lieb-Liniger}, integrability allows the functions $A(k,\varepsilon)$ and $S(q,\omega)$ to be singular within the spectral continuum of excitations, in addition to the ``mandatory'' non-analytical behavior at the thresholds of the continuum. If integrability is violated, the singularities within the continuum vanish even at zero temperature. At small $\varepsilon$ or $\omega$, the singularities are smeared but not washed out completely, being rather replaced by some finite-width peaks. Like in the theory of Fermi liquids, these widths may be associated with the inverse lifetimes of the quasiparticle states which approximately diagonalize the many-body Hamiltonian of the nonlinear LL. Peak broadening of $A(k,\varepsilon)$ may be measured, in principle, in a tunneling experiment. We will identify some important elementary relaxation processes specific for various 1D systems in Sec.~\ref{sec:QP_Relax}.

Equilibration processes and the dynamic density responses of the liquid determine some of its transport properties. The most studied of those are the linear conductivity and the linear conductance of electron liquids subject to an external electric field. In a 1D system, the relation between the conductivity and the conductance is not trivial. The conductivity is well-defined in a contactless setting, for a homogeneous liquid filling the entire 1D space. Contrary to that, the conductance is determined as the current flowing through a system attached to leads biased with some small voltage.  The conductance does depend on the properties of the leads. In fact, the linear LL theory predicts that the dc conductance is determined by the properties of the leads and is independent of the parameters of the LL~\cite{safi95,ponomarenko95,maslov95}.

The conductivity $\sigma(q,\omega)$ of a homogeneous liquid is related by the Kubo formula to the current-current correlation function \cite{mahan81}. The real part $\sigma'(q,\omega) = \Re \sigma(q,\omega)$ of the conductivity can be expressed, with the help of the continuity relation and the fluctuation-dissipation theorem, in terms of the DSF. The expression for the DSF obtained in the {\it linear} LL theory at any temperature results then in $\sigma'(0,\omega)\propto\delta(\omega)$, commonly referred to as the Drude peak. For Galilean-invariant systems it is not destroyed by spectrum curvature or finite temperatures, regardless of the interactions between particles~\cite{sirker11}. However, predicting its fate at finite temperatures and in the presence of a lattice is beyond the linear LL description. The umklapp processes which are caused by the lattice and are formally irrelevant at $T=0$, may smear the $\delta$-function singularity in $\sigma'$ at finite temperatures. We briefly review this question in Sec.~\ref{sec:Drude}.

Equilibration processes do affect the conductance $G$ of a 1D electron liquid. These processes, absent in the linear LL, make the conductance temperature-dependent. We review various elementary processes leading to equilibration and their effect on the conductance and other transport characteristics in Sec.~\ref{sec:Conductance}.

Concluding the introduction to this section, we wish to emphasize that all of the questions raised here are beyond of the realm of the linear LL theory. The latter is trivially integrable and easily mapped onto a system of free bosons or free fermions, so one does not expect to find any finite relaxation.

\subsection{Relaxation processes of excitations in a nonlinear Luttinger liquid}

Like in higher dimensions, it is instructive to start the consideration of relaxation processes in 1D by discussing the case of almost-free spinless fermions. At zero interaction, the single-fermion excitations are the true eigenstates with no degeneracies in the single-particle sector, and the ground state is not degenerate or pathological (unlike in the case of free bosons). This is helpful in building a theory of relaxation processes using perturbation theory in the interaction strength. The main part of Sec.~\ref{sec:WeakQP_Relax} is devoted to the identification and evaluation of the elementary relaxation processes for spinless fermions. We will see that the lack of particle-hole symmetry leads to drastically different relaxation rates for particles and holes at low temperatures. We will also investigate the peculiarities of the energy and particle number transfer between the left- and right-moving species.

Similar to the relaxation rates in higher dimensions, the perturbatively evaluated relaxation rate in 1D vanishes when the particle's excess energy tends to zero. Due to phase space constraints, the rate is proportional to a higher power of the excess energy, see \eg,  Eq.~(\ref{II.C.3}). This should help in building a full analogue of the Fermi liquid and the kinetic theories of the quasiparticles emerging in the universal description of the nonlinear LL, see Sec.~\ref{sec:universal}. Such a program for spinless fermions has not been performed yet.

A step in that direction for an actually more complicated case of spin-$1/2$ fermions is described in Sec.~\ref{sec:holon_relax}. The complication arises from the spin degeneracy of the free-fermion single-particle states. A harbinger of the difficulties is already seen within the perturbation theory: the scattering cross-section evaluated in the basis of free fermions is divergent at low energies, leading to a relatively slow dependence of the relaxation rate on the particle's energy, see Eq.~(\ref{spinfull-tau-p}) in Sec.~\ref{sec:WeakQP_Relax}. We will see in Sec.~\ref{sec:holon_relax} that upon proper removal of the degeneracy and introduction of spinons and holons, the decay of the latter  branch is efficiently suppressed.

Methods built in Sec.~\ref{sec:WeakQP_Relax} to consider the relaxation of fermions help in the investigation of the relaxation in a 1D Bose liquid. We move to 1D bosons in Sec.~\ref{sec:relax_bosons}. For a weakly interacting gas, the relaxation of particle-like excitations can be understood with the help of perturbation theory, with some improvements required for taking care of the strong modification of the low-energy excitation spectrum. The relaxation of the other important branch of excitations -- dark solitons -- turns out to be similar to the relaxation of holes in a Fermi gas near the bottom of the band, but requires a non-perturbative treatment.

\label{sec:QP_Relax}
\subsubsection{Weakly interacting fermions}
\label{sec:WeakQP_Relax}

In order to identify processes important in relaxation, we first turn to the case of spinless weakly interacting fermions. The curvature of the dispersion relation $\xi(k)$ introduces particle-hole asymmetry into the problem. For relaxation processes, the importance of particle-hole asymmetry is already seen within perturbation theory. Indeed, it follows from Eq.~(\ref{eq:freefermions}) that the hole velocity is smaller than the velocity of low-energy excitations, \ie, particle-hole pairs near the Fermi points. Therefore, according to the Cherenkov radiation criterion, a hole introduced into the system cannot emit these excitations and consequently cannot relax at zero temperature. On the other hand, a particle moves faster than the low-energy excitations. The emission of particle-hole pairs by a moving particle is therefore allowed by energy and momentum conservation laws. The emission of a single particle-hole pair is identical to a two-particle collision. In this case, energy and momentum conservation can only be satisfied if the two incoming particles with momenta $k_1$ and $k_2$ either keep their initial momenta, $(k_1,k_2) \to (k_1,k_2)$, or switch their momenta with the other particle, $(k_1,k_2) \to (k_2,k_1)$. Neither of these options can cause relaxation.

\begin{figure}
\includegraphics[width=0.99\columnwidth]{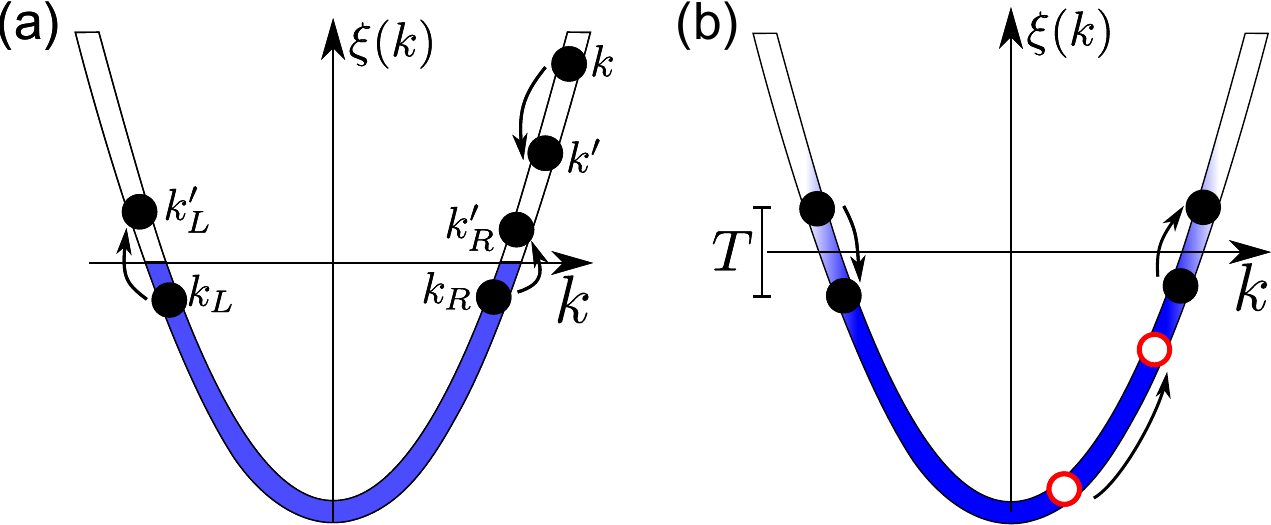}
\caption{(Color online)
(a) Relaxation of a high-energy particle due to three-particle scattering (b) Relaxation process for high-energy holes at nonzero temperatures. Filled states are depicted in blue (darker), empty states in white (lighter) color.
}
\label{Fig_rate}
\end{figure}

Scattering processes that result in a \textit{redistribution} of momenta and thus potentially lead to a finite relaxation rate must involve at least three particles. In such three-body collisions three particles with momenta $k = k_F + p$, $k_R = k_F + p_R$, and $k_L = -k_F + p_L$ in the initial state $|i\rangle$ end up in a final state $|f\rangle$ with \textit{different} momenta $k' = k_F + p'$, $k_R' = k_F + p'_R$, and $k_L' = -k_F + p'_L$, see Fig.~\ref{Fig_rate}a. For a \textit{generic} interaction the transition $|i\rangle\to|f\rangle$ has a nonvanishing momentum-dependent amplitude $\mathcal A$.

In order to evaluate the $T=0$ relaxation rate of an extra right-moving particle with momentum $k = k_F + p$ ($0 < p \ll k_F$) due to three-body collisions, we note that the single-particle states $p_R,p_L$ in the initial state of the transition
$|i\rangle$ are below the Fermi level, while all three single-particle states in the final state $|f\rangle$ are above it. Applying now Fermi's golden rule, we find
\begin{eqnarray}
\frac{1}{\tau_p(k)} &\propto\!\!& \int_0^\infty\!dp' dp_R^\prime dp_L
\!\! \int_{-\infty}^0 \!dp_L^\prime dp_R \,|\mathcal{A}|^2
\label{II.C.1}
\\
&\times\,&\delta\bigl[\left(p+p_R+p_L\right)
- \left(p'+p_R^\prime+p_L^\prime\right)\bigr]
\nonumber\\
&\times\,&\delta\Big\{
\left[\xi(k_F+p) + \xi(k_F + p_R) + \xi(-k_F + p_L) \right]\notag \\
&-&
\left[\xi(k_F+p') + \xi(k_F + p_R') + \xi(-k_F + p_L') \right]
\Big\},
\nonumber
\end{eqnarray}
where $\mathcal{A}$ is the three-body collision amplitude introduced above, and the $\delta$-functions express the energy and momentum conservation.

In writing Eq.~(\ref{II.C.1}) we took into account that for $p = k - k_F \ll k_F$ the conservation laws cannot be satisfied unless the collision involves both right- and left-moving particles.\footnote{The relaxation rates presented below in Eqs.~(\ref{II.C.3})-(\ref{spinfull-tau-p}) assume $|\xi(k_F + p)|\ll\epsilon_F$. In addition, we set a constraint $T\ll\epsilon_F$ for Eqs.~(\ref{II.C.4})-(\ref{tau-N}).} Further analysis shows that the conservation laws allow a small ($\lesssim\! p^2\!/m$) energy transfer to the left-movers. Such a solution can be found by iterations. To zero order in $p_L-p_L^\prime$, the momentum conservation gives $p-p'=p_R^\prime -p_R$. The energy released in the collision of two right-moving particles then is $\xi(k_F + p) + \xi(k_F + p_R) - \xi(k_F + p') - \xi(k_F + p_R') \lesssim p^2\!/m$. This energy is transferred to the left-movers,
$\xi(-k_F + p_L') - \xi(-k_F + k_L) \!\lesssim p^2\!/m$, which corresponds to the momentum transfer $p_L-p_L^\prime\lesssim p^2\!/(mv_F) \ll p$. Accordingly, energy and momentum conservation restrict the range of the momenta contributing to the integral in Eq.~(\ref{II.C.1}) to
\begin{equation}
p', p_R^\prime,|p_R|\lesssim p,
\qquad
p_L,|p_L^\prime|\lesssim p^2\!/(mv_F).
\label{II.C.2}
\end{equation}
The $\delta$-functions in Eq.~(\ref{II.C.1}) remove the integrations over $p_R^\prime$ and $p_L^\prime$. The remaining phase space constraints yield integration domains $\sim p^2/(mv_F)$ for $p_L$ and $\sim p$ for $p^\prime$ and $|p_R|$, see Eq.~(\ref{II.C.2}). These three factors (one $\propto p^2$ and two $\propto p$) yield the estimate \cite{khodas07a}
\begin{equation}
\frac{1}{\tau_p(k)} \propto |\mathcal{A}|^2 [\xi(k)/v_F]^4.
\label{II.C.3}
\end{equation}
For a weak generic interaction, the nonvanishing three-particle collision amplitude $\mathcal A$ appears already in the second order in the interaction strength. In the case of a long-range potential, allowing one to neglect terms proportional to $V(2k_F)$, the relaxation rate is \cite{khodas07a}
\begin{equation}
\frac{1}{\tau_p(k)}
= C\bigl[\nu^2V_0(V_0-V_{k - k_F})\bigr]^2\frac{[\xi(k)]^4}{(mv_F^2)^3}
\label{1dselfenergy}
\end{equation}
where $C = 3^3\pi/(5\cdot 2^8) \approx 0.06$ and  $\nu$ is the density of states. For a potential falling off faster than $1/x^2$ in real space, $V_0-V_{k - k_F}\propto (k - k_F)^2$, which leads to $1/\tau_p(k)\propto (k - k_F)^8$. Perturbation theory in the irrelevant interactions Eqs.~(\ref{eq:LL_Hint1}) and (\ref{eq:LL_Hint2}) to the universal-limit Hamiltonian hints that such behavior persists beyond the perturbation theory.

In the special case of an integrable model, one may expect $\mathcal{A}$ to be identically zero \cite{sutherland04}. Within the lowest-order perturbation theory, it was checked by \textcite{lunde07} for the Cheon-Shigehara model \cite{cheon98,cheon99} and by \textcite{khodas07a} for the Calogero-Sutherland model that indeed $1/\tau_p(k)=0$ for these models.

A vanishing relaxation rate would entail the presence of a power-law singularity in the spectral function $A(k,\varepsilon)$ at the energy spectrum of a particle excitation. That singularity lies {\it within} the spectral continuum. Apart from integrable models, however, $1/\tau_p(k)\neq 0$ at finite $k$ and therefore the particle peak in $A(k,\varepsilon)$ is broadened within the energy range defined by $1/\tau_p(k)$. We notice here, that in a generic case this range scales to zero faster than $(k - k_F)^4$ with $k\to k_F$, cf.~Eq.~(\ref{II.C.3}), while the deviations from the linear spectrum occur on the scale $(k-k_F)^2/m$. This justifies the consideration of power-law singularities in the limit $k\to k_F$.

A finite temperature trivially broadens the singularities in $A(k,\varepsilon)$ even within the linear LL description \cite{giamarchi04}. It would also broaden the singularities in $S(q,\omega)$ even in the absence of relaxation mechanisms (with a possible exception of the finite-temperature behavior of $S(q,\omega)$ at $q\to 0$ which we briefly review in Sec.~\ref{sec:Drude}). This broadening comes from the smearing of the edge of the Fermi distribution. Relaxation would manifest itself in the time evolution of the distribution function of excitations (thermalization) and in a number of transport phenomena. Here we concentrate just on the elementary processes of relaxation of particles ($p$) and holes ($h$).

Turning to the case of small finite temperatures, we notice that the above consideration of the zero-temperature particle relaxation rate remains valid as long as the particle energy $\xi(k) \gg\sqrt{\epsilon_F T}$, where $\epsilon_F=k_F^2/(2m)$ is the Fermi energy. At smaller $\xi(k)$, the phase space of left-moving excitations participating in the collision ($p_L$, $|p_L^\prime|$) is not controlled any more by the small transferred momentum of Eq.~(\ref{II.C.2}), but rather by thermal smearing $\sim T/v_F$ of the momentum distribution function. As a result, the factor $\propto (k-k_F)^2$ coming from integration over $p_L$ is replaced by a factor $\propto mT$, yielding
\begin{equation}
\frac{1}{\tau_p(k,T)} \propto |\mathcal{A}|^2 mT [ \xi(k)/v_F]^2\,,\quad T\ll\xi(k)\ll\sqrt{\epsilon_F T} \nonumber
\label{II.C.4}
\end{equation}
instead of Eq.~(\ref{II.C.3}).

The finite-temperature effect is more dramatic for holes, since it makes their relaxation possible in the first place, see Fig.~\ref{Fig_rate}b. Due to thermal smearing, a counterpropagating particle can give up an energy of order of $T$. Thus, a hole can relax its energy with a characteristic energy loss of $\Delta\epsilon\sim \epsilon_F T/|\xi(k)|$. It means that an energetic hole ``floats'' towards the Fermi level in many steps small compared to $|\xi(k)|$, as long as the hole energy remains large compared to $\sqrt{\epsilon_F T}$. Under this condition, application of Fermi's golden rule yields the rate
\begin{equation}
\frac{1}{\tau_h(k,T)} \propto |\mathcal{A}|^2 m^2\epsilon_F \frac{T^3}{|\xi(k)|^2},\quad |\xi(k)| \gg \sqrt{\epsilon_F T},
\label{tau-h}
\end{equation}
for a single step of the relaxation process; this rate defines the lifetime of a state with given energy $\xi(k)$.

Within the perturbative treatment, the evaluation of the relaxation rates was generalized to the case of spin-$1/2$ fermions by \textcite{karzig10}. Targeting the experiment by \textcite{barak10}, the evaluation was performed for electrons in a quantum wire of a small width $a\ll 1/k_F$ interacting via a Coulomb potential which is screened by a gate at some large distance compared to $a$ and $1/k_F$. The relaxation rates were found to be
\begin{equation}
\frac{1}{\tau_p(k,T)} = \frac{9\epsilon_F}{32\pi^3\hbar} \left( \frac{e^2}{\kappa\hbar v_F} \right)^4
\lambda^2[\xi(k)] |\xi(k)/\epsilon_F|^2, \quad T=0
\label{spinfull-tau-p}
\end{equation}
for particle excitations. Here, $\kappa$ is the dielectric constant of the host material, and $\lambda(\xi)=\ln |1/2k_Fa|\ln |\xi/4\epsilon_F|$. At finite temperatures and excitation energies $|\xi(k)|\ll\sqrt{\epsilon_F T}$, particles and holes relax with the same rate,
\begin{align}
\frac{1}{\tau_p(k,T)} &\approx \frac{1}{\tau_h(k,T)} \approx \frac{3 c_1 \epsilon_F}{4\pi^3\hbar} \left(\frac{e^2}{\kappa\hbar v_F}\right)^4
\lambda^2[\xi(k)] (T/\epsilon_F), \notag \\
&\quad |\xi(k)| \ll \sqrt{\epsilon_F T},
\label{tau-ph}
\end{align}
where the numerical constant is $c_1=4\ln 2-1$.

 Note that the rate Eq.~(\ref{tau-ph}) with $|\xi(k)|\sim T$ also determines the thermalization time of the system towards a boosted equilibrium distribution function [cf. Eq.~(\ref{Galilean-f})]. For the relaxation of spinless fermions such thermalization process was addressed rigorously by solving a linearized quantum Boltzmann equation exactly in Ref.~\cite{micklitz11}. Interestingly, the equilibrium is not characterized by an equipartitioning of the injected excitation energy. Most of the injected energy rather stays within the, say, right moving branch because the energy transfer between right and left movers is suppressed as can be seen from Eq.~(\ref{II.C.2}) and Fig.~\ref{Fig_rate}: if a right moving excitation loses an energy $\sim |\xi(k)|$, only a fraction $\sim |\xi(k)/\epsilon_F|$ of that energy is transferred to the left moving branch.

The perturbative treatment of scattering of a spin-$1/2$ electron off an electron in the Fermi sea requires that the incoming electron has energy $\xi(k) \gg mv_FV(q\to 0)/\hbar$. This is the applicability condition for the Born approximation. One may view this condition as the one allowing the electron to preserve its integrity without separating into spin and charge modes in the collision process. Indeed, in the weak-coupling limit, the difference between holon and spinon velocities is $v_c-v_s\simeq V(q\to 0)/\hbar$, so one may recast the condition for the applicability of the perturbation theory as $v_c-v_s\ll\Delta v$, where $\Delta v=\xi(k)/(mv_F)$ is the difference of the velocities of the colliding particles; in other words, holon and spinon have no time to separate in the course of the electron collision \cite{karzig10}. We note that the perturbative result  for particles at the boundary of its applicability, $\xi(k)\sim mv_FV(q\to 0)/\hbar$, matches the estimate of the holon relaxation rate evaluated in the limit of low energies, see Sec.~\ref{sec:holon_relax}  \cite{schmidt10b}.

\begin{figure}[t]
\includegraphics[width=0.99\columnwidth]{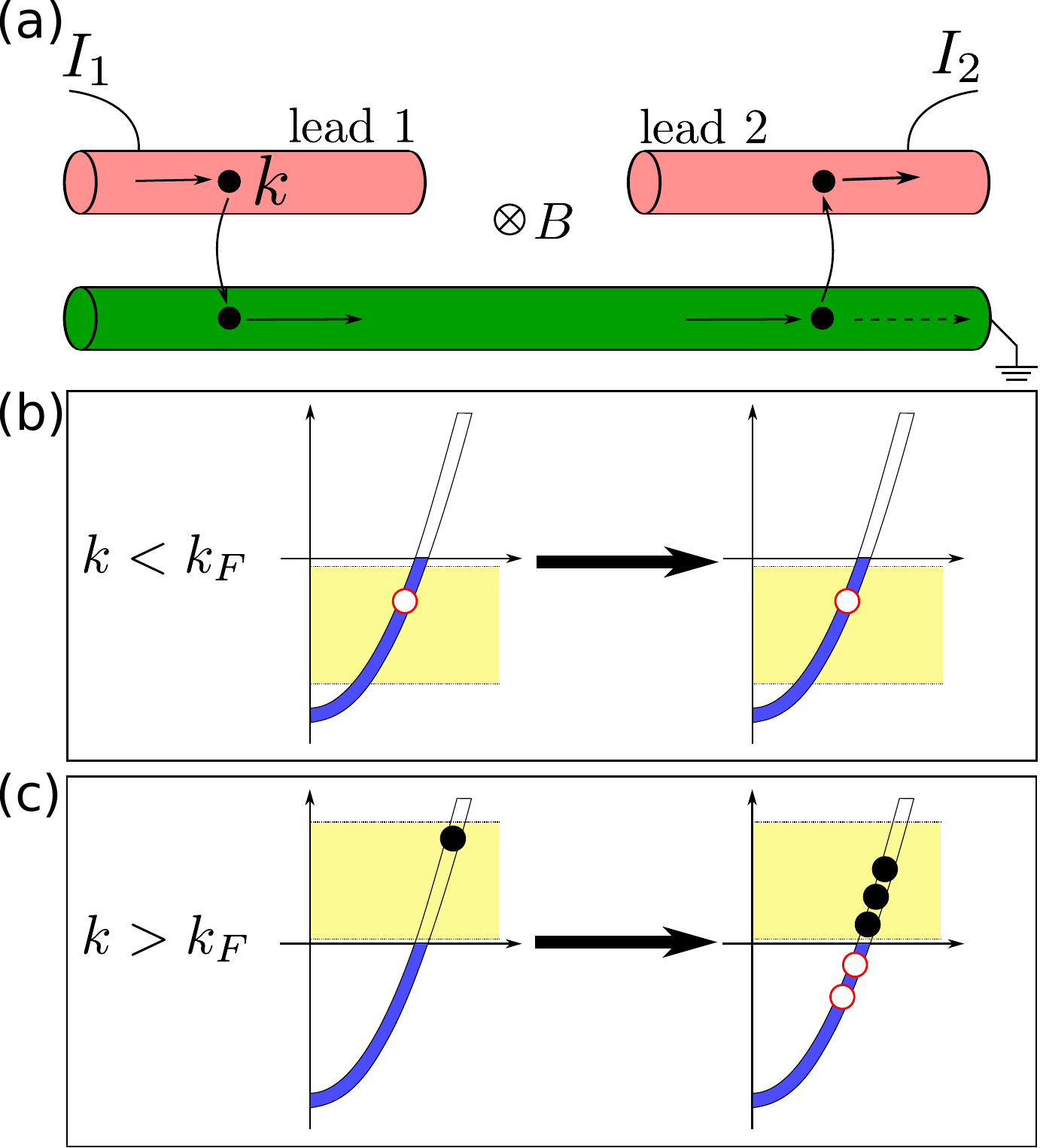}
\caption{(Color online)
(a) Schematic representation of the experimental setup used by \textcite{barak10}. A bias applied to lead 1 injects the current $I_1$ through the left tunnel junction into the grounded wire; the current $I_2$ through the right tunnel junction is collected in lead 2. Depending on the bias polarity, particles or holes are injected. The injection occurs in a window of momenta (marked by shaded regions in panels (b) and (c)) around a value $k$ controlled by the magnetic field $B$. (b) A hole ($0 < k < k_F$) injected from lead 1 cannot relax, and will be collected in lead 2. (c) The relaxation of a particle ($k > k_F$) injected from lead 1 results in the formation of additional particle-hole pairs. Since only the particles are extracted into lead 2, the collected current exceeds the injected current; the difference, drawn from the ground corresponds to the hole current sinking into the ground, see the dashed line in panel (a).}
\label{fig:barakSetup}
\end{figure}

\begin{figure}[t]
\includegraphics[width=0.99\columnwidth]{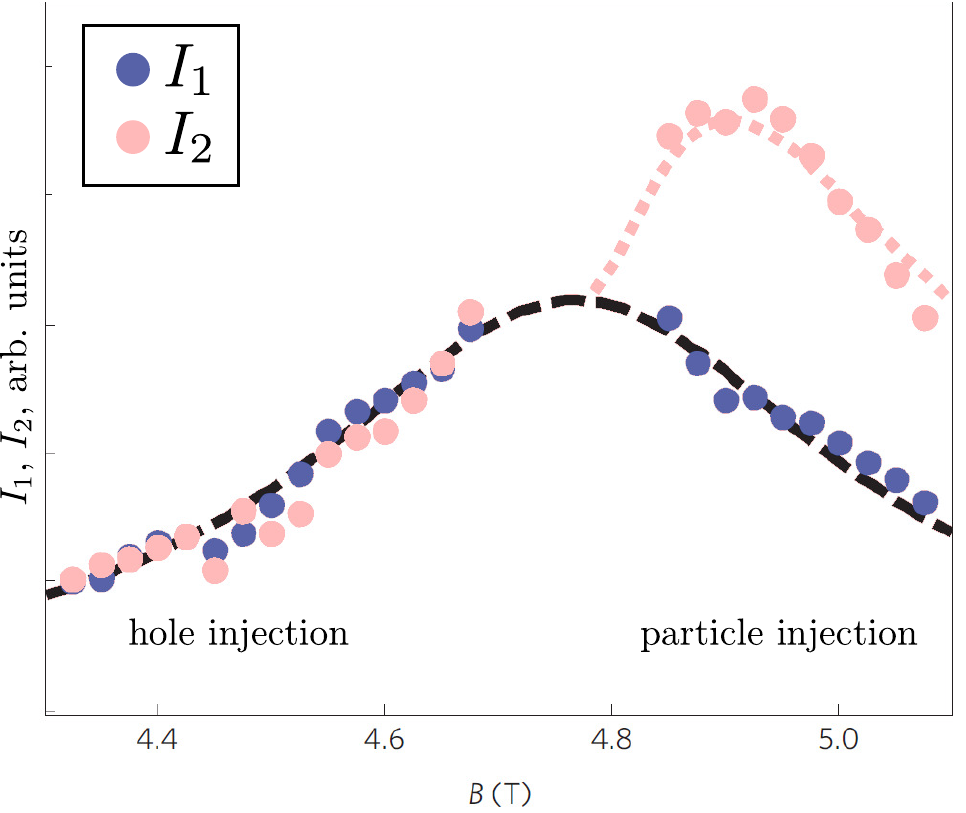}
\caption{(Color online) Injected current $I_1$ (dark dots) and collected current $I_2$ (light dots) as a function of the magnetic field, which controls the momentum of injected carriers. The two currents coincide in the case of hole injection. If particles are injected, on the other hand, the collected current exceeds the injected one due to relaxation. The experimental setup and physical explanation are shown in Fig.~\ref{fig:barakSetup}. Adapted from \textcite{barak10}.}
\label{barak}
\end{figure}

The asymmetry in the relaxation rates of particles and holes is a direct consequence of the nonlinearity of the excitation spectrum. It naturally explains the results of the experiment \cite{barak10} in which electrons were injected in and extracted from a quantum wire. In the experiment, two tunnel junctions designed to have a momentum-dependent tunneling rate, were attached to a grounded quantum wire, see Fig.~\ref{fig:barakSetup}a. Because of the device constraints, it was possible to inject particles or holes within some band of momenta, with the center of the band controlled by a magnetic field applied perpendicular to the wires comprising the device. When holes were injected through the left junction, the current collected by the right junction was equal to the injected current (blue and red dots follow the same curve in the left portion of Fig.~\ref{barak}). That is naturally explained by the absence of hole relaxation: a single hole injected through the left junction is extracted with the right one, see Fig.~\ref{fig:barakSetup}b. However, once the junctions (and the applied injection bias) are tuned to inject and collect particles, the collector current {\it exceeds} the injected one. This striking behavior can be explained by the relaxation of an injected particle, which creates a number of particle-hole pairs. Particles of these pairs are ``scooped'' by the collector, while holes are allowed to sink into the ground, see Fig.~\ref{fig:barakSetup}c. A simple set of rate equations 
explained quantitatively the observations \cite{barak10}.

\begin{figure}[t]
\includegraphics[width=0.49\columnwidth]{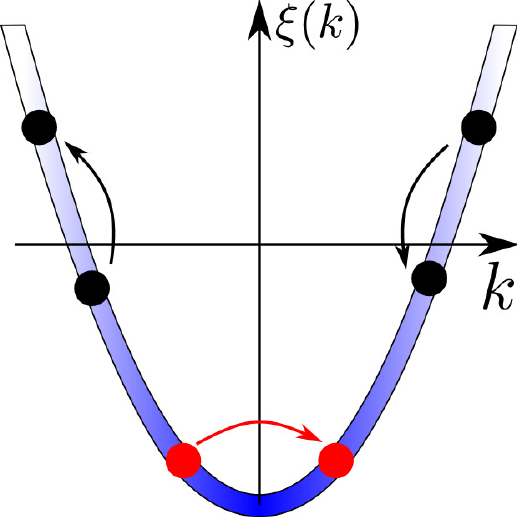}
\caption{(Color online)
A small-momentum relaxation process leading to a change in the numbers of left- and right-movers
}
\label{bottom-back}
\end{figure}

The relaxation processes considered above involve only low-energy excitations and do not change the numbers $N_L$ and $N_R$ of left- and right-moving particles. Changing those numbers bears consequences for the conductance, the thermopower, and the thermal conductance \cite{karzig10,matveev10,	 micklitz10a,levchenko10,rech09,rech08,levchenko11,levchenko11b}. At low temperatures, the relaxation of the difference $N_R-N_L$ involves states close to the bottom of the band, see Fig.~\ref{bottom-back} \cite{lunde07,matveev11}. We define the corresponding relaxation time $\tau_N$ by relation $d(N_R-N_L)/dt=-(N_R-N_L)/\tau_N$, assuming that the temperature is the same for the left- and right-movers, while their chemical potentials are slightly different. Because a ``deep'' hole is involved in the relaxation, the rate is exponentially small at low temperature, $1/\tau_{N}\propto\exp(-\epsilon_F/T)$. The pre-exponential factor scales as a power of temperature, with an exponent depending on the type of interaction potential and the presence of spin degeneracy. If one assumes a smooth (in real space) potential and sets $V_q=V_0(1-q^2/q_0^2)$ at small $q$, while $V_{k\gtrsim k_F}=0$, then \cite{lunde07,micklitz10a}
\begin{equation}
\frac{1}{\tau_N}\sim \epsilon_F\left(\frac{V_0}{v_F}\right)^4\left(\frac{k_F}{q_0}\right)^4\left(\frac{T}{\epsilon_F}\right)^7\exp\left(-\frac{\epsilon_F}{T}\right).
\label{tau-N}
\end{equation}
The $T^7$ temperature dependence of the pre-exponential factor comes from the phase space constraints on the scattering event (yielding a factor $\propto T^3$), and from the partial cancelation of the direct and exchange contributions to the scattering amplitude, similar to the one occurring in Eq.~(\ref{1dselfenergy}), which provides an additional factor $\propto T^4$. Note that the latter factor is not present in higher-order terms with respect to the inter-particle interaction potential. Therefore, in the generic case $1/\tau_N\propto T^3\exp[\varepsilon_{\rm th}(0)/T]$, where $\varepsilon_{\rm th}(0)<0$ is the energy of a hole at the bottom of the band, renormalized by interactions.

We should emphasize that the above estimates of $\tau_N$ refer to an ``elementary act'' of changing $N_R-N_L$. In that act, a hole in the fermion distribution near the bottom of the band changes the direction of its motion. The exponential factor in $1/\tau_N$ comes from the probability for the existence of such a hole, and the prefactor comes from the inverse lifetime $1/\tau_{\rm dh}\propto T^3$ of the existing deep hole, see Eq.~(\ref{tau-h}). The characteristic variation of the hole momentum in the scattering event depicted in Fig.~\ref{bottom-back} is $\Delta p\sim T/v_F$, while its energy variation $\sim T^2/\epsilon_F$ is small compared to the characteristic change of energy ($\sim T$) in each of the involved particle-hole pairs near the Fermi levels. This is why the hole dynamics may be viewed as diffusion in momentum space with the diffusion constant
\begin{equation}
B\sim (\Delta p)^2/\tau_{\rm dh}\propto T^5
\label{hole-diff}
\end{equation}
and is described by a Fokker-Planck equation \cite{castroneto96}. The proportionality coefficient missing in Eq.~(\ref{hole-diff}) was found for the case of weak and strong interactions in a system of spinless fermions, respectively, by \textcite{micklitz10a} and \textcite{matveev10}. For arbitrary interaction strength, the results are discussed in Sec.~\ref{sec:Conductance}.

To conclude the discussion of relaxation processes within the perturbative treatment of interactions, we mention here a peculiarity of the scattering processes for spinless fermions on a lattice \cite{pereira09}. The free-particle spectrum $\xi(k)\propto-\cos k$ of a tight-binding model allows for two particles with momenta $k$ and $k_3=\pi -k$ to scatter into two other states, $k_1$ and $k_2=\pi-k_1$.
If the chemical potential is shifted from the middle of the band, this process apparently yields a finite decay rate for a range of possible particle momenta $k$ within first-order perturbation theory. However, it is not fully clear if the lowest-order perturbative treatment is applicable in that special case: for a given state $k$, the mentioned states $k_1,k_2,k_3$ involved in the relaxation process are also involved in the formation of two-particle bound states \cite{pereira09}.

\subsubsection{Spinful fermions at arbitrary interaction strength: holon lifetimes}
\label{sec:holon_relax}

In this section, we will go beyond the weakly interacting limit for spinful fermions, and consider the lifetimes of holons in the low-energy limit.

As we already mentioned, the phase space argument applied to interacting spinless fermions in 1D leads to the estimate of the lifetime in Eq.~(\ref{II.C.3}). This estimate is valid at small energies of excitations whose dispersion relation resides within the particle-hole continuum. A similar argument applied to a decay of a holon into two spinons in a 1D spin-$1/2$ fermionic system would lead to a decay rate $\propto |k-k_F|$, possibly contradicting the notion of a well-defined holon branch at small $k-k_F$.

A combination of the methods described in Secs.~\ref{sec:universal} and \ref{sec:phenomenology} should allow us to express the proportionality coefficient in Eq.~(\ref{II.C.3}) in terms of higher derivatives of the dispersion relation with respect to momentum and particle density; similarly, these methods should allow one to reliably evaluate the decay rate of a holon. Such a program was not implemented yet for spinless fermions, but an attempt was made to evaluate the broadening of the holon branch of excitations in a spin-$1/2$ 1D fermion system. There is some disagreement in the conclusions of \textcite{schmidt10b} and \textcite{pereira10}. We find that the decay rate of a holon close to a Fermi point  ($+k_F$, for definiteness) scales to zero faster than $|k-k_F|^3$, as we discuss next.

In order to elucidate the possible decay processes for holons, it is convenient to start again from a description in terms of refermionized quasiparticles. The band curvature of the physical fermions leads to interactions between the quasiparticles. Away from the Fermi points, it is advantageous to classify the interaction processes by their relevance in the RG sense and to consider all possible interaction operators which are allowed by $SU(2)$-symmetry and Galilean invariance. Due to its built-in $SU(2)$-symmetry, non-Abelian bosonization \cite{gogolin98} is a convenient tool to achieve this. Expressed using the left- and right-moving holon densities $J_\alpha(x)$ and spinon densities $\VJ_\alpha(x)$ ($\alpha = L,R$), the Hamiltonian of the linear LL reads $H_0 = H_c + H_s$, where
\begin{align}
    H_c &= 2\pi v_c \int dx [ J_R^2(x) + J_L^2(x) ], \notag \\
    H_s &= \frac{2\pi v_s}{3} \int dx [ \VJ_R^2(x) + \VJ_L^2(x) ].
\label{quadratic}
\end{align}
The holon and spinon densities are related to the physical fermion operators by
\begin{align}
 J_{\alpha}(x) &= \frac{1}{2} \sum_{\sigma} \psi^\dag_{\alpha\sigma}(x) \psi_{\alpha\sigma}(x), \notag \\
 \VJ_\alpha(x) &= \sum_{\sigma\sigma'} \psi^\dag_{\alpha\sigma}(x) \vec{S}_{\sigma\sigma'} \psi_{\alpha\sigma'}(x),
\end{align}
and $\vec{S}_{\sigma\sigma'}$ denotes the vector of spin matrices (half of the Pauli matrices for spin-$1/2$). The operators $J_\alpha(x)$ are related to the physical charge density by $\rho_c(x) = 2 \sqrt{K_c} \expct{J_L(x) + J_R(x)}$. This Hamiltonian emerges at the low-energy RG fixed point and is valid in the narrow-band limit. The leading correction for increased bandwidth is an interaction between left-moving and right-moving spin densities, \cite{gogolin98}
\begin{align}
    H_g = - 2 \pi v_s g \int dx\ \VJ_R(x) \cdot \VJ_L(x).
\end{align}
Note that when expressed in terms of the Abelian spinon fields $\tilde{\phi}_s$ and $\tilde{\theta}_s$, the operator $H_g$ generates the sine-Gordon term \cite{giamarchi04}. The band curvature of the physical fermions leads to interaction operators which are cubic in spin and charge densities,
\begin{align}\label{Hekz}
    H_\eta &= \frac{4\pi^2}{3} \int dx  \left[ \eta_- (J_R^3 + J_L^3) - \eta_+ (J_R^2 J_L + J_L^2 J_R) \right], \notag \\
    H_\kappa &= \frac{4\pi^2}{3} \int dx \Big[ \kappa_- (J_R \VJ_R^2 + J_L \VJ_L^2) \notag \\
    &+ \kappa_+ ( J_R \VJ_L^2 + J_L \VJ_R^2) \Big], \notag \\
    H_\zeta &= \frac{4\pi^2 \zeta}{3} \int dx\ (J_L + J_R) \VJ_R \cdot \VJ_L.
\end{align}
Note that these operators represent all cubic terms which are compatible with $SU(2)$-symmetry. In particular, this symmetry prohibits terms linear in the vector operators $\VJ_\alpha(x)$. Interaction operators containing quartic and higher-order terms in $\VJ_\alpha(x)$ and $J_\alpha(x)$ do exist but their contribution is subleading for small bandwidths.

The prefactors $g$, $\zeta$, $\kappa_\pm$ and $\eta_\pm$ can be fixed phenomenologically by relating them to other observable quantities. The modification of the constants of the Hamiltonian Eq.~(\ref{quadratic}) in response to a small density variation yields the relations \cite{schmidt10b,pereira10,pereira06}
\begin{align}
    \kappa_- + \kappa_+ = \frac{v_c}{\sqrt{K_c}} \frac{\partial v_s}{\partial \mu}, \;\;
    \zeta = - \frac{3}{2} \frac{v_c }{\sqrt{K_c}} \frac{\partial(v_s g)}{\partial \mu}. \label{zeta_phen}
\end{align}
The difference $\kappa_- - \kappa_+$ can be related to the mass $m$ of the physical fermions by considering a charge current variation of the Galilean-invariant system. One finds \cite{nayak01,pereira10}
\begin{align}
    \kappa_- - \kappa_+ = \frac{1}{m \sqrt{K_c}}.
\end{align}

It is known that upon a bandwidth reduction $g$ flows logarithmically to zero \cite{gogolin98}. Assuming the initial bandwidth to be of order $k_F$, for a smaller bandwidth of order $k-k_F$ the effective coupling constant will flow to $g(k) = 1/\ln[k_F / (k-k_F)]$. As the chemical potential $\mu$ is proportional to $k_F$, the derivative $\partial g/\partial \mu \propto - g^2/k_F$, as noted by \textcite{schmidt10b}. The derivative $\partial v_s/\partial \mu$, on the other hand, remains finite for small bandwidths. Therefore, in leading logarithmic approximation, $\partial g/\partial\mu$ can be neglected and the coupling constants $\kappa_\pm$ and $\zeta$ can be related as
\begin{align}\label{zeta_kappa_phen}
    \zeta \approx -\frac{3}{2} g \frac{v_c}{\sqrt{K_c}} \frac{\partial v_s}{\partial \mu} = -\frac{3}{2} g ( \kappa_- + \kappa_+).
\end{align}

Holons can relax via the creation of low-energy spinons. Let us investigate the decay of an initial state $\ket{i} = \ket{k}_c \ket{0}_s$ which contains an additional holon with momentum above the Fermi edge and no spinon excitations. Relaxation of the holon to a momentum $k' < k$ can happen via the creation of two spinon density excitations with momenta $p_L < 0$ and $p_R > 0$. This final state will be labeled $\ket{f} = \ket{k'}_c \ket{p_L,p_R}_s$. For momenta $k$ close to the Fermi point, momentum and energy conversation for this process read
\begin{align}
    k = k' + p_R + p_L,  \;
    v_c k = v_c k' + v_s (p_R - p_L), \notag
\end{align}
and have nontrivial solutions ($k \neq k'$) for $v_c > v_s$.

The holon lifetimes associated with this decay channel can be calculated using Fermi's golden rule. Two combinations of operators from the interaction terms (\ref{Hekz}) have a nonzero matrix element between the states $\ket{i}$ and $\ket{f}$. To first order in the interaction, $\bra{f} H_\zeta \ket{i}$ is the only such term. To second order, only $\bra{f} H_g H_\kappa \ket{i}$ and $\bra{f} H_\kappa H_g \ket{i}$ are nonzero.

For the first-order matrix element $T_\zeta = \bra{f} H_\zeta \ket{i}$, one finds \cite{schmidt10b}
\begin{align}\label{T_zeta}
    T_\zeta
 = \frac{\pi \zeta}{2 L} \delta_{k-k'-p_L-p_R} \sqrt{|p_L p_R|}.
\end{align}
The matrix elements $\bra{f} H_g \ket{i}$ and $\bra{f} H_\eta \ket{i}$ vanish because $H_g$ and $H_\eta$ do not couple spinons and holons. The remaining first-order matrix element $\bra{f} H_\kappa \ket{i} = 0$ because it contains only terms of the form $\VJ_\alpha^2(x)$, which do not create spinons on opposite branches.

To the second order, cross-terms of the operators $H_g$ and $H_\kappa$ may couple the same initial and final states as above, yielding the amplitude
\begin{equation}\label{T_kappa_g}
     T_{\kappa g}=
 \frac{3 \pi g}{4 L} (\kappa_- + \kappa_+) \delta_{k-k'-p_L-p_R}\sqrt{|p_L p_R|}.
\end{equation}
Other second-order terms exist but they contain higher powers of $p_L$ and $p_R$ and are therefore subleading compared to $T_{\kappa g}$ for holon momenta $k$ near the Fermi points. According to Fermi's golden rule the rate is
\begin{align}\label{FGR}
    \frac{1}{\tau_{\rm holon}} = 2\pi \sum_{\ket{f}} |T_\zeta  + T_{\kappa g} |^2 \delta(\epsilon_f - \epsilon_i),
\end{align}
where $\epsilon_i$ and $\epsilon_f$ are the energies of the initial state $\ket{i}$ and the final state $\ket{f}$, respectively. The sum over all final states $\ket{f}$ translates to a summation over the momenta $p_L < 0$, $p_R > 0$ and $k' \in [k_F, k]$. It can be seen from Eqs.~(\ref{T_zeta}) and (\ref{T_kappa_g}) that each of the decay channels taken individually would lead to a decay rate $1/\tau_{\rm holon} \propto (k - k_F)^3$. However, Fermi's golden rule (\ref{FGR}) contains the square of the sum of the probability amplitudes $T_{\zeta}$ and $T_{\kappa g}$. The prefactors of both amplitudes are related according to Eq.~(\ref{zeta_kappa_phen}) and one finds $T_{\zeta} + T_{\kappa g} = 0$. Therefore, the decay rate vanishes\footnote{This conclusion of \textcite{schmidt10b} differs from the one of \textcite{pereira10}.} up to terms proportional to $g^2 (k-k_F)^3$, in the calculation of $1/\tau_{\rm holon}$ performed to the second order in $g=1/\ln[k_F/(k-k_F)]$. Retaining in Eq.~(\ref{zeta_phen}) the derivative $\partial g/\partial\mu\propto g^2/\epsilon_F$ exceeds the accuracy of our calculation. It is not clear if the evaluation of $1/\tau_{\rm holon}$ to order $g^4$ would yield zero. Possibly, in that order the distinction between integrable and non-integrable systems emerges.

In the limit of weak backscattering, $V(2k_F) \ll V(0) \ll v_F$, the universal logarithmic dependence for $g(k-k_F)$ is reached only at very low energies, while its bare value $g \propto V(2k_F)/v_F$ is applicable as long as $[V(2k_F)/v_F]\ln[k_F/(k- k_F)]\ll 1$. In that limit, which includes weak Coulomb repulsion, Eqs.~(\ref{zeta_phen}) and (\ref{T_kappa_g}) yield
\begin{align}
\frac{1}{\tau_{\rm holon}} \propto \epsilon_F \frac{V(0)}{v_F} \left[\frac{V(2k_F)}{v_F}\right]^2 \left[\frac{k-k_F}{k_F}\right]^3.
\end{align}
This estimate should be viewed as the result of perturbation theory in $V(2k_F)$ in the basis of well-defined holon and spinon modes with linear spectrum, which sets a limit on the holon momenta, $k - k_F \lesssim mV(0)$ (we also used $v_c-v_s\sim V(0)$ in the derivation). Curiously, the latter estimate for $1/\tau_{\rm holon}$ at the limit of its applicability, $k - k_F \sim mV(0)$, matches the estimate of the relaxation rate of a spinful fermion evaluated in the basis of free fermions perturbatively, see Eq.~(\ref{spinfull-tau-p}).

\subsubsection{Relaxation of excitations in a weakly interacting 1D Bose gas}
\label{sec:relax_bosons}

Within the integrable Lieb-Liniger model, see Eq.~(\ref{eq:LiebLiniger}), the excitations of a 1D Bose gas do not relax. The DSF exhibits a power-law singularity at the Lieb-I mode and a power-law behavior converging to zero at the Lieb-II mode, see Sec.~\ref{sec:bosonic}. As discussed in Sec.~\ref{sec:Lieb-Liniger}, in the limit of weak interactions the dispersion relation for the Lieb-I mode approaches the Bogoliubov quasiparticle spectrum, while the dispersion of the Lieb-II mode corresponds to the spectrum of ``dark'' solitons. A weak perturbation breaking the integrability leads to finite lifetimes of the excitations \cite{muryshev02}. At zero temperature, the finite relaxation rate of the Bogoliubov quasiparticles smears the singularity in the response functions at the Lieb-I mode \cite{tan10}. At $T\neq 0$, the relaxation rate of a dark soliton prepared in some high-energy state also becomes finite \cite{mazets08,muryshev02,gangardt10}. The relaxation rates of Bogoliubov quasiparticles and dark solitons in 1D strongly depend on temperature. The theory of dark soliton relaxation was also extended to include the dissipative dynamics of the so-called depletons forming around an impurity imbedded in a 1D Bose gas or a spin-flipped particle in a spinor 1D Bose gas \cite{gangardt09,schecter11}.

The leading corrections to the Lieb-Liniger model (\ref{eq:LiebLiniger}) have the form of a three-body interaction term,
\begin{equation}
V=-\frac{\alpha}{9m}\int dx :\rho^3(x):\,.
\label{three-body}
\end{equation}
These terms of the Hamiltonian $H=H_{\rm LiLi}+V$ can be derived explicitly by a projection onto the lowest subband of transverse quantization in a confining potential with cylindrical symmetry. For a model in which the interaction in 3D is described by a pseudopotential $V_{\rm 3D}(r)=4\pi (a/m)\delta({\bf r})$, where $a$ is the scattering length \cite{pitaevskii03}, and with the amplitude of radial zero-point motion $a_r=(m\omega_r)^{-1/2}\gg a$ one finds~\footnote{The value of $\alpha$ in \cite{mazets08} contains a spurious factor of $4$ compared to Eq.~(\ref{alpha}).} \cite{olshanii98,muryshev02,mazets08,tan10}
\begin{equation}
\gamma=2mc/\rho=2a/(\rho a_r^2)\,,\, \alpha=18\ln(4/3)(a/a_r)^2\,.
\label{alpha}
\end{equation}
Here, $\rho$ is the density of 1D Bose gas. The limit of weak interaction means $\gamma\ll 1$.

A finite three-particle scattering amplitude which leads to a damping of the Bogoliubov mode appears already in the first order in $\alpha\ll 1$. The evaluation of the corresponding relaxation rate is especially simple for quasiparticles with energies $\varepsilon_1(q)\gg\gamma m \rho^2$, so that $\varepsilon_1(q)\approx q^2/2m$. In addition, we assume $\varepsilon_1(q)\gg T$. To the lowest order in $\alpha$, the differential rate of inelastic scattering in which a quasiparticle with momentum $q$ loses energy $\omega$ is given by
\begin{equation}
 \sigma_q(\omega)=\frac{\alpha^2}{2\pi m^2}\int dp\delta[\omega-\varepsilon_1(q)+\varepsilon_1(q-p)]{\cal G}(p,\omega)\,.
\label{sigma-q}
\end{equation}
The Fourier transform ${\cal G}(p,\omega)=\int dx dt e^{i\omega t -ipx}{\cal G}(x,t)$ of the correlation function
\begin{equation}
{\cal G}(x,t)=\langle :\rho^2(x,t)::\rho^2(0,0):\rangle
\label{rho-squared}
\end{equation}
should be evaluated for the Lieb-Liniger model, Eq.~(\ref{eq:LiebLiniger}). In terms of $\sigma_q(\omega)$, the inverse lifetime for a given momentum $q$ is given by
\begin{equation}
\frac{1}{\tau_q}=\int d\omega\sigma_q(\omega)\,.
\label{tau-q}
\end{equation}
The set of equations (\ref{sigma-q})--(\ref{tau-q}) allows one to evaluate the temperature dependence of $1/\tau_q$. The characteristic temperature scale for the variation of $1/\tau_q$ is the quasi-condensation temperature $T_s=\gamma^{1/2}\rho^2/m$, accessible experimentally~\cite{armijo11}. By the order-of-magnitude, this is the temperature at which the chemical potential of the 1D interacting bosons crosses zero.

In the limits of low ($T \ll T_s$) and high ($T \gg T_s$) temperatures, one may use the proper asymptotes of the correlation function Eq.~(\ref{rho-squared}) to evaluate $\sigma_q(\omega)$ and $1/\tau_q$. It turns out that in these limits the lifetime given by Eq.~(\ref{tau-q}) is controlled by scattering processes with energy transfer in a broad range, $\omega\lesssim\varepsilon_1(q)/2$ and is independent of $q$ \cite{tan10}
\begin{equation}
\frac{1}{\tau_q^{(0)}}=\frac{2}{3\sqrt{3}}\frac{\alpha^2 \rho^2}{m} g_2(T)\,.
\label{limits-tau}
\end{equation}
Here $g_2=\langle :\rho^2(0,0):\rangle/\rho^2$ is the two-particle correlation function (normalized by $\rho^2$). For weak interactions $g_2$ decreases monotonically with $T$ from $g_2=2$ at $T\gg T_s$ to $g_2=1$ at $T\ll T_s$ \cite{kheruntsyan03}. The presence of $g_2$ in Eq.~(\ref{limits-tau}) is due to the fact that the two particles receiving the energy $\omega$ in the three-particle collision must be at the same spot. 

A more detailed analysis actually indicates that Eq.~(\ref{limits-tau}) yields the dominant contribution to the relaxation rate only outside the range of temperatures
\begin{equation}
\gamma^{3/8}\left(\frac{\varepsilon_1(q)}{T_s}\right)^{1/4}\lesssim \frac{T}{T_s}
\lesssim\gamma^{-3/4}\left(\frac{T_s}{\varepsilon_1(q)}\right)^{1/2}\,.
\label{range}
\end{equation}
That range is broad as long as the energy of incoming quasiparticle is not too high, $\varepsilon_1(q)/T_s\ll 8/\gamma^{3/2}$, and includes some parts of temperature intervals where, respectively, $T\ll T_s$, and $T\gg T_s$. Within the range given by Eq.~(\ref{range}), relaxation is dominated by processes with small energy transfer, $|\omega|\lesssim\max\{T,\gamma \rho^2/m\}$. These processes are Bose-enhanced by the high occupation factors of the final states of the quasiparticles receiving the energy $\omega$. Finding the full dependence of $1/\tau_q$ on temperature within this range would require a full knowledge of the correlation function Eq.~(\ref{rho-squared}), which is still not available. However, matching the results obtained at low and high values of $T/T_s$, one may see that $1/\tau_q$ reaches a maximum \cite{tan10}
\begin{equation}
1/\tau_q^{\rm max}\approx \frac{\alpha^2 \rho^2}{m}\sqrt{\frac{T_s}{\gamma^{-3/2}\varepsilon_1(q)}}
\label{rate-max}
\end{equation}
at $T_{\rm max}\approx 1.6 T_s$. Under the assumed condition on $\varepsilon_1(q)$, the temperature dependence of $1/\tau_q$ is not monotonic, the maximal rate (\ref{rate-max}) significantly exceeds the limits given in Eq.~(\ref{limits-tau}). The kinetic equation accounting for the small-energy transfers effective in the temperature interval (\ref{range}) was considered by \textcite{mazets11}.

The relaxation of dark solitons in the interacting 1D Bose gas is very similar to the relaxation of holes in the interacting Fermi gas which we considered in Sec.~\ref{sec:WeakQP_Relax}. Dark solitons correspond to the excitations over the ground state with the minimal energy $\varepsilon_2(q)$ at given momentum $q$. The soliton velocity $v_s(q)$ at any $q$ is smaller than the sound velocity in the Bose gas, $v=(\rho/m)\sqrt\gamma$, see Eq.~(\ref{eq:boson_GP}). Therefore, the relaxation of solitons is possible only at finite temperatures in a Raman-like process in which two phonons are created (each of the two participating phonons replaces a low-energy particle-hole pair in the case of the fermionic hole relaxation, see Figs.~\ref{Fig_rate} and \ref{bottom-back}.)

The two-phonon processes lead to a typical momentum transfer $\Delta q\sim T/v$. The soliton's velocity is zero at $q=\pi\rho$, so the transferred energy in the elementary act of relaxation of a dark soliton is on the order of $T^2/(|M^*|v^2)\sim T^2/T_s$ and much smaller than $T$. Here $M^*=-4\rho/v$ is the soliton's (negative) effective mass; following \textcite{gangardt10}, we consider $T\ll T_s$. The smallness of the energy allows one to use the Fokker-Planck equation \cite{landau80} to describe the time evolution of the momentum distribution function $f(q)$ of a dark soliton. The problem is similar to the previously considered kinetics of a heavy particle in a linear LL~\cite{castroneto96} and to the diffusion of a deep hole in an interacting electron gas, see \eg, \textcite{matveev11b}. Using the notations of the latter work, we write the Fokker-Planck equation in the form of a continuity condition for the distribution function in the momentum space,
\begin{equation}
\partial_t f=-\partial_q J\,,\,\,\, J=-\frac{B(q)}{2}\left[\frac{1}{T} \frac{dE}{dq}+\partial_q\right]f
\label{FP-soliton}
\end{equation}
with $J$ being the corresponding current~\footnote{The function $B(q)$ at arbitrary interaction strength was recently evaluated \cite{matveev12b}.}.

The applicability of Eq.~(\ref{FP-soliton}) is confined to the vicinity of $q=\pi \rho$ in order to satisfy the requirement of the smallness of the energy transfer. Thus, one may use the expansion $\varepsilon_2(q) = \varepsilon_2(0) +(q-\pi \rho)^2/(2M^*)$ for the soliton's energy, and replace $B(q)$ by a constant, $B=B(q=\pi \rho)$. After that, the meaning of Eq.~(\ref{FP-soliton}) becomes quite clear: it describes the motion of a ``particle'' (dark soliton)  subject to a Langevin random force (yielding the second term in the brackets) and a viscous force
\begin{equation}
F=-\chi v\,,\,\,\, \chi=B/2T\,,\,\,\, v=\partial_q \varepsilon_2(q)\,.
\label{viscous}
\end{equation}
The viscous force leads to a particle acceleration as $M^*<0$ for the dark soliton. Once the soliton's velocity $v_s(q)$ becomes of the order of $v$, Eq.~(\ref{FP-soliton}) becomes invalid. However, if the initial velocity was on the order of the thermal one, $v_s(q)\sim\sqrt{|M^*|T}$, then the Fokker-Planck equation describes the longest part of relaxation process, which takes a time $\tau\sim |M^*|/\chi$.

The viscosity coefficient was evaluated by \textcite{gangardt10},\footnote{The numerical coefficient in Eq.~(\ref{chi}) corrects an error in the original publication \cite{kamenev_private}.}
\begin{equation}
 \chi (T)=\frac{128\pi^3}{1215}\frac{\alpha^2 \rho^2 m^2}{\gamma}\left(\frac{T}{ \sqrt{\gamma} T_s}\right)^2, \, T\ll \sqrt{\gamma} T_s\,.
\label{chi}
\end{equation}
To obtain this result one needs to include not only linear but also quadratic terms in $\partial_x\varphi$ and $\partial_x\theta$ in the interaction Hamiltonian of the quantum impurity with the LL describing the low-energy excitations. Such terms are necessary because they account for the three-particle collisions which are needed to capture two-phonon processes and the resulting soliton relaxation. At $q=\pi \rho$ the terms allowed by symmetries are
\begin{align}
 H_3=\int  dx \left [ V_{\theta \theta} (\partial_x \theta)^2 +  V_{\varphi \varphi} (\partial_x \varphi)^2 \right] d^\dag(x) d(x).
\end{align}
An extension of the phenomenological approach of Sec.~\ref{sec:phenomenology} leads to the coupling strengths
\begin{align}\label{eq:boson_relax_phen}
V_{\theta \theta} &=\frac12\left(\frac1m + \frac{\partial^2 \varepsilon_2}{\partial q^2}\right) ,\\
 V_{\varphi \varphi} &= \frac1{2\pi^2} \left(\frac{\partial^2 \varepsilon_2}{\partial \rho^2}+\frac{\partial^2 \mu}{\partial \rho^2}\right).
\end{align}
The evaluation then proceeds by removing from the Hamiltonian the terms linear in $\partial_x \varphi$ and $\partial_x \theta$ using the unitary transformation (\ref{eq:U_phen}), and then treating the remainder within perturbation theory. As was shown explicitly by \textcite{gangardt10}, the coefficients of the interaction Hamiltonian yield a vanishing relaxation rate for the integrable Lieb-Liniger model. In the weakly interacting regime, the lowest-order correction appears in the order $\alpha^2$ due to corrections to $\varepsilon_2(q)$ coming from three-particle interactions (\ref{three-body}). Later on, the outlined approach to the relaxation of dark solitons was generalized to the case of a depleton, the dressed impurity state in a quantum liquid \cite{schecter11}.

\subsection{Conductivity and Drude weight for interacting particles}
\label{sec:Drude}

So far in this section we concentrated on the relaxation of excitations. A related question which has attracted a lot of attention recently is the effect of such relaxation on transport, and specifically the relation between transport properties and integrability~\cite{sirker09,sirker11,jung05,jung07,grossjohann10,wu11,
znidaric11,heidrich-meisner03,roch00,herbrych11, prosen11,karrasch11}.

The current response $j(x,t)$ to a force $f(x,t)=f_{q,\omega}\exp (i\omega t-qx)e^{+0t}$ applied to a {\it linear} LL is easily evaluated by solving the corresponding classical equation for the density waves in the liquid. It yields for the Fourier components of the current
\begin{align}\label{jqomega}
j(q,\omega)=-i\frac{Kv}{\pi}\frac{f_{q,\omega}}{\omega-qv-i0}.
\end{align}
In a Galilean-invariant system, the factor $Kv/\pi$  equals  $\rho/m$, where the average particle density $\rho$ and the mass $m$ are independent of interactions. The linear response of the current to a force field applied to a spatially-homogeneous system is characterized by the conductivity, $j(q,\omega) = \sigma(q,\omega) f_{q,\omega}$, which is a complex function, $\sigma(q,\omega)=\sigma'(q,\omega)+i\sigma''(q,\omega)$. Its real component, $\sigma'(q,\omega)$, is the dissipative part of the response. The response of a Galilean-invariant system at $q=0$ is purely inertial and independent of interactions, since it is nothing but a center-of-mass motion caused by an applied force uniform in space. According to Eq.~(\ref{jqomega}), the corresponding dissipative part of conductivity, $\sigma'(\omega)=\sigma'(q,\omega)|_{q=0}$, is
\begin{align}\label{eq:drudepeak}
 \sigma'(\omega)=2\pi D\delta(\omega).
\end{align}
The magnitude of the ``Drude peak'' $D$ in the conductivity $\sigma'(\omega)$ is given by $D=\rho/(2m).$ In the presence of a periodic lattice potential, $\sigma(q,\omega)$ is still well-defined
for wave vectors much smaller than the size of the Brillouin zone~\cite{sirker11}.
By continuity equations, the conductivity is related to the susceptibility; the dissipative part of the latter is related to the DSF by the fluctuation-dissipation theorem~\cite{doniach98}. Therefore, for small $q$
\begin{equation}
S(q,\omega)=\frac{2q^2}{\omega(1-e^{-\beta\omega})}\sigma'(q,\omega).
\label{real-sigma}
\end{equation}
As we have seen in the previous parts of the review, the delta peak in $S(q,\omega)$ at finite wave vectors $q$ becomes broadened in a nonlinear LL, and according to Eq.~(\ref{real-sigma}) leads to a finite width of the peak in the dissipative conductivity. At $T=0$, the peak at $\omega=vq$ has a width which scales as a higher power of wave vector, $\propto q^2$ in the absence of particle-hole symmetry or $\propto |q|^3$ in the presence of the symmetry~\cite{khodas07b,pereira06}. In either case, taking the limit $q \to 0$ one recovers the Drude peak in the conductivity in the limit $q\to 0$.

At $T \neq 0$, the universal nonlinear LL theory leading to Eq.~(\ref{eq:S_temp}) is insufficient for understanding the fate of the Drude peak. Indeed, the applicability of Eq.~(\ref{eq:S_temp}) in the limit $q\to 0$  requires that $T$ scales to zero not slower than $q$; one is not allowed to take the limit $q\to 0$ at fixed $T$. However, in a Galilean invariant system, the existence of a Drude peak is protected even at finite temperatures, since the constant uniform external field causes the same center-of-mass motion irrespective of the temperature.

The real part of the conductivity (at $q=0$) can be written as \cite{sirker11}
\begin{equation}
\sigma^\prime(\omega) =2\pi D\delta(\omega)+\sigma_{\rm reg}(\omega).
\label{drude-weight}
\end{equation}
Invoking the notion of a finite relaxation time $\tau$, one may expect $\sigma^\prime(\omega)\propto (1/\tau)/[\omega^2+(1/\tau)^2]$. In order to reproduce the correct zero-temperature limit, we have to set $1/\tau(T)=0$ at $T=0$. In a generic system, one expects $1/\tau(T)$ finite at $T\neq 0$ which means zero Drude weight, $D(T)=0$ at any finite temperature. In the special case of {\it integrable} models, one may conjecture $1/\tau(T)=0$ and consequently $D(T)\neq 0$ even at finite temperature.

One way of checking this conjecture relies on the rigorous Mazur inequality \cite{zotos97,mazur69}
\begin{equation}
D\geq\frac{1}{2LT}\sum_k\frac{\langle{\cal I}Q_k\rangle^2}{\langle Q_k^2\rangle^2}\,.
\label{zotos}
\end{equation}
Here $L$ is the length of the system, ${\cal I}$ is the spatial integral of the current density operator, and the operators $Q_k$ form a set of commuting conserved quantities, orthogonal to each other ($\langle Q_l Q_k\rangle\propto\delta_{kl}$). Moreover, if the set includes {\it all} conserved quantities $Q_k$, then equality is reached in Eq.~(\ref{zotos}) \cite{suzuki71}. Finding at least one conserved quantity $Q_n$ with a nonzero overlap with ${\cal I}$ allows one to prove $D(T)\neq 0$ at finite temperature. This is the case, for example, for charge transport in the Hubbard model away from half-filling and spin transport in the $S=1/2$ XXZ model at finite magnetic field, where a proper local conserved quantity can be found. At zero field, all local conserved quantities for the XXZ model are even under the transformation $S_j^z \to - S_j^z$, $S_j^\pm \to S_j^\pm$, while $\cal{I}$ is odd, and the corresponding overlaps equal zero. It seems that a conserved quantity which has finite overlap with $\cal{I}$ was recently constructed \cite{prosen11}, supporting the claim of finite $D(T)$ for the XXZ model at zero field. Recent numerical results seem to support this claim \cite{karrasch11}.

Alternatives to the investigation methods based on Mazur's inequality are reviewed in a recent excellent paper by \textcite{sirker11}.

\subsection{Conductance of interacting fermions in 1D}
\label{sec:Conductance}

The conductance of ballistic quantum wires adiabatically connected to leads is quantized in units of $e^2/(\pi\hbar)$ at low temperatures. This experimental observation \cite{vanwees88,wharam88} was first understood in terms of adiabatic transport of free fermions \cite{glazman88}. It was realized later that interactions, taken into account within the framework of the linear LL theory, do not affect the quantization of adiabatic transport \cite{maslov95,safi95,ponomarenko95}. Finite-temperature corrections to the quantized conductance, whether in the picture of free or interacting fermions, are associated with the electron states near the bottom of the conduction band. In the case of free fermions the correction is easily evaluated and shows no dependence on the wire length $L$, but an activated temperature dependence $\delta G\propto e^{-\epsilon_F/T}$, see Eq.~(\ref{G-ballistic}) below. Interactions do not alter the activated nature of the temperature dependence for relatively short wires \cite{lunde07}. In sufficiently long wires, however, equilibration facilitated by the scattering of holes near the bottom of the band, see Fig.~\ref{bottom-back}, ultimately leads to a much bigger correction, $\delta G\propto T^2$ \cite{rech09}. We will mostly concentrate on the transport of spinless fermions below; one may think of fully spin-polarized electrons, the corresponding conductance quantum is $e^2/(2\pi\hbar)$.

\begin{figure}[t]
\includegraphics[width=0.99\columnwidth]{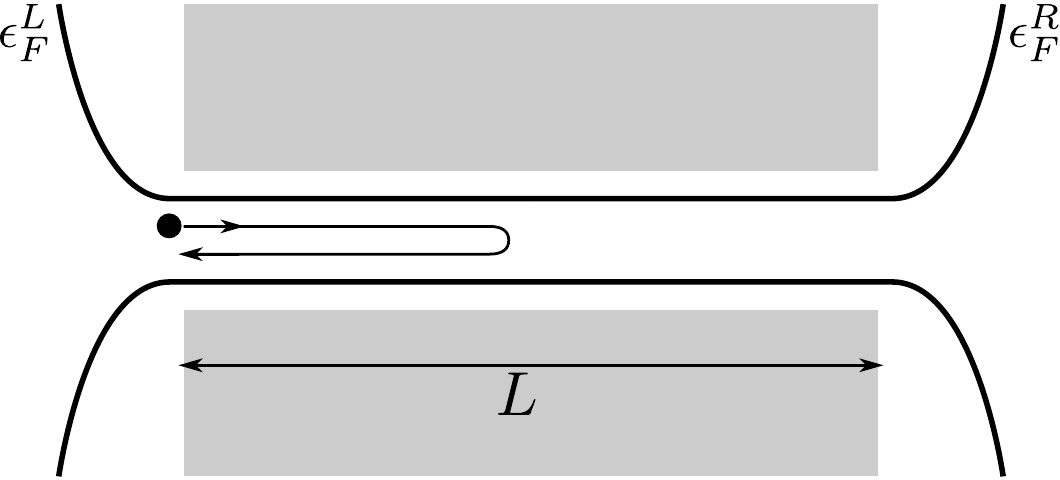}
\caption{Schematic picture of the quantum wire of length $L$ which is formed by confining a 2D  electron gas with gates (dark regions). Electrons in the left and right leads are described by Fermi distribution functions characterized by a temperature $T$ and chemical potentials $\epsilon_F^L$ and $\epsilon_F^R$, respectively. As particles propagate from the left lead to the right one, some get reflected due to the momentum-conserving electron-electron interaction, creating $dN_R/dt\neq 0$, see Eq.~(\ref{NR-dot}). Adapted from \textcite{micklitz10a}.}
\label{wire}
\end{figure}

In the absence of interactions, the distribution functions of fermions in the wire, see Fig.~\ref{wire}, keep the memory of the distribution in the lead they originated from,
\begin{equation}
f^{(0)}(k)=\frac{\theta(k)}{\exp(\xi^L(k)/T)+1}+\frac{\theta(-k)}{\exp(\xi^R(k)/T)+1}.
\label{free-el}
\end{equation}
Here $\xi^{L,R}(k)=k^2/(2m)-\epsilon_F^{L,R}$ are the energies of electrons coming from the left ($L$) or right ($R$) leads, respectively. The difference between the chemical potentials is determined by the bias $V$ applied to the wire, $\epsilon_F^L=\epsilon_F^R+eV$. Evaluating the current $I=e\int (dk/2\pi\hbar)(k/m)f^{(0)}(k)$ at small bias ($V\to 0$), one easily finds
\begin{equation}
G_0=\frac{e^2}{2\pi\hbar}\frac{1}{e^{-\epsilon_F/T}+1},
\label{G-ballistic}
\end{equation}
which is equal to the quantum $e^2/(2\pi\hbar)$, up to a correction proportional to $e^{-\epsilon_F/T}$ , which is exponentially small at low temperatures ($\epsilon_F^L=\epsilon_F^R=\epsilon_F$ at $V=0$).

The presence of a current $I\neq 0$ means that there is some finite average velocity of electrons in the wire. An equilibrium distribution function in the rest frame, at given drift velocity, chemical potential, and temperature ($u$, $\tilde T$, and $\tilde\mu$, respectively) would be
\begin{equation}
f(k)=\frac{1}{\exp [k^2/(2m)-uk-\tilde\mu]/\tilde T+1}
\label{Galilean-f}
\end{equation}
At $T=\tilde T=0$, the distribution function (\ref{free-el}) may be brought to the form of Eq.~(\ref{Galilean-f}). Therefore, we do not expect equilibration (caused by electron-electron interaction) to bring any corrections to the ballistic conductance at $T=0$. Besides, each of the two parts of Eq.~(\ref{free-el}) represents a distribution describing equilibrium separately within the left- and right-movers with $N_L\neq N_R$. At $T=0$, the relaxation rate $1/\tau_N=0$, and the distribution (\ref{free-el}) is stable. However, at finite temperature the processes depicted in Fig.~\ref{bottom-back} cause a redistribution between $N_R$ and $N_L$. To the lowest order in $1/\tau_N$, the correction $\delta G\sim L/(v_F\tau_N)$ to the ballistic conductance (\ref{G-ballistic}) increases linearly with the wire length $L$ \cite{lunde07,micklitz10a}. This defines a new characteristic equilibration length, $l_{\rm eq}\sim v_F\tau_N \propto e^{\epsilon_F/T}$. The resulting resistance reduces $G$ by an $L$-independent amount, $\delta G\sim -(e^2/\hbar)(T/\epsilon_F)^2$ which can be found  essentially from particle number and energy conservation laws \cite{rech09}. The initial consideration of weak interactions \cite{rech09} was generalized later to the case of strong~\cite{matveev10} and arbitrary~\cite{matveev11b} interactions. We address next that latest development.

The Hamiltonian (\ref{eq:HLL}) in a finite-size system written in terms of individual bosonic modes reads~\cite{haldane81a}
\begin{equation}
H=\sum_q v|q|b_q^\dagger b_q + \frac{\pi}{2L}\left[v_N N^2+v_JJ^2\right],
\label{H-LL}
\end{equation}
while its momentum is
\begin{equation}
P=k_FJ+\sum_q qb^\dagger_q b_q,
\label{P-LL}
\end{equation}
where $v_J = v_F$  and  $v_N =v_F/K^2$ as a consequence of Galilean invariance. An eigenstate of the system is described by boson occupation numbers, and the total numbers of the left- and right-movers with respect to the ground state, $N_{R,L}=(N\pm J)/2$. It is worth noting that an excitation of the boson modes does not contribute to the current: a creation of bosons corresponds to exciting particle-hole pairs {\it within} the branches of left- or right-movers. So, the electric current $I$ is related by $I=ev_J(J/L)$ to the difference $N_R-N_L$ \cite{haldane81a}. On the other hand, conservation of energy and momentum determine the form of the equilibrium distribution, $e^{-(H-uP)/T}/Z$ with some parameters $u$ and $T$ (here $Z$ is the proper partition function). Using here Eqs.~(\ref{H-LL}) and (\ref{P-LL}), we find the average value of $J$ to be $\pi k_F L u/v_J$. Using that together with the relation between $I$ and $J$, we see that $u$ is nothing but the drift velocity:
\begin{equation}
u=\frac{I}{e} \frac{1}{\pi k_F}\,.
\label{u-drift}
\end{equation}
In equilibrium, the very same velocity ``shifts'' the boson distribution function:
\begin{equation}
n(q)=\langle b_q^\dagger b_q\rangle=\left[e^{(v|q|-uq)/T}-1\right]^{-1}.
\label{u-nq}
\end{equation}
That distribution does not carry any charge, but does create energy current,
\begin{equation}
j_E=(\pi/3)(T^2/v)u\,.
\label{jE}
\end{equation}

As we already discussed, {\it at zero temperature} the distribution $e^{-(H-uP)/T}/Z$ with a finite drift velocity can be viewed as two counter-propagating fluxes of particles with different chemical potentials, resulting in the quantized conductance $G_0=e^2/(2\pi\hbar)$ \cite{maslov95,safi95,ponomarenko95}. At finite temperatures, the accommodation of the distributions in the leads to the drifting one in the wire may cause backscattering, see Fig.~\ref{wire}, which leads to relation
\begin{equation}
I=G_0V+e\frac{dN_R}{dt}.
\label{NR-dot}
\end{equation}
The crucial observation is that $dN_R/dt$ is related to the energy redistribution between right- and left-movers \cite{rech09,matveev11}. Indeed, backscattering of a right-mover corresponds to a change $\Delta N_R=-1$, which in the limit of $u\to 0$ (linear response regime) does not affect the total energy, see Eq.~(\ref{H-LL}). At the same time, by momentum conservation, bosons must acquire the additional momentum of $2k_F$, see Eq.~(\ref{P-LL}). This is only possible if momenta and energies given by $\Delta P_{R,L}=k_F$ and $\Delta E_{R,L}=\pm vk_F$ are transferred to the left- and right-moving bosons. Therefore,
\begin{equation}
\frac{dE_R}{dt}=-vk_F\frac{dN_R}{dt}\,.
\label{ER-dot}
\end{equation}
By energy conservation, one has
\begin{equation}
j_E=\frac{dE_R}{dt}.
\label{jE-NR}
\end{equation}
Equations~(\ref{jE-NR}), (\ref{ER-dot}), and (\ref{jE}) relate $dN_R/dt$ to the drift velocity $u$. Using then relations (\ref{u-drift}) and (\ref{NR-dot}), one finds the corrected conductance, \cite{matveev11}
\begin{equation}
G=G_0\left(1-\frac{\pi^2}{3}\frac{T^2}{v^2k_F^2}\right)\,.
\label{G-eq}
\end{equation}

A full equilibration of the bosons to the distribution (\ref{u-nq}) requires energy equilibration and the adjustment of the velocity to the correct drift value (\ref{u-drift}). As we discussed for the example of weak interactions, see Eqs.~(\ref{II.C.3})--(\ref{tau-ph}), the energy equilibration rate scales as some power of temperature, while the adjustment of  the velocity requires a variation of $N_R-N_L$. The corresponding rate has an activated temperature dependence and happens on a much slower scale, see Eq.~(\ref{tau-N}) and the discussion around it. Therefore, there is an exponentially wide interval of wire lengths $L$ for which full equilibration of the energy does occur, but $u(L)$ does not reach the value (\ref{u-drift}). Considering $u(L)$ as an adjustable parameter replacing the distribution (\ref{u-nq}), and using the momentum conservation Eq.~(\ref{P-LL}), one finds \cite{matveev11}
\begin{equation}
\frac{1}{L}\frac{dN_R}{dt}=\frac{\pi}{3}\frac{T^2}{v^2k_F}\frac{u(L)-u}{l_{\rm eq}}
\label{uofL}
\end{equation}
which generalizes Eq.~(\ref{G-eq}) to finite values of $L/l_{\rm eq}$,
\begin{equation}
G=G_0\left(1-\frac{\pi^2}{3}\frac{T^2}{v^2 k_F^2}\frac{L}{L+l_{\rm eq}}\right)\,.
\label{G-neq}
\end{equation}

The evaluation of the equilibration length $l_{\rm eq}=2v\tau_{N}$ is beyond the linear LL theory, and should account for the processes involving holes at the bottom of the band, see Fig.~\ref{bottom-back}. Like in the case of dark solitons considered in Sec.~\ref{sec:relax_bosons}, this problem can be reduced to the one of the kinetics of a mobile impurity and entails a solution of the Fokker-Planck equation (\ref{FP-soliton}). It yields \cite{matveev11b}
\begin{equation}
\frac{1}{l_{\rm eq}}=\frac{(3/2)k_F^2B}{\pi^2v\sqrt{2\pi M^*T}}\left(\frac{v}{T}\right)^3
e^{\varepsilon_{\rm th}(0)/T}\,,
\end{equation}
where the parameters $\varepsilon_{\rm th}(0) < 0$ and $M^*>0$ are determined by the energy spectrum of the impurity, $\varepsilon_{\rm th}(k)\approx \varepsilon_{\rm th}(0) + k^2/(2M^*)$, and depend on the equilibrium electron density $\rho$ in the wire. The extension of the phenomenological treatment of Sec.~\ref{sec:phenomenology} aimed at including the Raman scattering processes, see Eqs.~(\ref{eq:boson_relax_phen}), allows to express the coefficient $B$ in terms of the functions $\partial \varepsilon_{\rm th} (0) / \partial \rho$ and $\partial v / \partial \rho$:
\begin{align}
B&=\frac{4\pi^3\rho^2T^5}{15 m^2v^8}\Big(- \frac{d^2\varepsilon_{\rm th}(0)}{d\rho^2}
+\frac{2}{v}\frac{dv}{d\rho}\frac{d\varepsilon_{\rm th}(0)}{d\rho} \notag \\
&+\frac{(d\varepsilon_{\rm th}(0)/d\rho)^2}{M^*v^2}\Big)^2.
\end{align}

The approach leading to Eq.~(\ref{G-neq}) was also applied to spin-$1/2$ fermions \cite{matveev11}. In the most interesting case of strong interactions ($v_s\ll v_c$) the result is
\begin{equation}
G=\frac{e^2}{\pi\hbar}\left(1-\frac{\pi^2}{6}\frac{T^2}{v_s^2 k_F^2}\frac{L}{L+l^{(s)}_{\rm eq}}\right)\,,
\notag
\end{equation}
but the corresponding equilibration length $l^{(s)}_{\rm eq}$ for spinons has not been evaluated yet.

\section{CONCLUSIONS}

The linear Luttinger liquid (LL) theory has been in use for decades by now, and provides an effective tool to describe the low-energy properties of 1D quantum liquids
in terms of quantized linear sound modes. Despite its spectacular successes, the linear LL theory
has limitations constraining its applicability even in the
low-energy physics of quantum 1D systems. Replacing
the generic spectrum of particles with a linear one does affect
qualitatively the momentum-resolved dynamic responses and introduces an artificial particle-hole symmetry. Furthermore, being a free-field theory, the linear LL description is devoid of any intrinsic mechanisms of relaxation and equilibration.

This review exhibits ways of studying 1D quantum liquids outside
these limitations and beyond low energies. For this purpose,
the representation of the linear LL theory in terms of the fermionic quasiparticles introduced by~\textcite{mattis65} turns out to be more beneficial than the standard bosonization treatment.
The quasiparticles share many features with their counterparts in the Fermi liquid theory.
If the constituent particles of the liquid have a nonlinear dispersion relation, so do the fermionic quasiparticles. Similarly to the Fermi liquid theory, the interactions between the quasiparticles are weak.
That, in principle, should allow one to build a full kinetic theory valid at low energies. The difference from the Fermi liquid theory comes in the relation between the measurable degrees of freedom and the
quasiparticles: the corresponding transformation is rather nonlinear. While a
nonperturbative kinetic theory based on the fermionic quasiparticle representation has not been developed yet, some elementary relaxation processes in a nonlinear LL are understood.

The realization of links between the physics of a nonlinear LL and the Fermi liquid theory
makes available an arsenal of methods existing in the latter. One of them, the theory of the Fermi edge singularity, facilitated the development of new methods for the evaluation of the singularities in the dynamic response functions of 1D quantum liquids.
The new paradigm which emerged is the description of the many-body dynamics in terms of effective models of quantum impurities moving in linear LLs.

We reviewed the existing tools for the investigation of nonlinear LLs and some results obtained with these tools. As we mentioned, building a kinetic theory of 1D liquids remains an open question. Other questions closely related to the review include the kinetic theory of weakly-nonintegrable systems relevant for cold atomic gases, and the dynamics of edge states (chiral and helical) relevant for electrons in solids. The field remains wide open beyond these few problems, with a variety of practically and conceptually important questions to be answered.

\acknowledgments

The authors wish to thank A.~V.~Andreev, V.~V.~Cheianov, A.~Kamenev, T.~Karzig, A.~Levchenko, K.~A.~Matveev, and R.~G.~Pereira for discussions and comments which were helpful in writing this review. We acknowledge the support of the NSF under Grants No.~0906498 and 1066293, and the hospitality of the Aspen Center for Physics and of the Nanosciences Foundation (Grenoble, France). AI also acknowledges support from the Texas NHARP Grant No.~01889 and the A.~P.~Sloan Foundation. TLS acknowledges support from the Swiss NSF.

\bibliographystyle{apsrmp4-1}
\bibliography{imambekov}

\end{document}